\newcommand*{\ATLASLATEXPATH}{}
\documentclass[PAPER, cernpreprint, atlasdraft=false, texlive=2016, USenglish, english, block=none]{\ATLASLATEXPATH atlasdoc}
 
\usepackage[subfig]{\ATLASLATEXPATH atlaspackage}
\usepackage{amsmath}
 
\usepackage{\ATLASLATEXPATH atlasbiblatex}
 
\usepackage{\ATLASLATEXPATH atlasphysics}
 
\graphicspath{{logos/}{figures/}}
 
\usepackage{ANA-EXOT-2017-14-PAPER-defs}
\usepackage{afterpage}

\AtlasTitle{Search for pair production of heavy vector-like quarks
decaying into hadronic final states in $pp$ collisions at $\sqrt{s}$ =
13 TeV with the ATLAS detector}
 
\AtlasVersion{3.0}
 
\AtlasAbstract{
A search is presented for the pair production of heavy vector-like
quarks, $T\bar T$ or $B\bar B$, that decay into final states with jets
and no reconstructed leptons. Jets in the final state are classified
using a deep neural network as arising from hadronically decaying
$W/Z$ bosons, Higgs bosons, top quarks, or background. The analysis
uses data from the ATLAS experiment corresponding to 36.1 fb$^{-1}$ of
proton--proton collisions with a center-of-mass energy
of $\sqrt{s}$ = 13 TeV delivered by the Large Hadron
Collider in 2015 and 2016. No significant deviation from the Standard
Model expectation is observed. Results are interpreted assuming the
vector-like quarks decay into a Standard Model boson and a
third-generation-quark, $T\rightarrow Wb,Ht,Zt$ or $B\rightarrow
Wt,Hb,Zb$, for a variety of branching ratios.  At 95\% confidence
level, the observed (expected) lower limit on the vector-like
$B$-quark mass for a weak-isospin doublet ($B, Y$) is 950 (890) GeV,
and the lower limits on the masses for the pure decays $B\rightarrow
Hb$ and $T\rightarrow Ht$, where these results are strongest, are 1010
(970) GeV and 1010 (1010) GeV, respectively.  }
 
\author{The ATLAS Collaboration}
 
\AtlasRefCode{ANA-EXOT-2017-14}
 
\PreprintIdNumber{CERN-EP-2018-176}

\AtlasJournalRef{\PRD 98 (2018) 092005}
\AtlasDOI{10.1103/PhysRevD.98.092005}

\AtlasCoverSupportingNote{ATL-COM-PHYS-2017-675}{https://cds.cern.ch/record/2266580}
 
\AtlasCoverCommentsDeadline{26 June 2018}
 
\AtlasCoverAnalysisTeam{Ece Akilli, Tobias Golling, Joseph Haley, Clement Helsens, David Olivier Jamin, Kevin Kr\"oninger, Madhuranga Madugoda Ralalage Don, Allison McCarn Deiana, Daniel Marley, Garrett Merz, Olaf Nackenhorst, Thomas Andrew Schwarz}
 
\AtlasCoverEdBoardMember{Cigdem Issever~}
\AtlasCoverEdBoardMember{Peter Loch (chair)}
\AtlasCoverEdBoardMember{Lo\"ic Val\'ery (chair)}

\AtlasCoverEgroupEditors{atlas-ANA-EXOT-2017-14-editors@cern.ch}
 
\AtlasCoverEgroupEdBoard{atlas-ANA-EXOT-2017-14-editorial-board@cern.ch}
\hypersetup{pdftitle={ATLAS document},pdfauthor={The ATLAS Collaboration}}
 
\begin{document}
 
\maketitle
 
\tableofcontents
 
\section{Introduction}
 
Many theories beyond the Standard Model (SM) are motivated by the
naturalness problem~\cite{natproblem}, and are intended to resolve the
quadratic divergences in the radiative corrections to the Higgs-boson
mass.  Several extensions to the SM, such as Little
Higgs~\cite{ArkaniHamed:2002qy,Schmaltz:2005ky} and Composite
Higgs~\cite{Kaplan:1983sm,Agashe:2004rs} models, have been proposed to
address this issue. A common feature of these models is the existence
of \TeV{}-scale vector-like quarks (VLQs) that couple preferentially
to third-generation SM quarks~\cite{Aguilar-Saavedra:2013qpa}.
 
VLQs are spin-$1/2$ colored fermions with left--right symmetric
transformation properties under the weak-isospin SU(2) gauge
group. Unlike chiral quarks, which obtain mass through electroweak
symmetry breaking \cite{Higgs:1964ia,PhysRevLett.13.321,
PhysRevLett.13.508, PhysRevLett.13.585,PhysRev.145.1156,
PhysRev.155.1554}, VLQs can have a gauge invariant mass term
$m\bar{\psi}\psi$. Therefore, VLQs are not subject to the constraints
from Higgs production which highly disfavor additional chiral
quarks~\cite{Djouadi:2012ae,Eberhardt:2012gv,HIGG-2012-27,CMS-HIG-12-028,}.
VLQs also couple to flavor-changing neutral currents, so a
charge\footnote{Electric charge is measured in units of $e$.} $+2/3$
vector-like partner of the top quark, $T$, could decay\footnote{It is
assumed that the VLQs decay only into SM particles, and couple to only
third-generation quarks.} into $Wb$, $Zt$ or $Ht$, while a charge
$-1/3$ bottom quark partner, $B$, could decay into $Wt$, $Zb$, or
$Hb$~\cite{VLFermions_1990, AguilarSaavedra:2009es, vlqmodel, Harigaya:2012ir}. The branching ratios depend on the VLQ
mass and weak-isospin multiplet. Vector-like $T$ and $B$ can occur
alone in a singlet scenario. Doublet and triplet scenarios also allow
for more exotic $X$ and $Y$ VLQs with charges $+5/3$ and $-4/3$,
respectively. Charge conservation requires these to decay only via
$X\to Wt$ and $Y\to Wb$. Because this search has not been optimized
for $X$ and $Y$ vector-like quarks, they will not be discussed in this
paper.
 
Many previous searches for pair-produced VLQs by ATLAS and CMS at
$\sqrt{s} =$~8\,\TeV{}~\cite{EXOT-2013-16,
EXOT-2013-17,
EXOT-2013-18,
EXOT-2014-10,
CMS-B2G-12-015,
CMS-B2G-12-012}
and $\sqrt{s} =$~13\,\TeV{}~\cite{EXOT-2016-15,
EXOT-2016-14,
ATLAS_WtX,
ATLAS_ZtX,
CMS-B2G-15-006,
CMS-B2G-17-003,
CMS-B2G-16-024,
CMS_VLQ13_lep}
have focused on final states with one or more leptons. Additionally,
previous results from CMS at $\sqrt{s} =$~8\,\TeV{} and ATLAS at $\sqrt{s} =$~13\,\TeV{}
have included fully hadronic as well as leptonic final states~\cite{CMS-B2G-13-005,CMS-B2G-13-006, EXOT-2016-13}.
The previous fully-hadronic search by ATLAS~\cite{EXOT-2016-13} only focuses on the high missing transverse momentum (\met{}) region ($\met{} >$~200~GeV).
The analysis presented in this paper searches for heavy VLQs produced in pairs
and decaying into fully hadronic final states in the low-\met{} region ($\met{} <$~200~GeV).  This channel is
complementary to those used in previous ATLAS VLQ searches and is
particularly powerful for the $B\rightarrow Hb$ decay mode, which is
difficult to probe with leptonic final states.
 
\section{ATLAS detector}
The ATLAS detector~\cite{PERF-2007-01} at the Large Hadron Collider
(LHC) is centered on the collision point and covers nearly the entire
solid angle.\footnote{\AtlasCoordFootnote} It consists of an inner
tracking detector surrounded by a \SI{2}{\tesla} superconducting
solenoid, electromagnetic and hadronic calorimeters, and a muon
spectrometer incorporating three large superconducting toroid magnets.
The inner detector, including the insertable B-layer installed in
2014~\cite{CERN-LHCC-2012-009,ATLAS_IBL}, provides charged-particle
tracking information from a pixel and silicon microstrip detector in
the pseudorapidity range $|\eta| < 2.5$ and a transition radiation
tracker covering $|\eta| < 2.0$.
 
The calorimeter system covers the pseudorapidity range $|\eta| < 4.9$
and measures the positions and energies of electrons, photons, and
charged and neutral hadrons.  Within the region $|\eta|< 3.2$,
electromagnetic calorimetry is provided by barrel and endcap
high-granularity lead and liquid-argon sampling calorimeters.  The
hadronic sampling calorimeter uses either scintillator tiles or liquid
argon as active material and steel, copper or tungsten as absorber.
 
The muon spectrometer comprises separate trigger and high-precision
tracking chambers measuring the tracks of muons in a magnetic field
generated by superconducting air-core toroid magnets.  The precision
chamber system covers the region $|\eta| < 2.7$ , while the muon
trigger system covers the range $|\eta| < 2.4$.
 
A two-level trigger system is used to select which events to save for
offline analysis~\cite{TRIG-2016-01}.  The first level is
implemented in hardware/firmware and uses a subset of the detector
information to reduce the event rate from \SI{40}{\MHz} to less
than \SI{100}{\kHz}.  This is followed by the software-based high-level
trigger that reduces the event rate to
approximately \SI{1}{\kHz}.
 
 
\section{Data and simulated events}
\label{sec:Samples}
 
The data analyzed correspond to $pp$ collisions with a center-of-mass
energy of $\sqrt{s}=13$\,\TeV{} recorded by the ATLAS detector in 2015
and 2016.  Data quality requirements ensure that all components of the
detector were functioning.  The full data set corresponds to an
integrated luminosity of \lumi{}.
 
The primary background for this search is multi-jet events, followed
by \ttbar{} events and minor contributions from single-top-quark
and \ttbarX{} (X=$W,Z,H$) events.  The multi-jet background is
estimated using a data-driven method (\Sect{\ref{sec:backgrounds}}),
while signal events and other background contributions were simulated
via Monte Carlo (MC) generation of LHC collisions that are then passed
through a \textsc{Geant}4 simulation~\cite{GEANT4} of the ATLAS
detector~\cite{SOFT-2010-01}.  All simulated events are reconstructed using
the same analysis chain as the data.  In all MC samples, the top-quark
and Higgs-boson masses were set to 172.5\,\GeV{} and 125.0\,\GeV{},
respectively, and the \textsc{EvtGen} v1.2.0 program~\cite{EvtGen} was
used to simulate the properties of bottom and charm hadron decays.

Simulated events of VLQ pair production, \QQbar{}, were produced with
the leading-order (LO) generator \PROTOS
v2.2~\cite{protos,AguilarSaavedra:2009es} using the NNPDF2.3 LO parton
distribution function (PDF) set~\cite{pdf3} and passed to
\PYTHIAV{8.186}~\cite{Pythia81} for parton showering and fragmentation.
The A14~\cite{ATL-PHYS-PUB-2014-021} set of tuned parameters is used.
VLQs were produced for the isospin singlet scenario with a narrow width
and for masses between 700 and 1200\,\GeV{} in steps of 50\,\GeV{},
with additional events produced at 500, 600, 1300, and 1400\,\GeV{}.
Additional samples were produced assuming a doublet scenario for VLQ
masses of 700, 950 and 1200\,\GeV{}, in order to study differences
from the different chirality of VLQs arising in singlet and
doublet models.

The pair production cross section varies from $3.38\pm0.25$ pb
($m_{Q}=500$\,\GeV{}) to $3.50\pm0.43$ fb ($m_{Q}=1400$\,\GeV{}),
computed using \texttt{top++} v2.0~\cite{Czakon:2011xx} at
next-to-next-to-leading order (NNLO) in QCD, including resummation of
next-to-next-to-leading logarithmic (NNLL) soft-gluon terms, and using
the MSTW 2008 NNLO set of PDFs~\cite{MSTW}.
Theory uncertainties are estimated by variations of the factorization and renormalization scales and by taking uncertainties of the PDF and strong coupling constant, $\alpha_\text{S}$, into account.
The latter two represent the largest
contribution to the overall theoretical uncertainty in the predicted
cross-section and are calculated using the PDF4LHC~\cite{pdf4lhc}
prescription with the MSTW 2008 68\% CL NNLO, CT10
NNLO~\cite{pdf1,pdf2} and NNPDF2.3 5f FFN PDF sets.
 
The \ttbar{} events were generated using \POWPYTHIAV{v2}{8.210}~\cite{Powheg, Powheg_tt}
with the CT10 NLO PDF set and the {Perugia2012} set of tuned
parameters~\cite{Perugia} for parton showering. The NLO radiation
factor, $h_\text{damp}$, was set to $1.5 m_\text{top}$. The \ttbar{}
background is split into \ttbar{}~+~light-flavor jets (\ttlight)
and \ttbar{} + $c$- or $b$-flavor jets (\tthf).  Single-top-quark production
($Wt$ and $t$-channel) was generated using
\POWPYTHIAV{v1}{6.428}~\cite{Powheg_Wt,Powheg_st,Pythia6} and the {Perugia2012} set of tuned
parameters for parton showering and the CT10 NLO PDF set. The \ttbarV{}
($V$=$W,Z$) and \ttbarH{} background was modeled using
\MGMCatNLO\,{v2.3.2}~\cite{Madgraph} as the
generator in LO precision with up to two additional partons and in NLO
precision, respectively.  The parton showering and fragmentation is
performed using \PYTHIAV{8.210}~\cite{Pythia81} (\PYTHIAV{8.186})
for \ttbar+$Z$ and \ttbarH{} (\ttbar+$W$).  The contribution from
single-top-quark and \ttbarX{ events is less than $6\%$ in all signal
regions.
 
Finally, although a data-driven method is used to estimate the
multi-jet background, a sample
of simulated multi-jet events is used for the training of an algorithm
employed to identify boosted objects
(\Sect{\ref{sec:ObjectDefinitions}}).  The simulated multi-jet events
were produced with \PYTHIAV{8.186} using the A14 set of tuned
parameters for the underlying event and the NNPDF2.3 LO PDFs.  The
renormalization and factorization scales were set to the average
transverse momentum (\pt{}) of the two leading jets.
 
 
\section{Object definitions}
\label{sec:ObjectDefinitions}
 
The main objects used in this search are small-radius (\smallR{}) jets
reconstructed from clusters of energy deposited in the calorimeter.  A
variable-radius re-clustering algorithm~\cite{Nachman2015,Krohn:2009zg} is then used to find groups
of \smallR{} jets that are consistent with the hadronic decays of
high-momentum bosons and top quarks.  To ensure orthogonality with
ATLAS VLQ searches that include
leptons~\cite{EXOT-2016-15,EXOT-2016-14,EXOT-2016-13,ATLAS_WtX,ATLAS_ZtX}, events containing
electrons or muons with \pt{} $>$ 20\,\GeV{} are vetoed using the same tight object
definitions as in those searches.  For a given reconstructed event,
the missing transverse momentum, with magnitude \etmiss{}, is calculated from the
negative vector sum of the \pt{} of all reconstructed jets, and any reconstructed electrons
and muons.  
A soft energy term is included to account for non-reconstructed particles originating from the hard scatter.
It is calculated using only charged tracks matched to the primary vertex to reduce contamination from
particles originating from other $pp$ interactions in the same or nearby bunch crossings
(pileup)~\cite{PERF-2016-07}. 
 
\SmallR{} jets are reconstructed from calibrated topological energy
clusters in the calorimeter using the \antikt{}
algorithm~\cite{antikt1,antikt2} with a radius parameter of 0.4.  They
are required to have $\pt{}>25$\,\GeV{} and $|\eta|<2.5$.  Low-$\pt{}$
jets produced in pileup interactions
are suppressed using the Jet Vertex
Tagger (JVT) algorithm~\cite{ATLAS-CONF-2014-018}.  A jet is removed from the event if
it has $\pt{}<60$\,\GeV{}, $|\eta|<2.4$, and a JVT value lower than
0.59. This requirement on the JVT value has an efficiency of 92\%
for jets of $\pt{}<60$\,\GeV{} and $|\eta|<2.4$ originating from the primary vertex.
In order to avoid misidentification and overlap of
objects, a jet is removed from the event if an electron or muon
selected with loosened identification criteria
is found within $\Delta R = 0.2$ or if a
loosely selected muon is found in a jet that is not well
matched to the primary vertex, as in
Refs.~\cite{EXOT-2016-15,EXOT-2016-14}.
 
A \smallR{} jet is \btagged{} if it satisfies the 77\% working point criterion of the
\btagalgo{} ATLAS $b$-tagging algorithm~\cite{ATL-PHYS-PUB-2016-012, ATLAS_btagging2018}.
Working points are defined by a requirement on the output discriminant
and are labeled by the $b$-jet efficiency they give on an
inclusive \ttbar{} sample. The 77\% working point has rejection
factors of 6.2 and 134 for jets containing charm hadrons ($c$-jets)
and jets containing light-quark hadrons or gluons (light-jets),
respectively.  Correction factors are applied to the simulated event
samples to correct for differences in the \btagging{} efficiencies for
$b$-jets, $c$-jets, and light-jets between data and simulation.  In
addition to using \btagging{} to select events with the 77\% working
point, three other working points (60\%, 70\%, 85\%) are used in the
context of the boosted-object tagging as described later in this
section. 
 
\SmallR{} jets are re-clustered~\cite{Nachman2015} using
the \antikt{} algorithm with a variable cone size~\cite{Krohn:2009zg}
to create variable-radius re-clustered jets (\vrcjets{}). Constituent \smallR{} jets are not allowed to be
shared by multiple \vrcjets{}.  Because the \smallR{} jets used in the
re-clustering are already calibrated, the \vrcjets{} are also
calibrated and their uncertainties are obtained directly from
the \smallR{} jet uncertainties~\cite{ATLAS-CONF-2017-062}.  A requirement on a \pt{}-dependent variable radius
reduces the overlap of boosted objects in the high-multiplicity final
state of this search and exploits the fact that the radius separation
$R$ between the decay products of a heavy, high-\pt{} particle of mass
$m$ can be approximated with $R\sim 2m/\pt$.  The radius
parameter threshold is chosen to be $\Reff = \rho/\pt$, with $\rho=315\,\GeV$,
within the restriction of $0.4 \le \Reff \le 1.2$. This results in a
good compromise between the accuracy and efficiency of the
reconstruction for the objects considered in the final state.  To
reduce contributions from low-energy pileup, a trimming
procedure~\cite{PERF-2012-02} removes \smallR{} jets from a \vrcjet{} if
their \pt{} is less than 5\% of the \vrcjet{} \pt{}.  The \vrcjets{}
are required to have mass greater than $40$\,\GeV,
$\pt{}>150$\,\GeV{}, and $|\eta|<2.5$.

A multi-class deep neural network (DNN) is trained to identify the
most likely parent particle of the \vrcjets{}, distinguishing between four
categories: $V$-boson ($W$- or $Z$-boson), Higgs-boson, top-quark,
and {\it background} jets.  In simulation the label for a reconstructed signal jet ($V$-boson,
Higgs-boson or top-quark jet) is obtained by matching the \vrcjets{} to a hadronically decaying boson or top quark
at generator level within a cone of $\Delta
R=0.75\cdot \rho/{\pt}$. For the Higgs boson, only direct decays into
quark pairs are considered. All \vrcjets{} matched to multiple
generator-level $V$ bosons, Higgs bosons or top quarks are
discarded. The {\it background} label is given to any \vrcjets{}
reconstructed from simulated multi-jet events.  The DNN is trained
using the mass, \pt{}, and number of constituent jets of
the \vrcjet{}, as well as the four-momentum vectors and $b$-tagging
information of the three highest-\pt{} constituent \smallR{} jets as
input.
 
The \textsc{Keras} software package~\cite{keras} is used to build and
train the DNN, using the \textsc{Theano} backend~\cite{theano}. The
DNN has four fully connected hidden layers and a four-dimensional
output layer, and is trained using
the \textit{Adam}~\cite{DBLP:journals/corr/KingmaB14} optimizer
algorithm.\footnote{For an introduction to DNNs and related
terminology, see Refs.~\cite{DBLP:Schmidhuber14,
Goodfellow-et-al-2016}.}  Hidden layers of the DNN use batch
normalization~\cite{DBLP:journals/corr/IoffeS15} and
rectified-linear-unit activation functions, whereas the output layer
uses a sigmoid function.  As the performance of the DNN tagger is
dependent on the architecture and training hyperparameters, DNNs with
different number of layers, learning rate, L1 regularizer and batch
size are tested to define the architecture and training
hyperparameters.  The \pt{} distribution of the background is
reweighted to match the signal distribution.  In this way the DNN is
prevented from learning the differences between the \pt{}
distributions of signal and background jets, while allowing for
learning relations between the \pt{} of the \vrcjets{} and other input
features.  By using only properties of the calibrated \smallR{} jets
as input to the tagger, all jet-related systematic uncertainties can
be propagated through the DNN by varying the corresponding properties
of the \smallR{} jets.  To reduce the four-dimensional DNN output
information (\DNN{}), the outputs of the different classes are
combined by building a discriminant function.
 
The discriminant function $P$ for a $V$ boson, Higgs boson and top
quark is given by
\begin{equation}
\begin{aligned}
P(V) &= \log_{10}\left( \frac{\DNN{V}}
{0.9\cdot\DNN{\text{background}} + 0.05\cdot\DNN{t} + 0.05\cdot\DNN{H} } \right), \\
P(H) &= \log_{10}\left( \frac{\DNN{H}}
{0.9\cdot\DNN{\text{background}} + 0.05\cdot\DNN{V} + 0.05\cdot\DNN{t} } \right) \; \rm{and} \\ \nonumber
P(t) &= \log_{10}\left( \frac{\DNN{t}}
{0.9\cdot\DNN{\text{background}} + 0.05\cdot\DNN{H} + 0.05\cdot\DNN{V} } \right),
\end{aligned}
\end{equation}
respectively.  The relative weighting factors of 0.9 for background
jets and 0.05 for $V$-boson, Higgs-boson or top-quark jets are chosen
as a compromise between background rejection and the ability to
discriminate amongst signal sources.  For each signal discriminant
$P$, an optimized working point is defined to obtain a boosted-object
tagger with a specified signal efficiency. The discriminant functions
and the corresponding thresholds for these working points are shown in
~\Fig{\ref{fig:objectdefs:dnn_loglikelihood_distributions}}, where
$|\eta|$, $\pt$, and $m$ refer to the pseudorapidity, transverse
momentum and mass of the \vrcjet{}.
The multi-peak behavior in some of the discriminant functions is a result of differences in
important \vrcjet{} properties used as input to the DNN, such as the mass, number of constituent
\smallR{} jets, and whether the constituent \smallR{} jets are $b$-tagged.  These properties relate
to, for example, whether or not all of the decay products of the $V$ boson, Higgs boson or top quark
are fully contained within the \vrcjet{}. For example, the double-peak structure in the distribution
of Higgs-boson jets in~\Fig{\ref{toref}} arises predominantly
from whether the \vrcjet{} contains the two expected sub-jets from the Higgs-boson decay or if it contains
additional hadronic energy.
 
The $V$- and Higgs-boson taggers use 70\% working points, which
correspond to the thresholds $P(V) > -0.2$ and $P(H) > 0.35$.  The
top-quark tagger operates at a 60\% working point using a threshold of
$P(t) > 0.1 $. The resulting signal efficiency and background
rejection (estimated from simulated multi-jet events) for each
boosted-object tagger is shown as a function of \pt{}
in~\Fig{\ref{fig:objectdefs:dnn_eff}}.
 
\begin{figure}[!ht]
\centering
\subfloat[]{\includegraphics[width=0.45\textwidth]{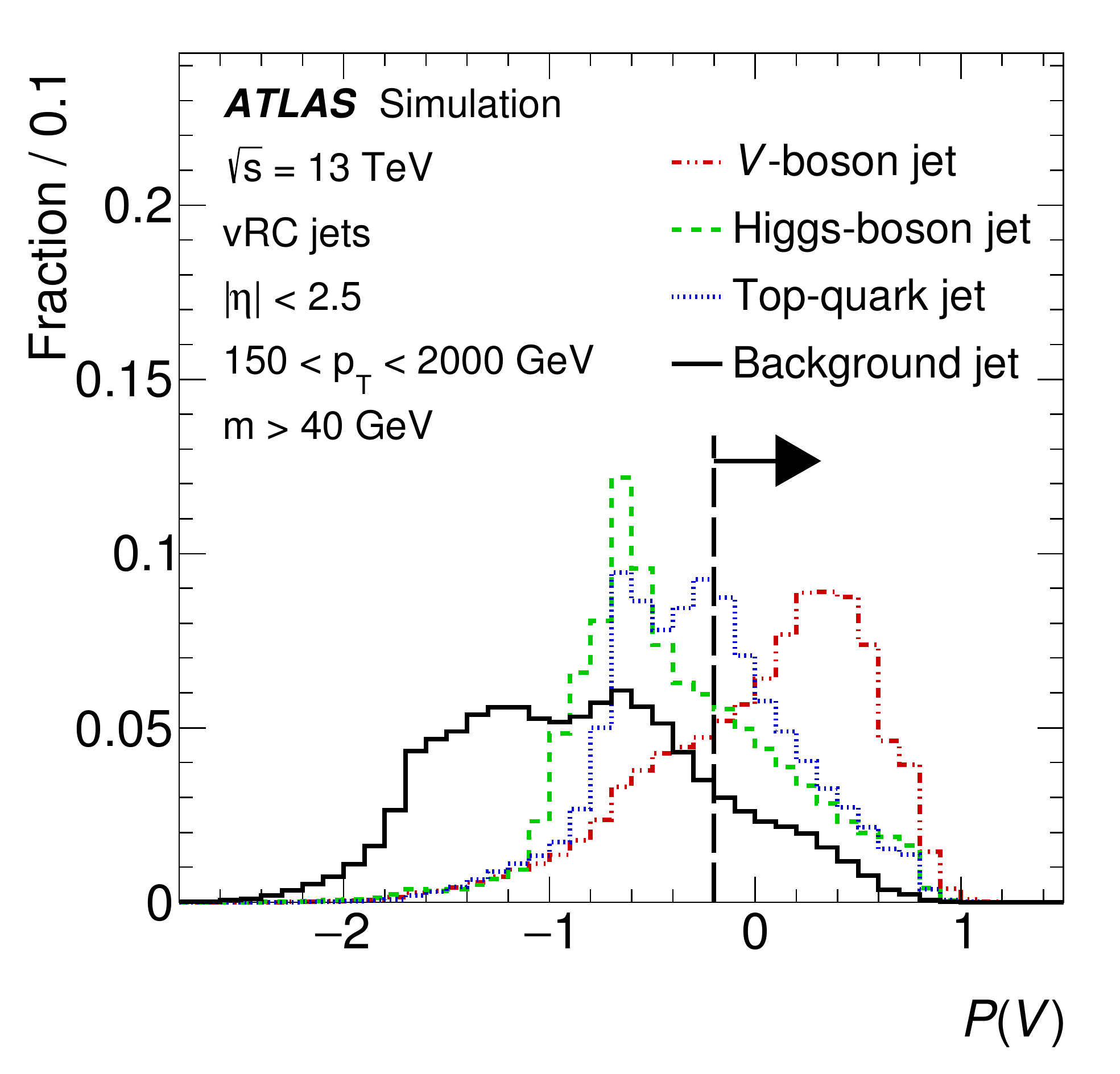}  }
\subfloat[]{\includegraphics[width=0.45\textwidth]{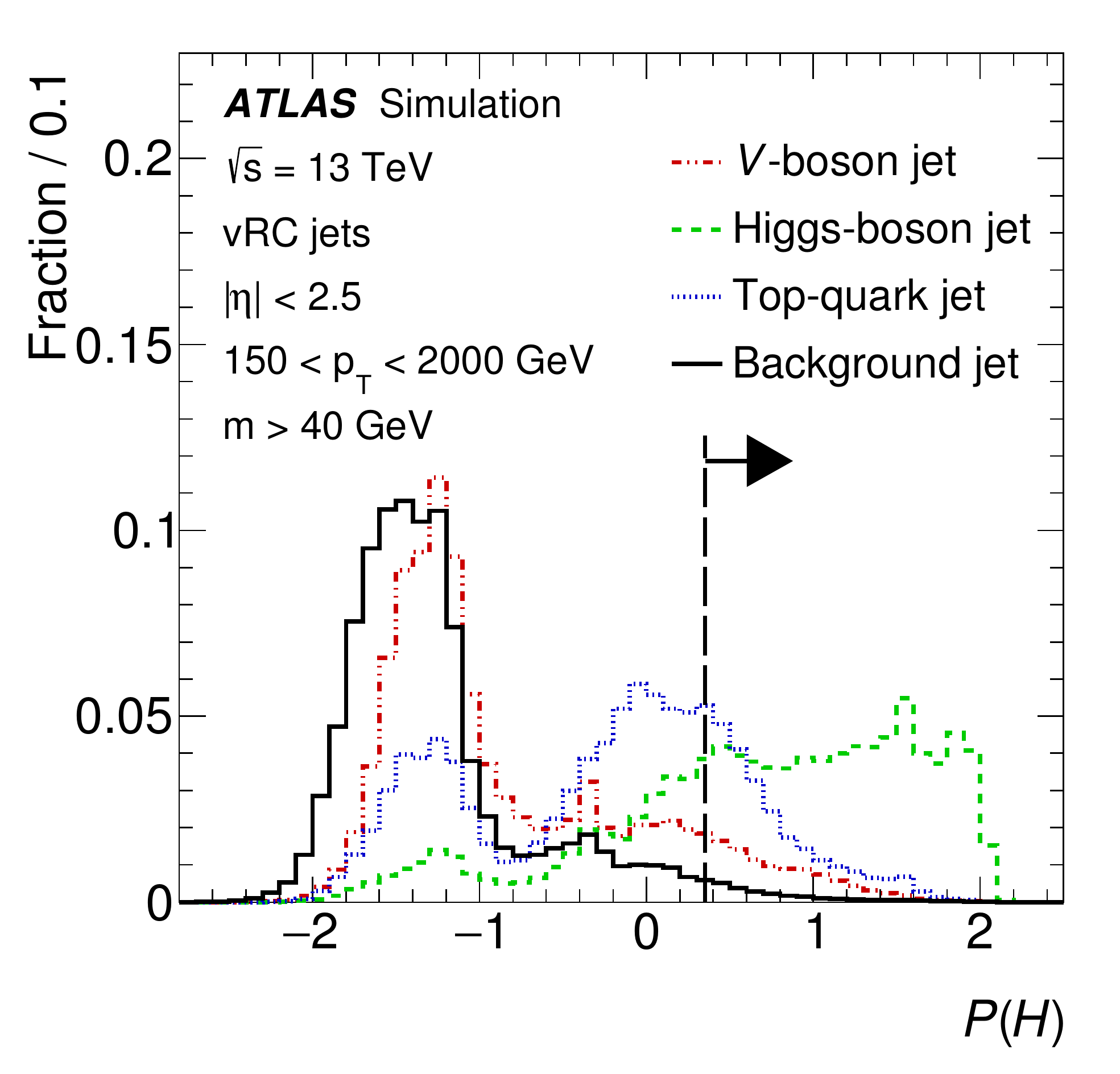}  } \\
\subfloat[]{\label{toref}\includegraphics[width=0.45\textwidth]{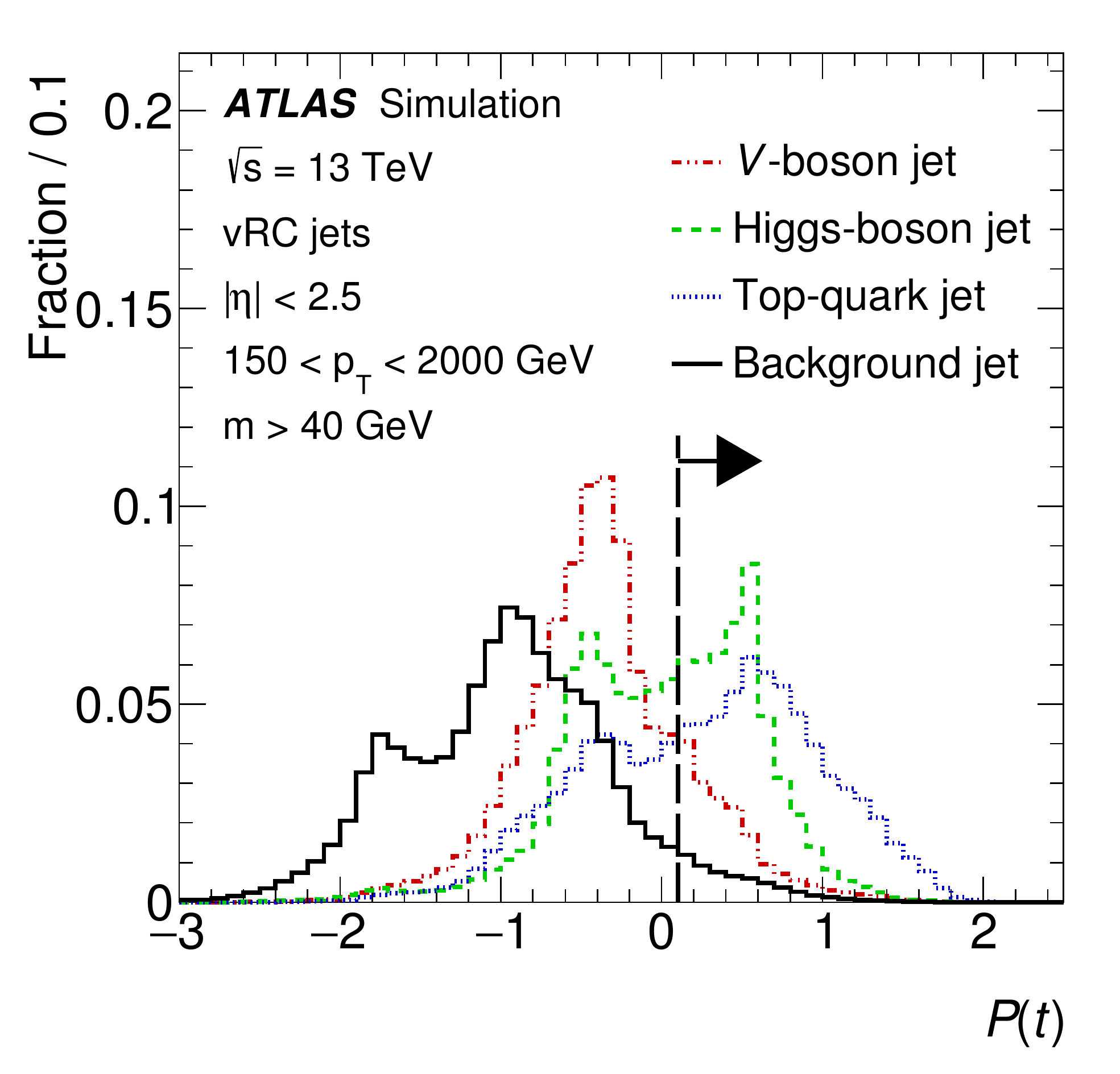}}
\caption{The discriminant function $P$ for the (a) $V$-tagger, (b)
Higgs-tagger and (c) top-tagger. Signal jets ($V$-boson, Higgs-boson, top-quark jets) are
defined by matching the \vrcjet{} to the corresponding object at
generator level. The distributions are made by merging all simulated
VLQ samples.  Background jets are taken from simulated multi-jet
events. The object selection applied to the \vrcjets{} is shown on
the left side of the figures. The dashed vertical line represents
the applied tagging selection.}
\label{fig:objectdefs:dnn_loglikelihood_distributions}
\end{figure}

\begin{figure}[!ht]
\centering
\subfloat[]{\includegraphics[width=0.45\textwidth]{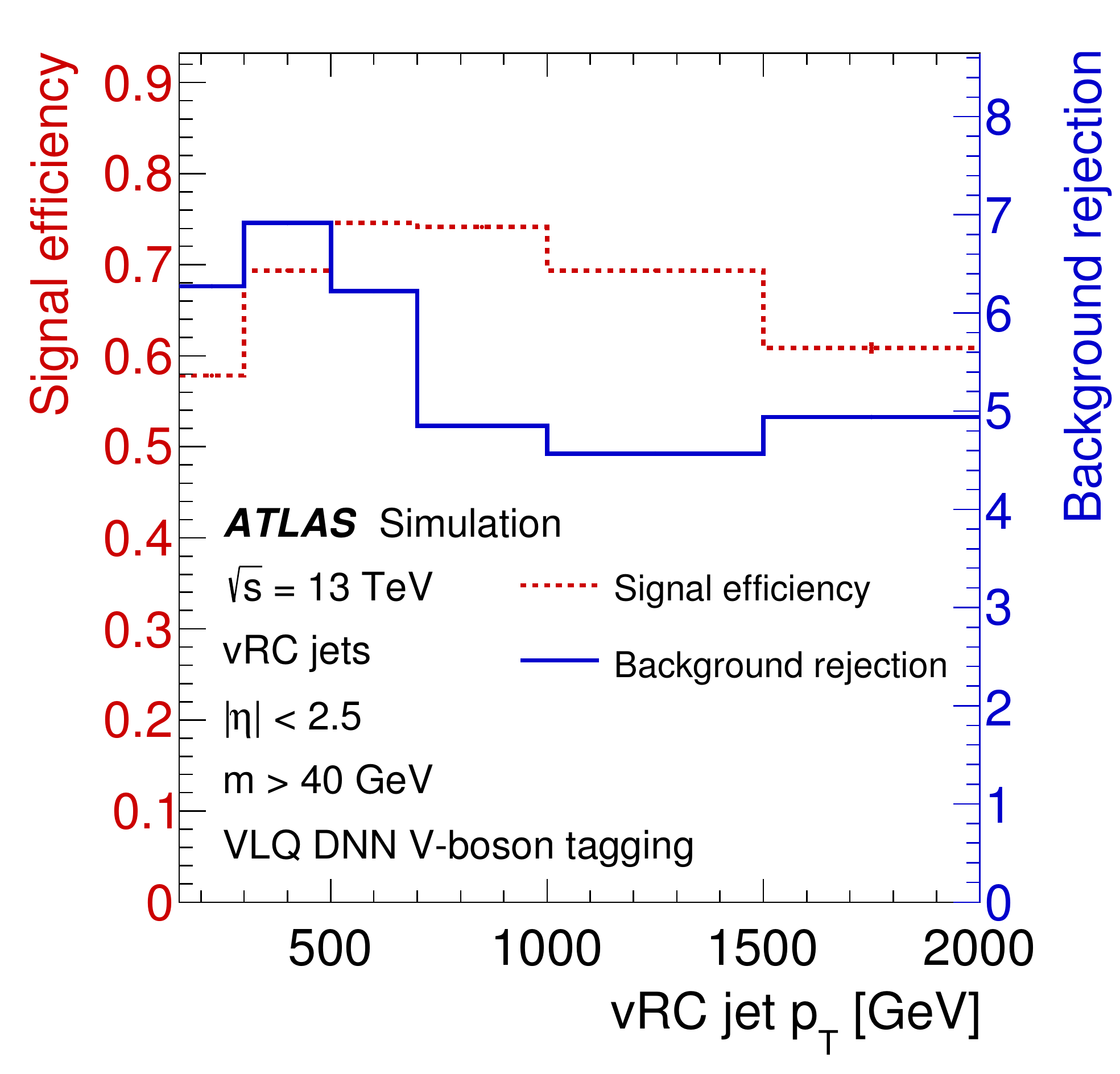}  }
\subfloat[]{\includegraphics[width=0.45\textwidth]{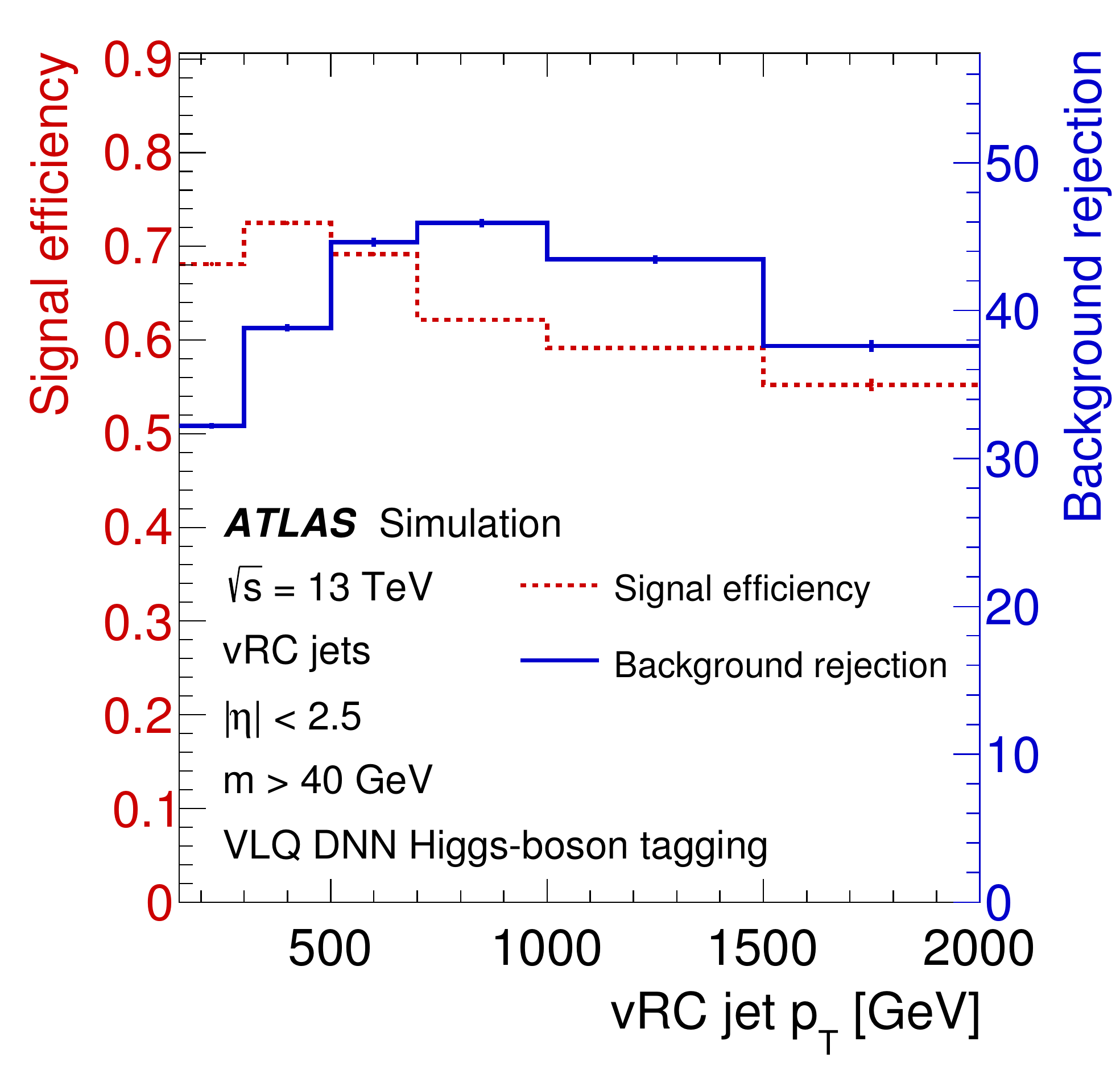}  } \\
\subfloat[]{\includegraphics[width=0.45\textwidth]{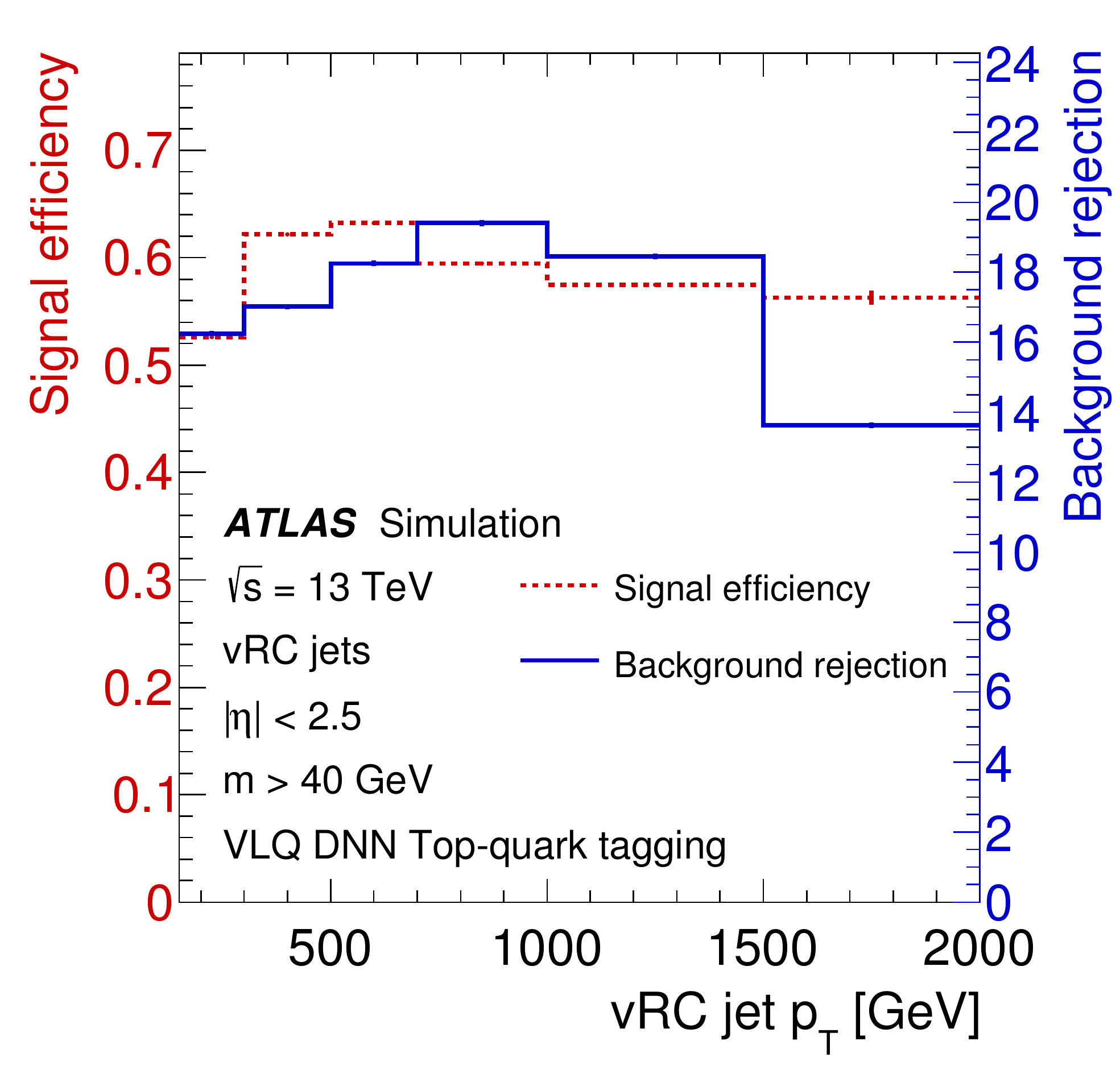}}
\caption{The signal efficiency (dashed) and background rejection
(solid) as a function of \vrcjet{} $\pt$ for the DNN VLQ (a)
$V$-tagger, (b) Higgs-tagger and (c) top-tagger. The dashed lines
refer to the left $y$-axis scale, and the solid lines refer to the
right. Signal jets ($V$-boson, Higgs-boson, top-quark jets) are defined by matching
the \vrcjet{} to the corresponding object at generator level. The
distributions are made by merging all simulated VLQ samples.
Background jets are taken from simulated multi-jet
events. Statistical uncertainties are shown for signal and
background.}
\label{fig:objectdefs:dnn_eff}
\end{figure}
 
To handle the ambiguities due to multiple-tagged \vrcjets{},
additional discriminant functions, shown
in~\Fig{\ref{fig:objectdefs:dnn_loglikelihood_distributions_ambiguity}},
are defined.  Optimized thresholds, shown in each sub-figure, are chosen
to resolve double-tagged \vrcjets{}. Higgs bosons are more frequently
triple-tagged than $V$ bosons or top quarks, so
triple-tagged \vrcjets{} are tagged as a Higgs boson. The \vrcjets{}
that are tagged as a $V$ boson, top quark, or Higgs boson are referred
to as $V$-tagged, top-tagged, and Higgs-tagged, respectively.
 
\begin{figure}[!ht]
\centering
\subfloat[]{\includegraphics[width=0.45\textwidth]{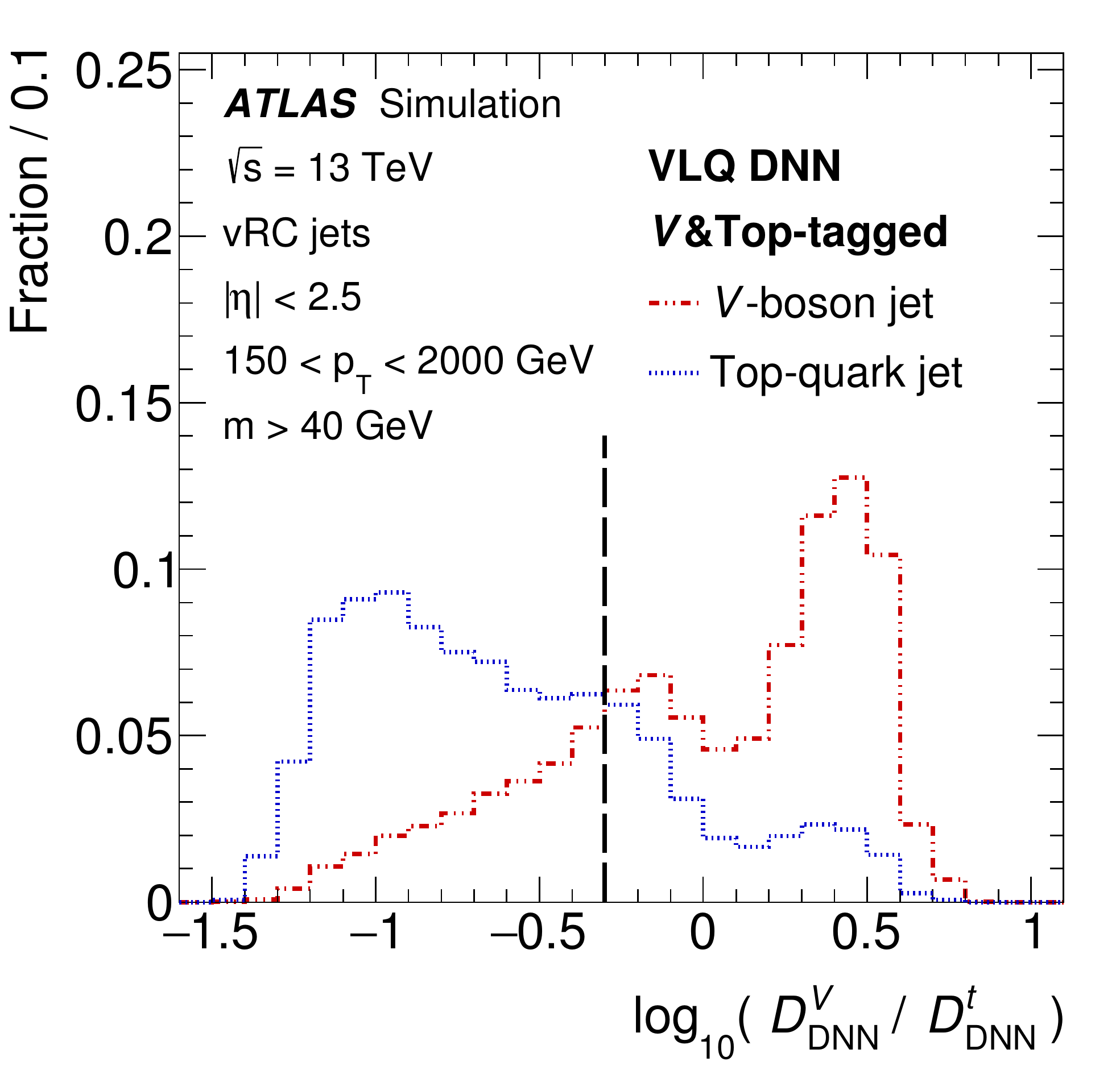}}
\subfloat[]{\includegraphics[width=0.45\textwidth]{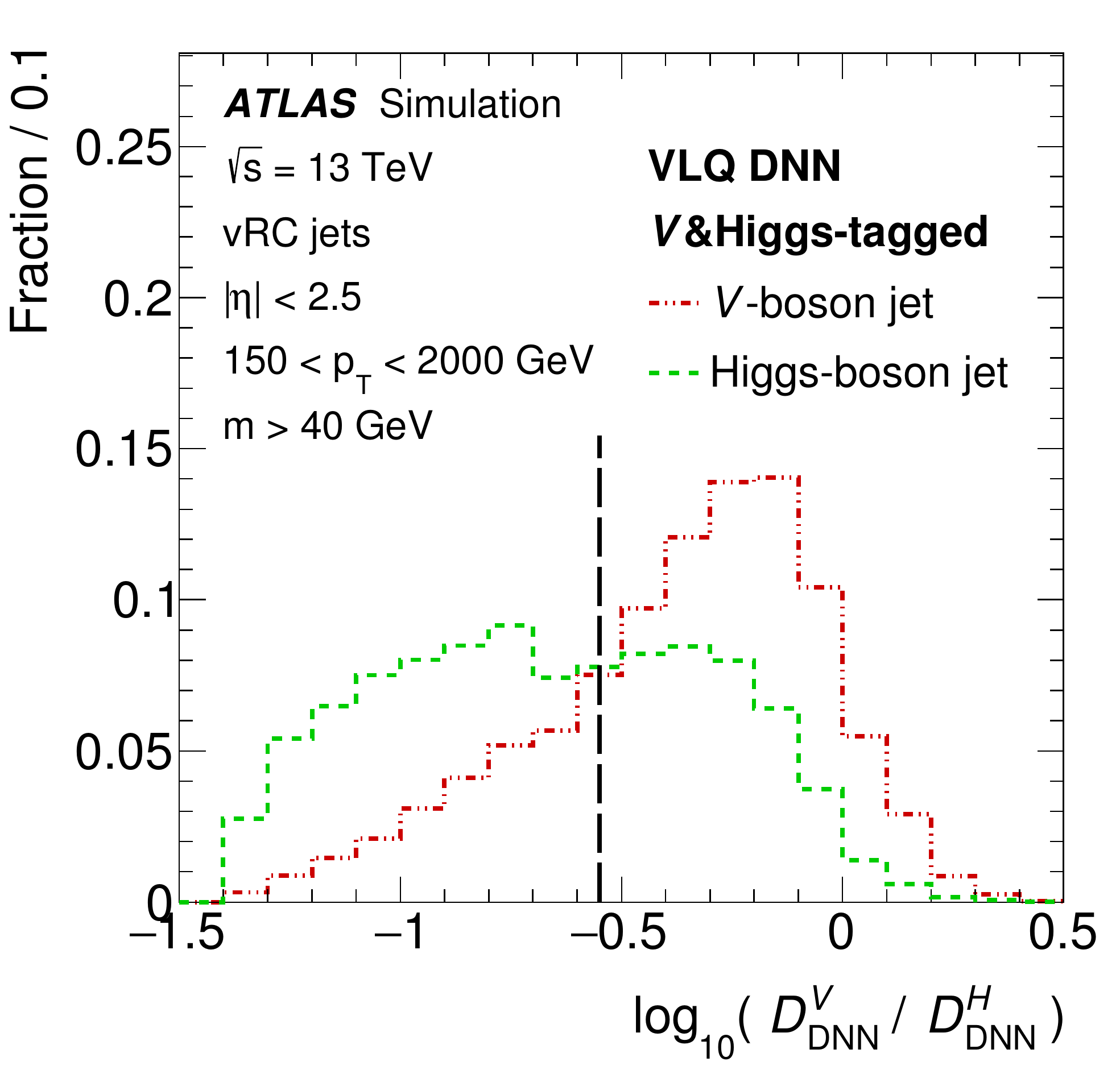}	}\\
\subfloat[]{\includegraphics[width=0.45\textwidth]{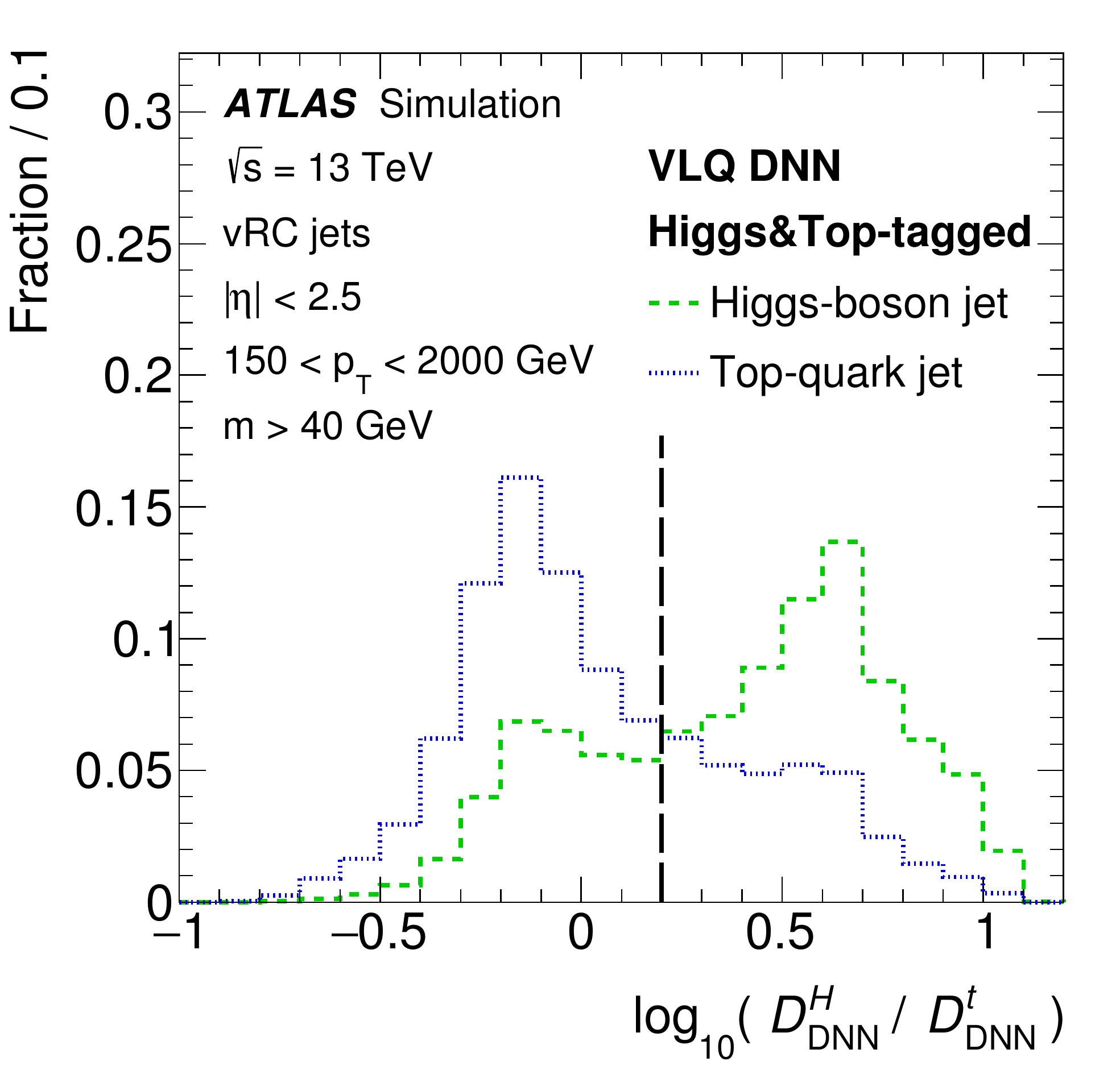}}
\subfloat[]{\includegraphics[width=0.45\textwidth]{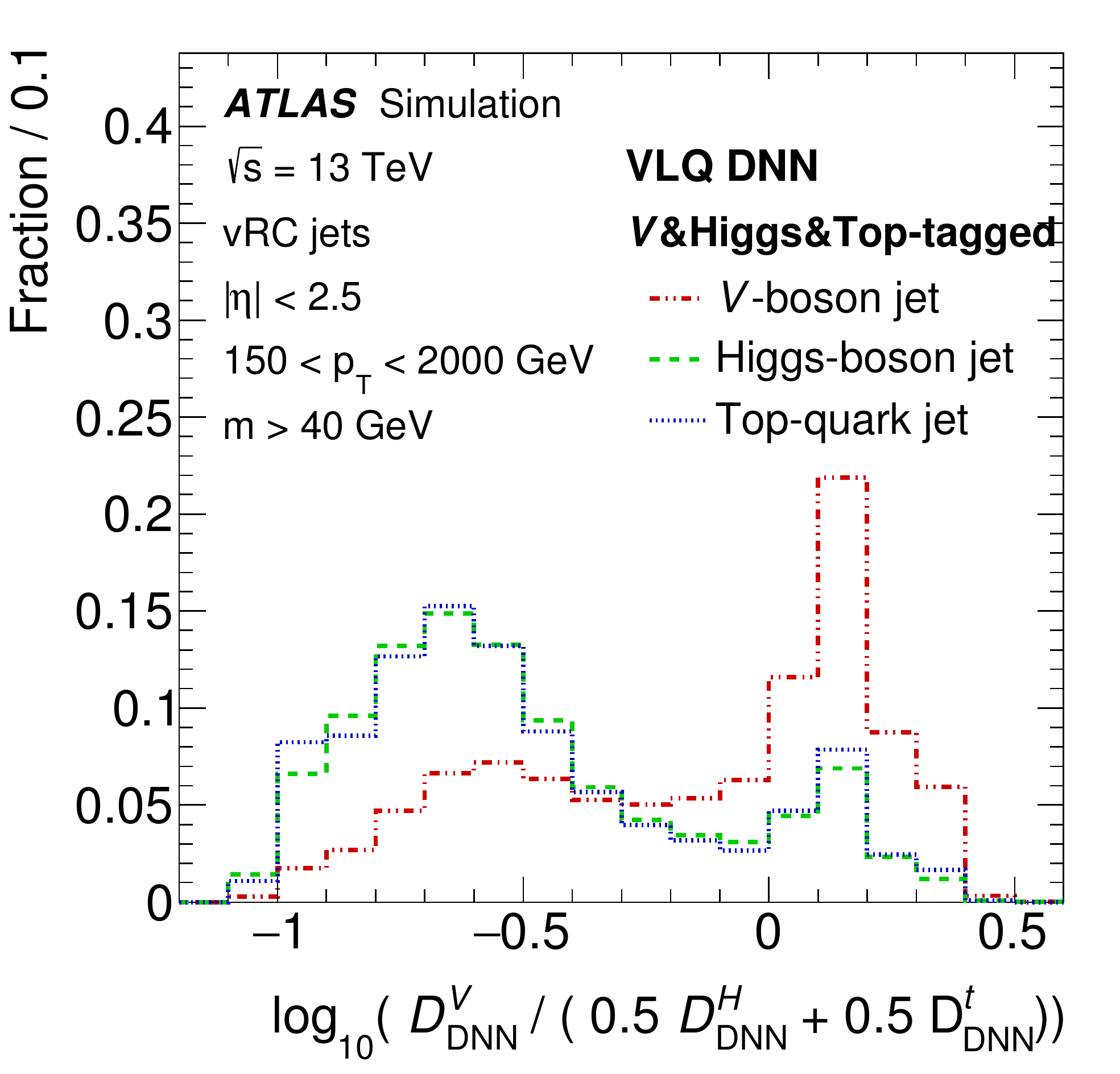}  }
\caption{The additional discriminant functions defined to resolve
multiple-tagged \vrcjets{} such as (a) DNN VLQ $V$- and top-tagged,
(b) DNN VLQ $V$- and Higgs-tagged, (c) DNN VLQ Higgs- and top-tagged
and (d) DNN VLQ $V$-, Higgs- and top-tagged. Signal ($V$-boson,
Higgs-boson, top-quark) jets are defined by matching the \vrcjet{}
to the corresponding object at generator level. The distributions
are made by merging all simulated VLQ samples.  The dashed vertical
line represents the applied selection.}
\label{fig:objectdefs:dnn_loglikelihood_distributions_ambiguity}
\end{figure}

The shape of the \vrcjet{} mass distribution before and after the
final boosted-object tagging is shown
in~\Fig{\ref{fig:objectdefs:dnn_mass_before_after}} for each jet type.
As expected, each tagger preferentially selects \vrcjets{} with a mass
near the mass of the desired particle. For the top quarks, \vrcjets{}
with a mass near the $W$-boson mass are generally $V$-tagged (dominant at
low \pt{}) and Higgs-tagged.
 
\begin{figure}[!ht]
\centering
\subfloat[]{\includegraphics[width=0.45\textwidth]{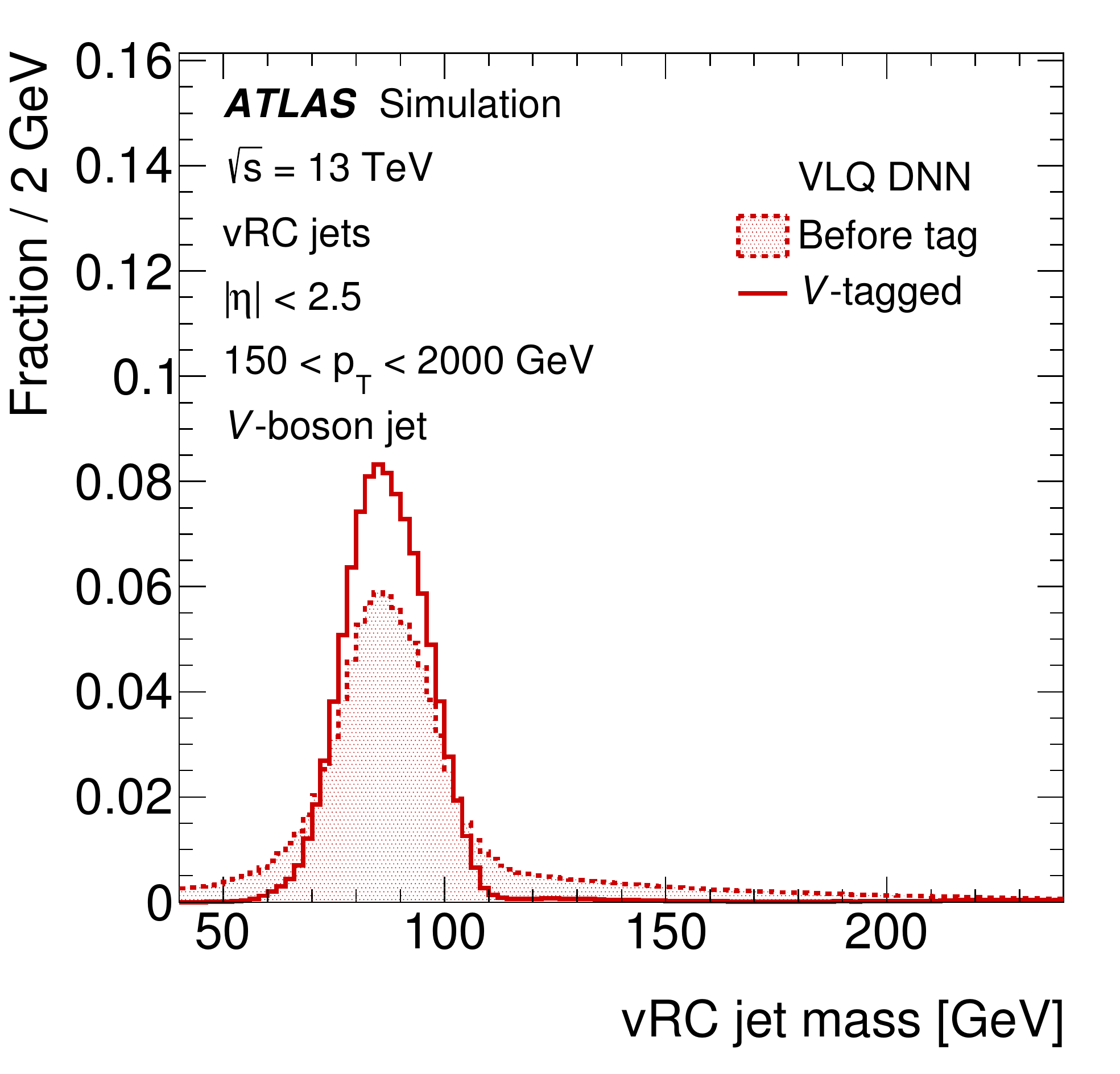}}
\subfloat[]{\includegraphics[width=0.45\textwidth]{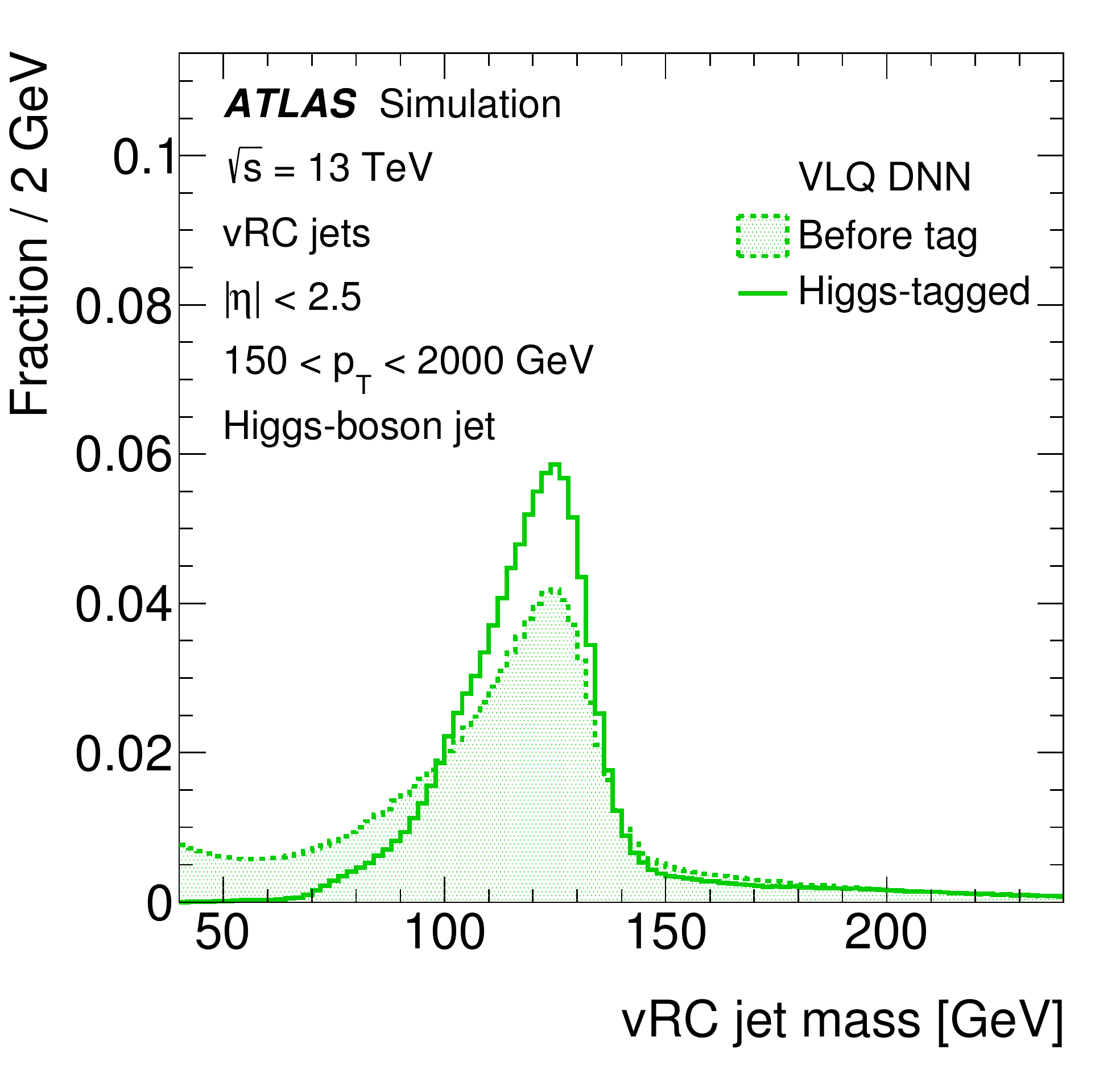}}\\
\subfloat[]{\includegraphics[width=0.45\textwidth]{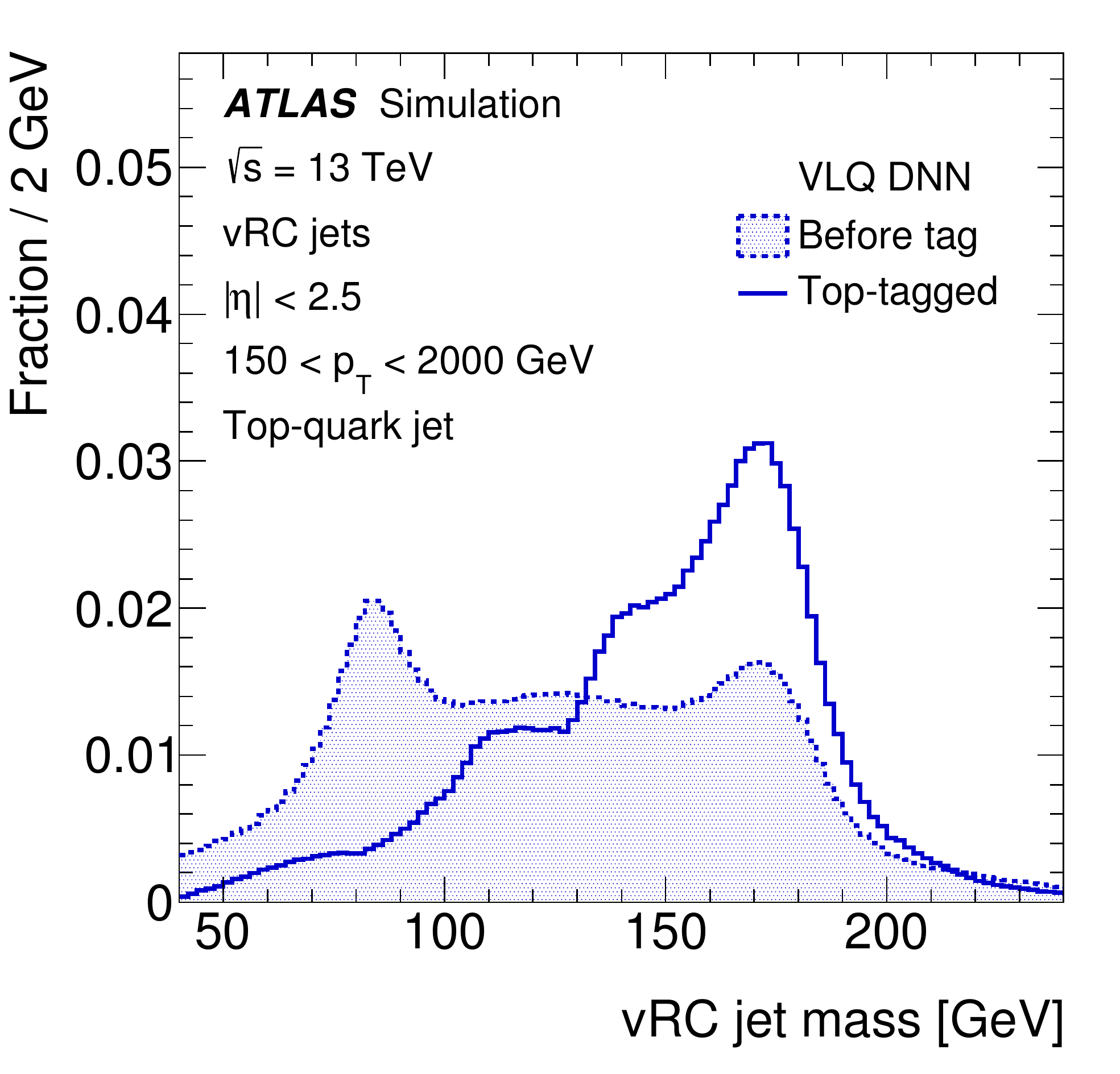}}
\subfloat[]{\includegraphics[width=0.45\textwidth]{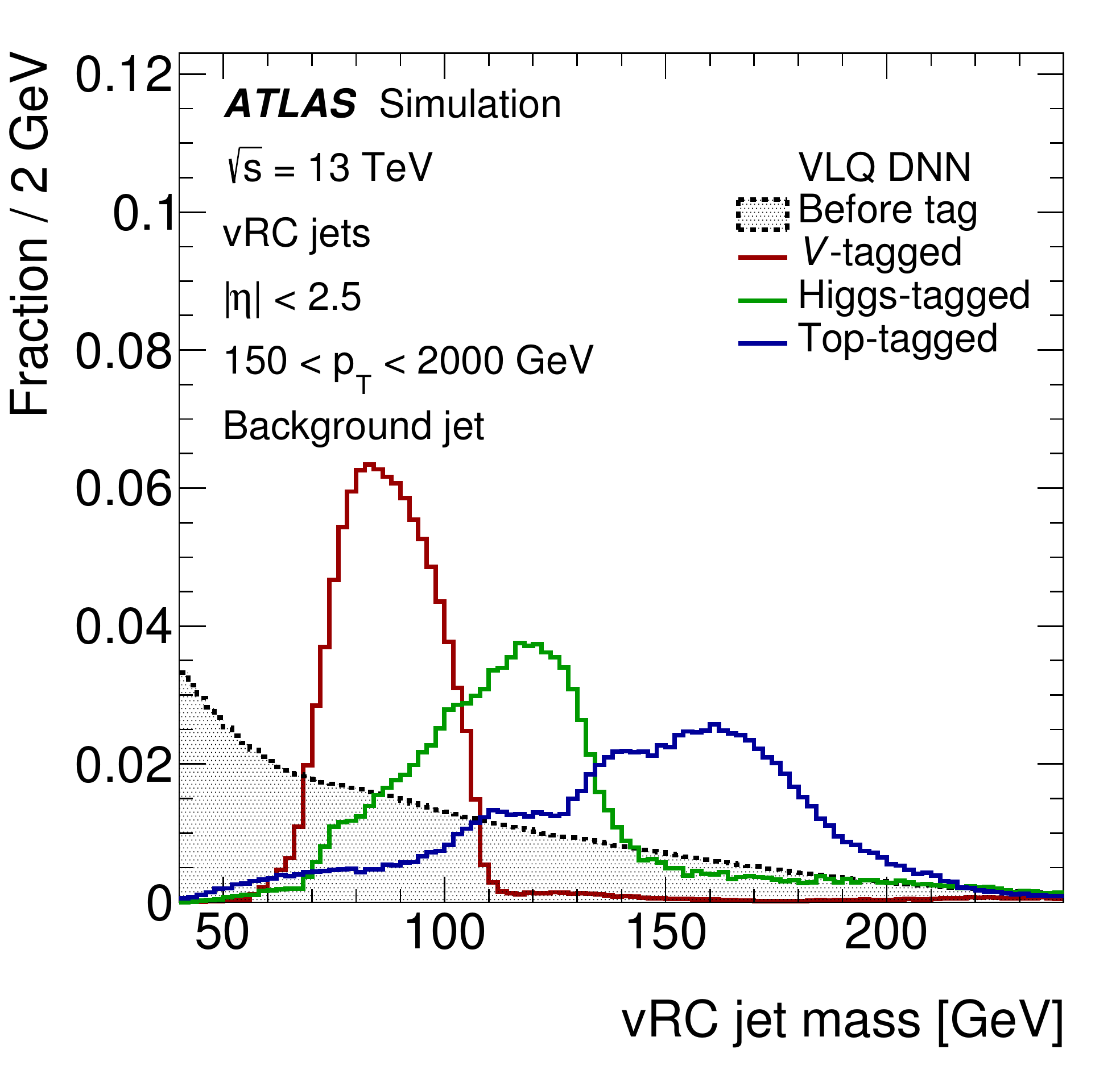}}
\caption{The mass distribution of (a) $V$-boson jets, (b)
Higgs-boson jets, (c) top-quark jets and (d) background jets are
shown before and after the final DNN VLQ boosted-object
tagging. Signal ($V$-boson, Higgs-boson, top-quark) jets are defined
by matching the \vrcjet{} to the corresponding object at generator
level. The distributions are made by merging all simulated VLQ
samples.  Background jets are taken from simulated multi-jet
events. For signal jets, only the impact of the correct tag is
shown, while for background jets the impact of each boosted-object
tag is shown.  All distributions are normalized to unit integral.}
\label{fig:objectdefs:dnn_mass_before_after}
\end{figure}
 
\FloatBarrier

\section{Analysis strategy}
The search presented in this paper focuses on all-hadronic final
states with small \met{}, which allows it to be sensitive to all
possible final states involving hadronic decays of $W$, $Z$, and Higgs
bosons and top quarks.  The key aspect of this search is to suppress
multi-jet background and accurately model multi-jet events that
satisfy the selection criteria.  As a first step, the multi-jet
background is reduced by requiring multiple high-\pt{}
and \btagged{} \smallR{} jets.  As a second step, events are rejected
if they do not contain \vrcjets{} that originate from either a $V$
boson, Higgs boson, or top quark as identified using the DNN
boosted-object tagger.  Events are then categorized according to the
numbers of $V$-tagged, Higgs-tagged and top-tagged vRC jets and of
$b$-tagged \smallR{} jets and are divided into twelve non-overlapping signal
regions, in order to be sensitive to all possible VLQ decays.
Finally, multi-jet events are distinguished from signal events by
calculating, for each signal region, a signal probability using the
matrix element method~\cite{kondo1}.  This signal probability is then
used in a binned profile-likelihood fit in order to extract the signal
strength and improve the background modeling.  The multi-jet
background is estimated in each signal region using a bin-by-bin
`ABCD' method, which is described in~\Sect{\ref{sec:backgrounds}}.
The analysis strategy is optimized while assuming pair production of
VLQs and considering all possible fully hadronic decay modes.
 
\subsection{Event selection and classification}
Data were collected using a combined trigger that requires a single
jet with $\pT>100$\,\GeV{} at the first trigger level and a total
scalar sum of the transverse momenta of all track particles and energy deposits $\HT >1000$\,\GeV{} at the
high-level trigger. An offline threshold of $\HT>1250$\,\GeV{} ensures
that this trigger is fully efficient.  Events are required to have
exactly zero leptons and \met{} $< 200$\,\GeV{} to remove background
and maximize the significance of the signal.  Events enter the
signal regions if they contain at least four selected \smallR\ jets
with descending \pt{} thresholds of 300, 200, 125, and 75\,\GeV\ and
at least two \smallR\ jets that are $b$-tagged, where individual jets
can satisfy one or both criteria.  In addition, the events must have
at least two \vrcjets{} tagged as a $V$ or Higgs boson and
satisfy \met{} $> 40\,\GeV{}$.  The \etmiss{} requirement rejects
significantly more background than signal.  For example, the \etmiss{}
requirement is 71--82\% efficient for the various decay modes of a signal
with a mass of $m_{VLQ} = 1$\,\TeV{}, but only $55\%$ efficient for
simulated multi-jet background events.  Sources of \etmiss{} in VLQ
pair production can include true sources, such as $Z\rightarrow\nu\nu$
decays or leptonic decays of $W$ bosons and top quarks with a soft or
misreconstructed lepton, as well as \etmiss{} from mismeasurement of
high-energy jets.
 
The events are then classified into twelve different signal regions
based on the number of $V$- and Higgs-tags ($VV$, $VH$, $HH$),
top-tags (0, 1, $\ge2$), and $b$-tags (2, $\ge3$), as shown
in \Tab{\ref{tab:eventselection:signalregions}}.  The regions are
designed to cover all of the possible VLQ decays and enhance the ratio
of signal events to SM background events.
\Fig{\ref{fig:eventselection:PieChart}} shows the fraction of
events from each background source that contribute to each signal
region after the full event selection and the background-only fit to data
described in~\Sect{\ref{sec:StatisticalAnalysis}}.  In addition to the signal
regions, nine validation regions are also defined in order to validate the multi-jet background estimation
and evaluate a closure uncertainty for the method. The two regions that are
used to validate the multi-jet background estimation are defined to have exactly
two $b$-tagged jets, two Higgs-tags, and no top-tags. The seven regions that are
used to evaluate the closure uncertainty require exactly one
$b$-tagged jet and the same number of $V$- Higgs- and top-tags as in
each of the signal regions.
 
\begin{table}[htbp]
\begin{center}
\caption{Summary of the definition of the twelve signal regions in
the analysis. The number of $b$-tags is based on all \smallR{} jets,
including those used to construct \vrcjets{} with $V$-, Higgs-, or
top-tags.  The last two signal regions require two bosons of any
type X ($V$ or Higgs boson). The rightmost column lists the matrix
element method (MEM) final states used to define the signal
hypothesis in Section \ref{sec:MatrixElement}.  }
\begin{tabular}{l r r r r c }
\toprule
Region Name & $V$-tags & $H$-tags & top-tags & $b$-tags & MEM final states \\
\midrule
\VVbb\     & 2        & 0        & 0        & 2         & WbWb,ZbZb \\
\VVbbb\    & 2        & 0        & 0        & $\ge$3    & WbWb,ZbZb \\
\VVtbb\    & 2        & 0        & 1        & 2         & ZtWb,WtZb \\
\VVtbbb\   & 2        & 0        & 1        & $\ge$3    & ZtWb,WtZb \\
\VHbb\     & 1        & 1        & 0        & 2         & WbWb,ZbZb\\
\VHbbb\	 & 1        & 1        & 0        & $\ge$3    & WbWb,ZbZb     \\
\VHtbb\ 	 & 1        & 1        & 1        & 2         & HtWb,WtHb\\
\VHtbbb\	 & 1        & 1        & 1        & $\ge$3    & HtWb,WtHb     \\
\HHbbb\	 & 0        & 2        & 0        & $\ge$3    & HbHb     \\
\HHtbbb\   & 0        & 2        & 1        & $\ge$3    & HtWb,WtHb\\
\XXttbb\   & $\ge$0   & $\ge$0   & $\ge$2   & 2         & HtHt,ZtZt,WtWt,HtZt\\
\XXttbbb\  & $\ge$0   & $\ge$0   & $\ge$2   & $\ge$3    & HtHt,ZtZt,WtWt,HtZt     \\
\bottomrule
\end{tabular}
\label{tab:eventselection:signalregions}
\end{center}
\end{table}
 
\begin{figure}[!ht]
\centering
\includegraphics[width=0.75\textwidth]{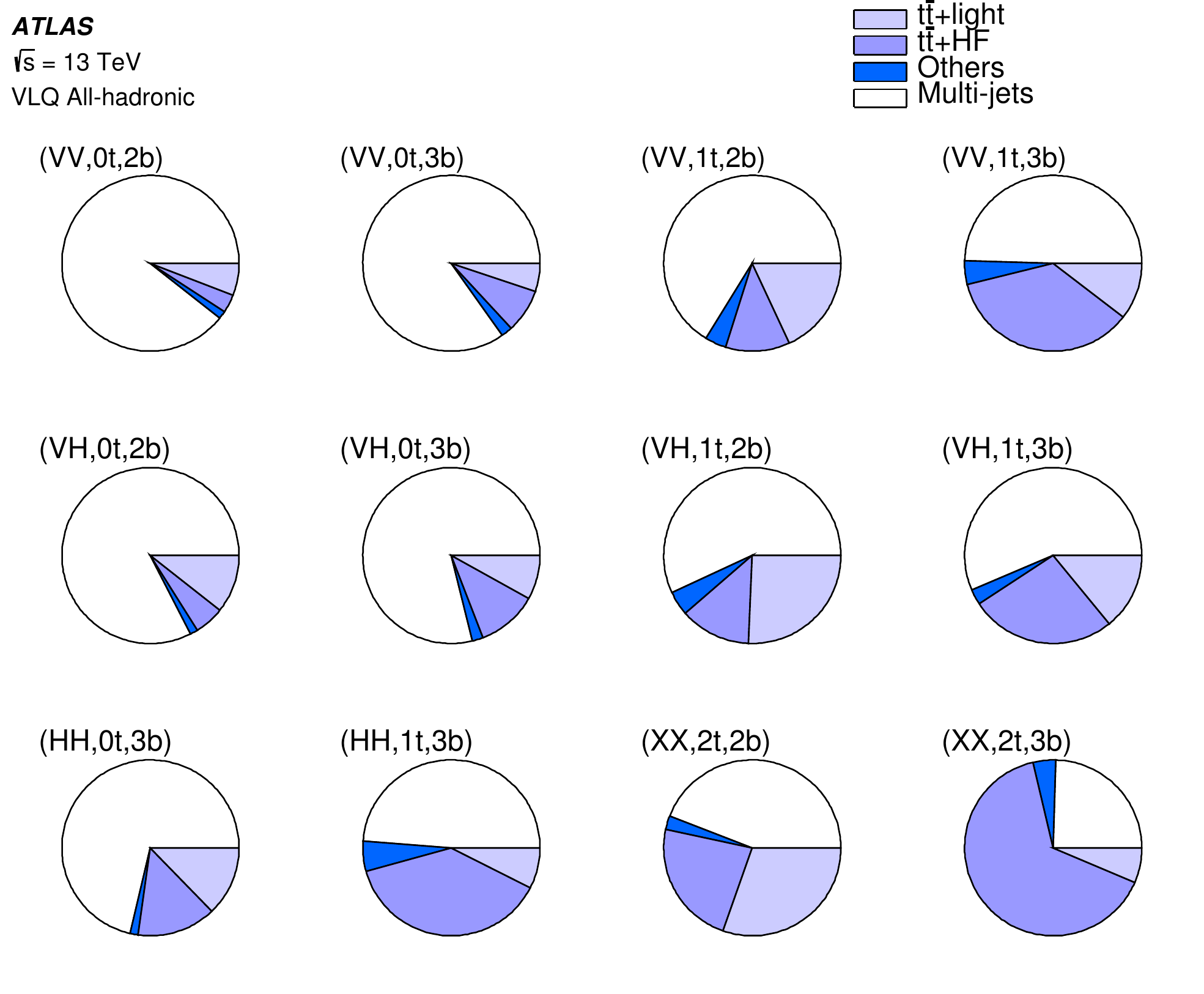}
\caption{Relative size of all background contributions in each
signal region after the background-only fit to data described
in~\Sect{\ref{sec:StatisticalAnalysis}}. `Others' refers to
backgrounds from single-top-quark and \ttX\
production.}  \label{fig:eventselection:PieChart}
\end{figure}
 
\FloatBarrier
 
\subsection{Matrix element method}
\label{sec:MatrixElement}
 
The matrix element method~\cite{kondo1} has been utilized for
measurements~\cite{d0_topmass,cdf_topmass,d0spincorr} and searches for
SM physics
processes \cite{d0_stop,cdf_stop,TOPQ-2015-01,
CDF_VH_MEM1,CDF_VH_MEM2,HIGG-2013-27,
CMS-HIG-14-010}.  This analysis applies the method to a search for physics
beyond the SM.  This method requires the calculation of an event-based
probability density function $P_i \left ( \MB{x} | \MB{\alpha} \right
)$ for a given physics process $i$ described by the matrix element of
the process and a set of theoretical and experimental parameters
$\MB{\alpha}$:
\EQ{MEM_process_final}{
P_i \left ( \MB{x} | \MB{\alpha} \right ) =
\frac{(2\pi)^4}{\sigma_i^{\mathrm{eff}}\left(\MB{\alpha}\right)} 
\int d{\Phi_N}\left(\MB{y}\right){ 
\MB{f}\left(\SUB{p}{A}\right) \MB{f}\left(\SUB{p}{B}\right) 
\frac{\left|\mathcal{M}_{i}\left(\MB{y}|\MB{\alpha}\right)\right|^2}{\mathcal{F}} 
W\left(\MB{y}|\MB{x}\right)} 
}{}
\noindent
The numerical integration is performed over the phase space of the
initial- and final-state particles and can be time consuming.  In this
equation, $\MB{x}$ and $\MB{y}$ represent the four-momentum vectors of
all initial- and final-state particles at reconstruction and parton
level, respectively.  The Lorentz-invariant flux factor
$\mathcal{F}$~\footnote{$\mathcal{F} = 4 \sqrt{
(p_\text{A}p_\text{B})^2 - m_\text{A}^2m_\text{B}^{2}}$} and
phase-space element $d{\Phi_N}$ describe the kinematics of the
process. The transition matrix element $\mathcal{M}_{i}$ is defined by
the Feynman diagrams of the hard-scattering process.  The functions
$\MB{f}\left(\SUB{p}{A}\right)$ and $\MB{f}\left(\SUB{p}{B}\right)$
are the PDFs for the initial-state partons with momenta $\SUB{p}{A}$
and $\SUB{p}{B}$.  The transfer functions
$W\left(\MB{y}|\MB{x}\right)$ map the detector quantities $\MB{x}$ to
the parton-level quantities $\MB{y}$.  Finally, the effective cross
section $\sigma_i^{\mathrm{eff}}$ normalizes $P_i$ to unity taking
acceptance and efficiency into account.
 
The reconstructed objects in an event can be combined to form multiple
candidate VLQ final states.  The process probability density is
calculated for each allowed assignment permutation of the jets to the
final-state quarks and bosons. A process likelihood function is then
built by summing the process probabilities of each allowed assignment
permutation.  The \vrcjets{} are assigned to final states according to
their DNN VLQ boosted-object--tag label ($V$-tagged, Higgs-tagged, or
top-tagged) and are permuted if they have the same label. If more than
two \vrcjets{} are tagged as a boson ($V$-tagged or Higgs-tagged),
only the two with the highest transverse momenta are used.  If
$b$-quarks occur in the hypothesized final state, up to five different
$b$-tagged \smallR{} jets are assigned to the final state and freely
permuted.  These $b$-tagged jets are allowed to overlap with
the \vrcjets{} and could have been used already in the reconstruction
of the \vrcjet.
 
The transition matrix element defines the hypothesis being tested and
is calculated using \MADGRAPH5 in LO precision.  The VLQ
pair-production matrix element calculation is performed using the
Feynrules~\cite{Christensen:2008py} model as defined in
Ref.~\cite{vlqmodel}.  In this analysis, only probabilities of signal
hypotheses are calculated, since the dominant background is from
multi-jet processes, for which it is difficult to define a model in
the matrix element method.  The second most important \ttbar+jets
background was studied as a background hypothesis, but its inclusion
does not improve the sensitivity of the search.  Top quarks, $V$
bosons, and Higgs bosons are assumed to be reconstructed as \vrcjets{}
and are hence not decayed in the matrix element calculation.
 
In each signal region, the signal hypothesis is computed from all
Feynman diagrams of vector-like $T$ or $B$ pair production resulting
in the same number of top quarks, $V$ bosons, and Higgs bosons as
defined in Table \ref{tab:eventselection:signalregions}. Following the
definition of the signal regions \XXttbb{} and \XXttbbb{}, all Feynman
diagrams resulting in final states with two top quarks are used
and no distinction is made based on the number of $V$ and Higgs bosons.
Combining these diagrams into a single hypothesis increases the
performance significantly and allows mis-tags of the $V$ and Higgs
bosons.  Because there is no direct decay of VLQs into a final state
with two Higgs bosons and one top quark, the same diagrams as used for
the \VHtbb{} and \VHtbbb{} signal regions are used in the \HHtbbb{}
region taking mis-tags into account. Preliminary studies indicated
that this analysis would be sensitive to VLQ masses around 900\,\GeV{},
therefore, in the calculation of the matrix elements, the masses of the
vector-like $B$ and $T$ quarks are set to 900 GeV.  The analysis sensitivity
becomes slightly degraded when considering signal samples with a significantly higher VLQ mass.
 
The transfer functions are parameterized as single-Gaussian functions,
which is a good compromise between separation power and reasonable
integration time.  For the modeling of the parton distribution
functions, the {\sc CTEQ6L1} set from the \textsc{LHAPDF}
package~\cite{lhapdf} is used.  The integration is performed
using \textsc{VEGAS}~\cite{VEGAS}.  Due to the complexity and high
dimensionality, adaptive MC techniques~\cite{gsl}, simplifications and
approximations are needed in order to perform the integration in a
reasonable time.  The matrix element calculation is accelerated by
evaluating only the most significant helicity states, which are
identified at the beginning of each integration.  The dimensionality
of the integration is reduced by assuming that the final-state object
directions in $\eta$ and $\phi$ are measured with negligible
uncertainty.  The total momentum conservation and the negligible
transverse momentum of the initial-state partons allow further
reduction.  No change of integration variables is performed in order
to allow a general treatment of all signal regions.  The integration
variables are the energies of the top quarks, $b$-quarks, $V$, and
Higgs bosons according to their numbers as defined for each region.
The total integration volume is restricted by requiring the difference
between the parton-level quantities and the observed values to be
within five standard deviations of the width of the transfer
functions.  Finally, the likelihood contributions of all allowed
assignment permutations are coarsely integrated and sorted by their
contribution, then the full integration is performed with a decreasing
precision.  The logarithm of the resulting signal likelihoods (\LLH{})
is used in each signal region as the final discriminating variable.
Normalized distributions of the \LLH{} for the total background and
signal simulations in the most sensitive signal regions assuming
exclusive $T\rightarrow Wb$, $B\rightarrow Zb$, and $B\rightarrow Hb$
decays are shown as examples in Figure \ref{fig:mem:separation}. The
binning in the \LLH{} distribution is the same as that shown for the
corresponding regions
in~\Figs{\ref{fig:results:VV_PostFitResult}}{\ref{fig:results:HH_PostFitResult}}. The
separation given in Figure \ref{fig:mem:separation} between signal and
background is defined by the formula
\begin{equation}
\label{separation}
\frac{1}{2} \int \frac{(S(x)-B(x))^2}{S(x)+B(x)} dx
\end{equation}
where $S(x)$ and $B(x)$ are the signal and background yields per bin and $S$ and $B$ are normalized
to unity.
 
\begin{figure}[!ht]
\centering
\subfloat[]{\includegraphics[width=0.45\textwidth]{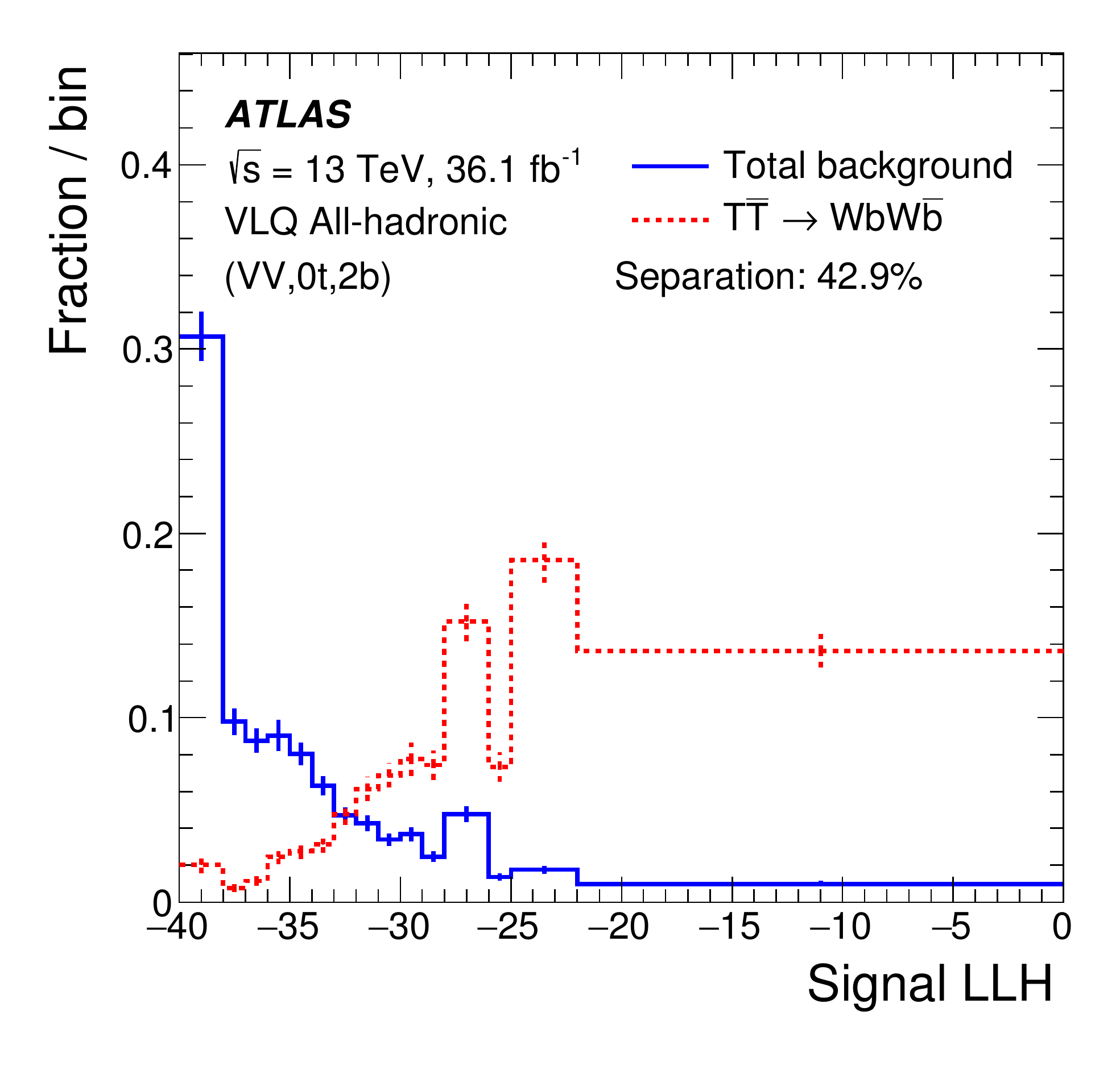}}
\subfloat[]{\includegraphics[width=0.45\textwidth]{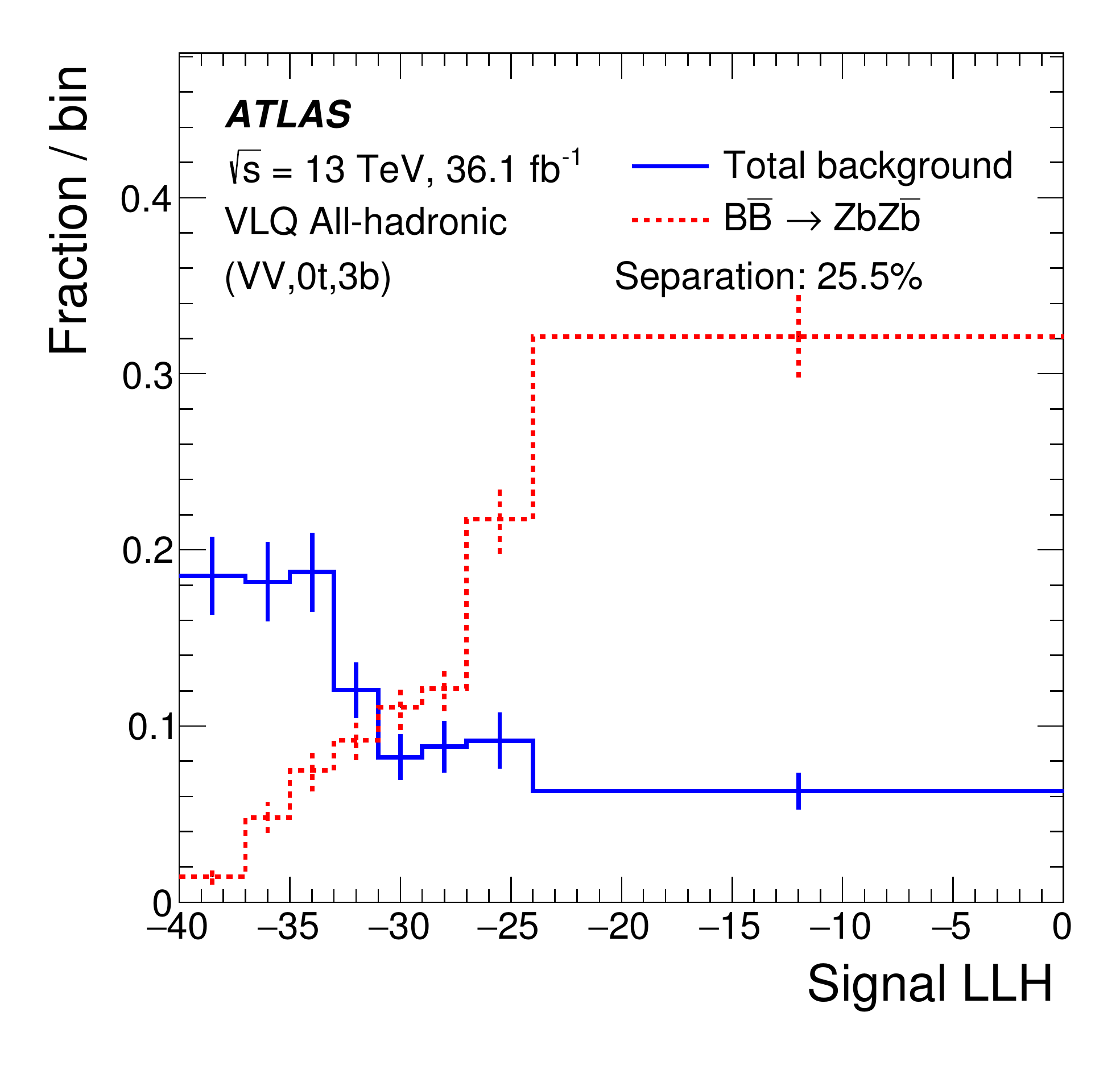}}\\
\subfloat[]{\includegraphics[width=0.45\textwidth]{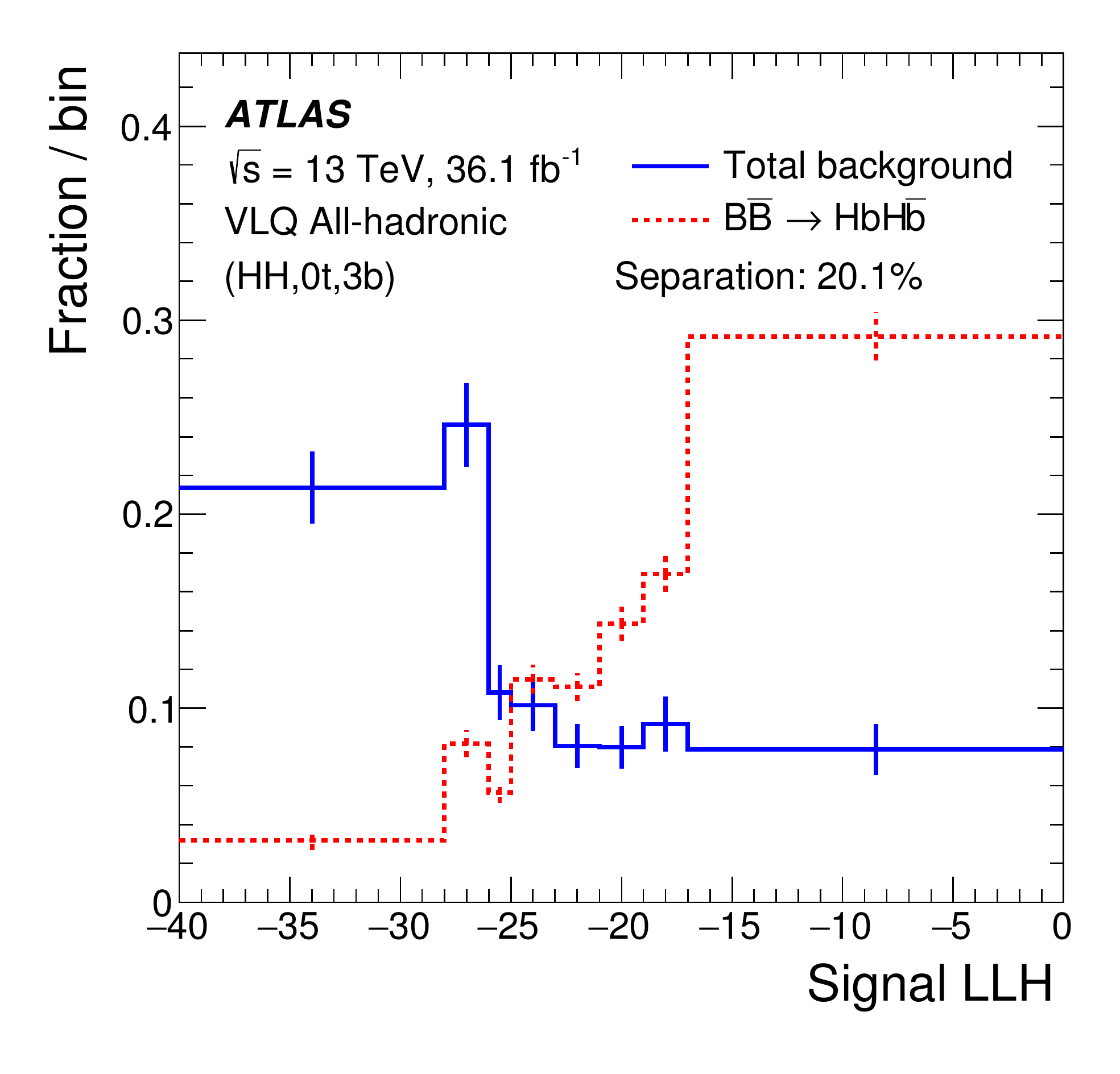}}
\caption{The normalized \LLH{} distributions of the total background
and vector-like \TTbar{} and \BBbar{} production ($m_{T/B} =
1\,\TeV{}$) assuming exclusive (a) $T\rightarrow Wb$, (b)
$B\rightarrow Zb$, and (c) $B\rightarrow Hb$ decays are shown for
the regions with the highest signal significance \VVbb{}, \VVbbb{},
and \HHbbb.  The separation is defined by the formula in
Eq.~\ref{separation}.}  \label{fig:mem:separation}
\end{figure}
 
 
\subsection{Background estimation}
\label{sec:backgrounds}
 
The dominant multi-jet background is estimated using a data-driven
double sideband method (`ABCD'). This method relies on three control
regions (A, B, and C), defined by inverting two uncorrelated selection
requirements, in order to predict the contribution of a background in
a signal region (D). The two selection requirements of this method are applied
on \etmiss{} and boson tagging.  In order to invert the boson tagging,
a `loose-tagged' boson selection is defined.  In contrast to the VLQ
DNN tagger, this selection consists of a simple mass window for
the \vrcjet{} of $69$--$104$\,\GeV{} for $V$ bosons and
$104$--$155$\,\GeV{} for Higgs bosons. The regions used in the method
are then defined as:
 
\begin{itemize}
 
\item Region A: $\ge 2$ \vrcjets{} that are \{$V$-tagged or
Higgs-tagged or `loose-tagged'\} and $< 2$ \vrcjets{} that
are \{$V$-tagged or Higgs-tagged\} and \etmiss{} $< 40$\,\GeV{},
 
\item Region B: $\ge 2$ \vrcjets{} that are \{$V$-tagged or
Higgs-tagged\} and \etmiss{} $< 40$\,\GeV{},
 
\item Region C: $\ge 2$ \vrcjets{} that are \{$V$-tagged or
Higgs-tagged or `loose-tagged'\} and $< 2$ \vrcjets{} that
are \{$V$-tagged or Higgs-tagged\} and \etmiss{} $\ge 40$\,\GeV{},
 
\item Region D: `Signal Region,' $\ge 2$ \vrcjets{} that
are \{$V$-tagged or Higgs-tagged\} and \etmiss{} $\ge 40$\,\GeV{}.
 
\end{itemize}
 
\noindent The four regions are orthogonal and there is no significant correlation
between boson tagging and \etmiss{}.  The level of correlation is
evaluated by checking the correlation factor between the two variables
in simulated multi-jet events, which is found to be consistent with zero.
 
In the control regions A, B, and C, the non--multi-jet contributions
are subtracted from the data using simulation. The relationship
between the yields, $N$, in the signal region, D, and the control regions is given by
$N_\text{D} = N_\text{C}\times(N_\text{B}/N_\text{A})$.  This simple scaling is
performed on a bin-by-bin basis in the \LLH{} distribution to produce
the expected multi-jet shape and normalization in the signal
region. This procedure is followed separately for each of the twelve
signal regions. Seven validation regions are also defined, with the
same $V$-, Higgs-, and top-tagging requirements as the signal regions,
but with exactly one $b$-tagged jet.  These regions are used to
evaluate a closure uncertainty, described
in~\Sect{\ref{sec:Systematics}}. Two examples of these validation
regions can be seen in~\Fig{\ref{fig:results:1b_VR_dist}}, where the
only uncertainties taken into account are those from statistical
sources and related to the detector simulation.
 
\begin{figure}[!ht]
\centering
\subfloat[]{\includegraphics[width=0.4\textwidth]{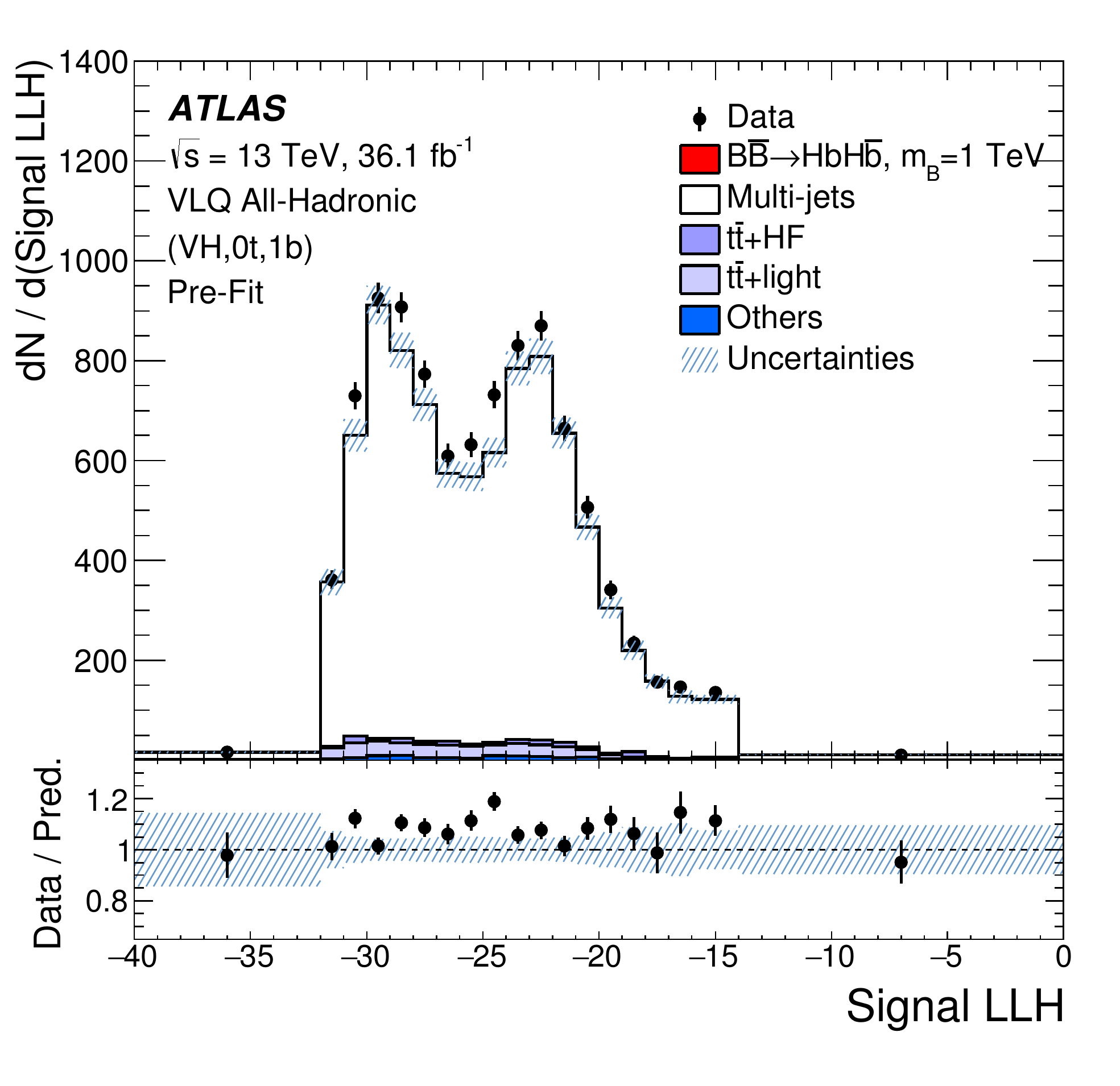}}
\subfloat[]{\includegraphics[width=0.4\textwidth]{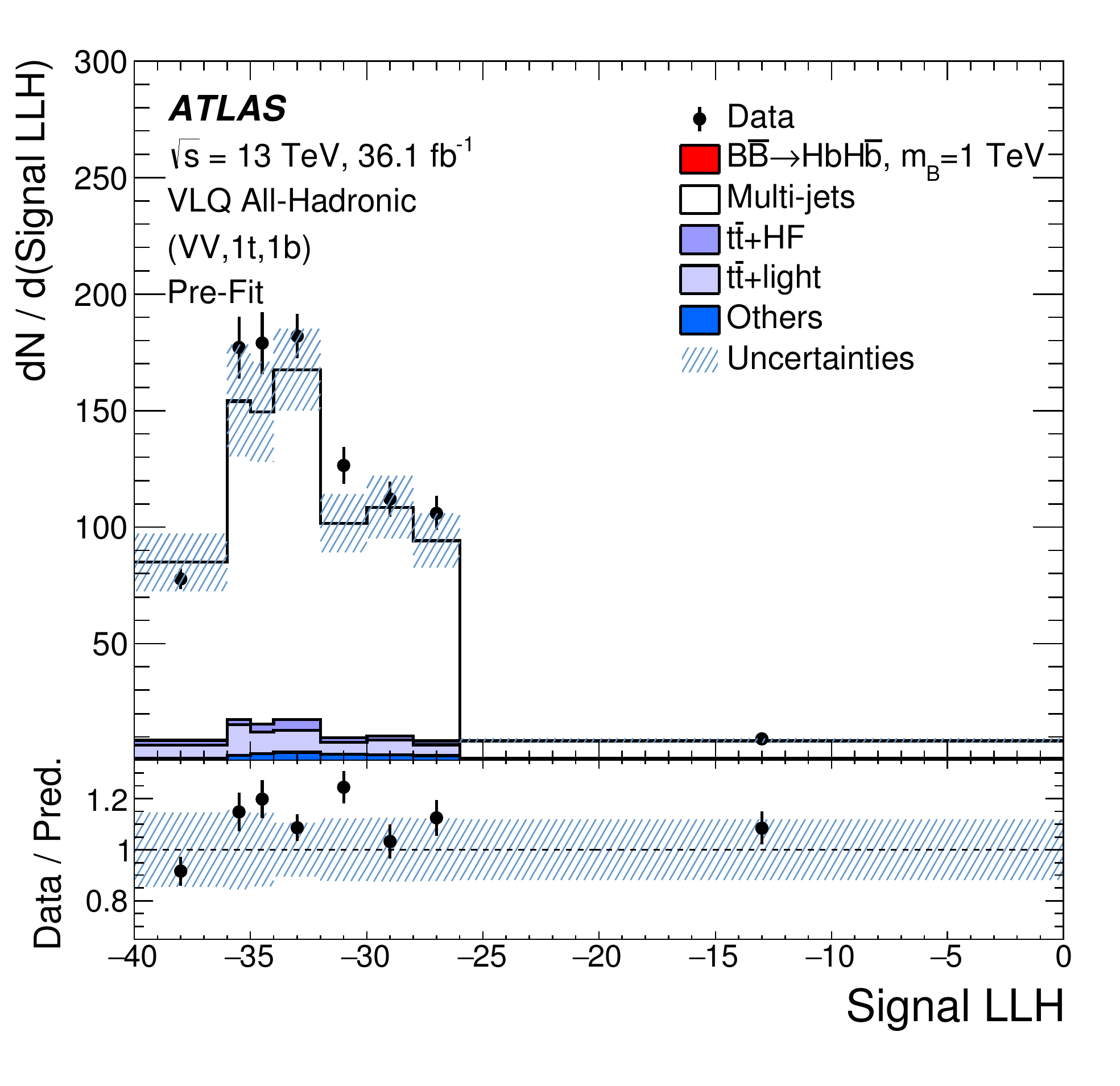}}
\caption{Comparison between data and prediction for the \LLH{} in the validation regions with
two boson-tagged \vrcjets{}, exactly one $b$-tagged \smallR{} jet,
and either (a) zero top-tagged \vrcjets{} or (b) exactly one
top-tagged \vrcjet{}. The distributions show the number of events
per width of 1.0 in the $x$-axis.  The hatched area represents the
statistical and detector-related uncertainties of the background,
added in quadrature. The deviation of the prediction from data is
taken as the multi-jet closure uncertainty. The underflow and
overflow are included in the first and last bins,
respectively. These figures do not include any information from the
fit described in~\Sect{\ref{sec:StatisticalAnalysis}}, and are
therefore described as `Pre-fit'.}
\label{fig:results:1b_VR_dist}
\end{figure}
 
The binning that is used for each region is determined by the number
of events in the A, B, and C control regions.  It is required that
there are a sufficient number of events in each bin of the control
regions (at least 50) to produce a sufficiently smooth distribution.
 
To evaluate the performance of the background estimation method with
all uncertainties, two regions kinematically close to the signal
regions, but with very low expected signal contribution, are also
defined. These regions have two Higgs-tagged \vrcjets{}, exactly two
$b$-tagged \smallR{} jets, and either zero or one
top-tagged \vrcjet{}. Good agreement is observed in these regions, as
shown in~\Fig{\ref{fig:results:HH_VR_dist}}.
 
\begin{figure}[!ht]
\centering
\subfloat[]{\includegraphics[width=0.4\textwidth]{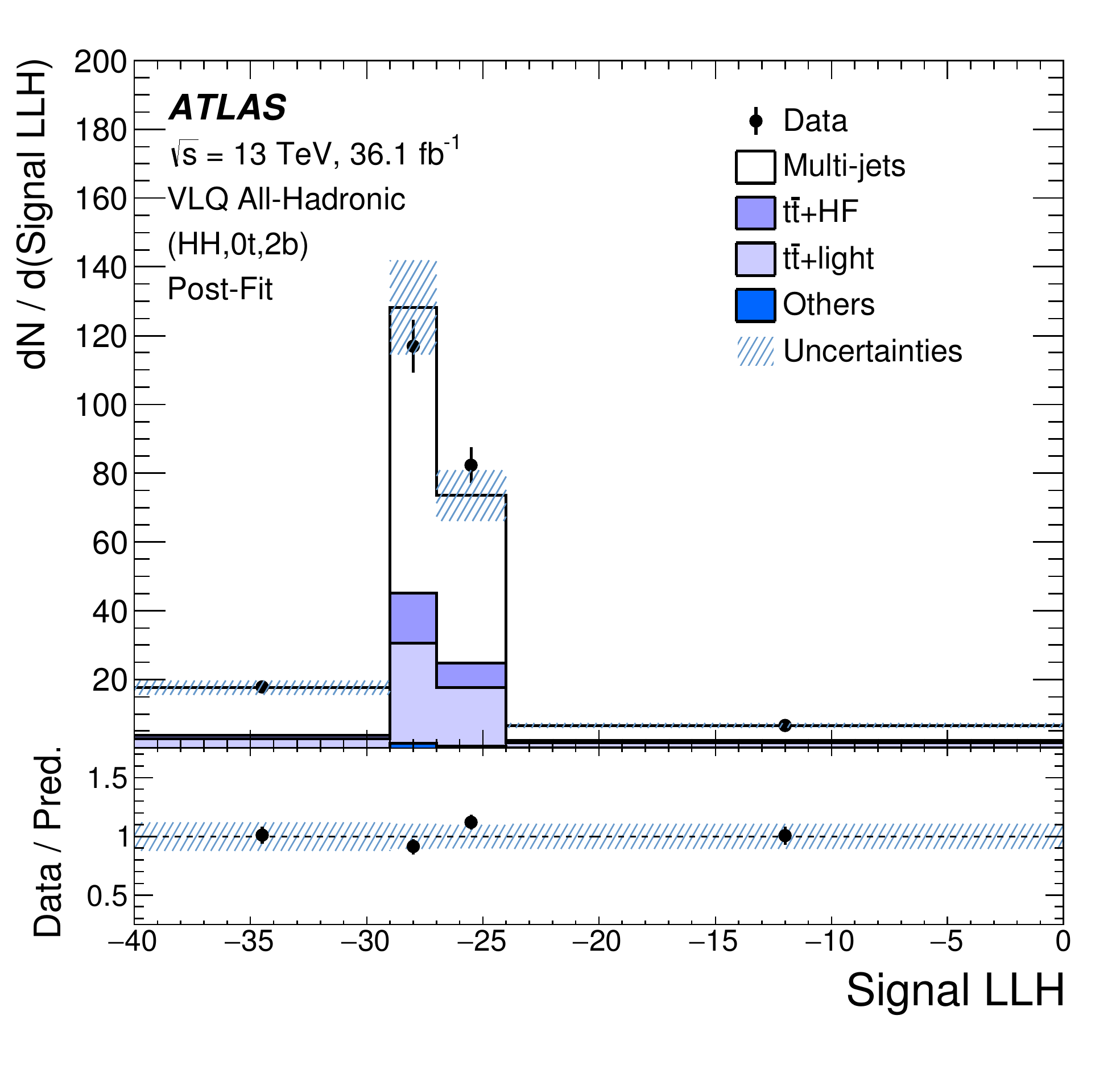}}
\subfloat[]{\includegraphics[width=0.4\textwidth]{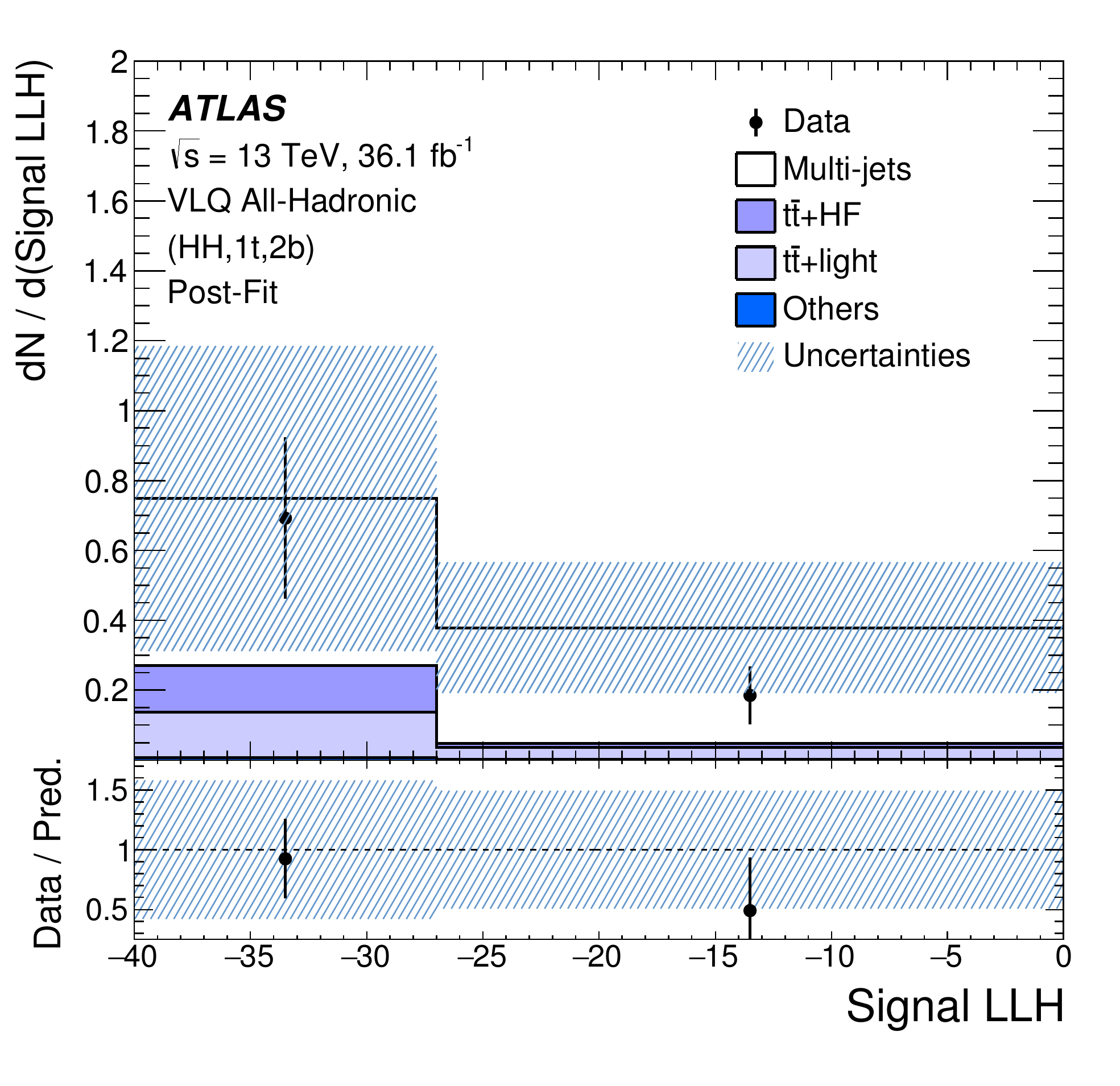}}
\caption{Comparison between data and prediction for the \LLH{} in the validation regions with
two Higgs-tagged \vrcjets{}, exactly two $b$-tagged \smallR{} jets,
and either (a) zero top-tagged \vrcjets{} or (b) exactly one
top-tagged \vrcjet{}. The distributions show the number of events
per width of 1.0 in the $x$-axis.  The hatched area represents the
uncertainty on the background from statistical uncertainty and all
sources of systematic uncertainty described
in~\Sect{\ref{sec:Systematics}}.  The background and uncertainty
take into account the constraints, pulls and correlations of the
background-only fit to data of the signal regions, described
in~\Sect{\ref{sec:StatisticalAnalysis}}, and are therefore described
as `Post-fit'. The underflow and overflow are included in the first
and last bins, respectively.}
\label{fig:results:HH_VR_dist}
\end{figure}

Standard Model backgrounds from \ttbar{}, single-top-quark, and \ttbarX{} processes
are estimated with simulated events, described
in \Sect{\ref{sec:Samples}}.  The normalization and shape are taken
directly from simulation for all of these processes.
 
\FloatBarrier

\section{Systematic uncertainties}
\label{sec:Systematics}
 
The systematic uncertainties considered in this analysis arise from
uncertainties in the treatment of the luminosity, object
reconstruction and background modeling.  Each source of uncertainty is
treated as a nuisance parameter in the final likelihood fit, as
described in~\Sect{\ref{sec:StatisticalAnalysis}}.  Different sources
of uncertainty are assumed to be uncorrelated; however, a given
uncertainty is assumed to be 100\% correlated across all regions and
samples. For each source of systematic uncertainty, the effect on the
analysis is evaluated by propagating a $\pm 1 \sigma$ variation of the
quantity in question.
 
\subsection{Luminosity and pileup}
 
The uncertainty in the integrated luminosity of the 2015 and 2016 data
set is 2.1\%.  It is derived, following a methodology similar to that
detailed in Ref.~\cite{DAPR-2013-01}, from a calibration of the
luminosity scale using $x$--$y$ beam-separation scans performed in
August 2015 and May 2016.  Because MC events are simulated
with different pileup conditions than observed in data, the events are
corrected to have the same pileup distributions as the data and an
uncertainty is assigned to account for the uncertainty in the ratio of
the predicted and measured inelastic proton--proton cross
section~\cite{STDM-2015-05}.

\subsection{Reconstructed objects}
 
Several systematic uncertainties in the simulated background and the
signal predictions arise from the reconstruction and identification of
the selected reconstructed objects, as described
in \Sect{\ref{sec:ObjectDefinitions}}, due to the determination of
correction factors applied to compensate for differences between data
and predictions.  The most important sources in this category are the
uncertainties associated with jets, missing transverse momentum, and
flavor tagging.  Other sources, such as lepton reconstruction
(affecting the lepton veto), are also considered, but have a
negligible impact on the results.  The impact on both shape and
normalization is taken into account for the following uncertainties.
 
\paragraph{Jets}
In case of the \smallR{} jet selection, uncertainties arise from the
jet reconstruction, the jet energy and mass scale calibrations, the JVT
requirement, and corrections to the jet energy and mass resolutions.
The most significant uncertainties associated with \smallR{} jets are
from energy scale and energy resolution.  The energy scale is
determined using the transverse momentum balance between a jet and a
reference object such as a photon, $Z$ boson, or another
jet~\cite{PERF-2016-04}.  The uncertainty in the energy scale ranges
from less than 1\% to around 5\% for $|\eta| < 0.8$ and \pt{} up to
500\,\GeV{}.  Jets with higher $|\eta|$ have an additional uncertainty
of up to 2\%.  The jet energy resolution is measured by studying dijet
events in data and simulation~\cite{PERF-2011-04}.  The jet energy
resolution in data and simulation are found to agree within 10\% and
the differences are used to determine the relative systematic
uncertainties, which range from 10\% to 20\%.  Additional
uncertainties are considered for the jet mass scale and mass
resolution, but are found to have little impact on the search
sensitivity. The uncertainties associated with \vrcjets{} are
inherited from the \smallR{} jet uncertainties.
 
\paragraph{Missing transverse momentum}
The \met{} is sensitive to changes in the momenta of the reconstructed
objects, namely the \smallR{} jets, as well as the additional soft
term that accounts for low-energy deposits not associated with a
reconstructed object.  Uncertainties from the reconstructed objects
are already accounted for. A soft-term uncertainty is assigned to
account for variations in the modeling of the underlying event that
change the amount of unclustered energy. The uncertainties in the
yields are in the range 0.0--18.7\% for simulated samples and
0.0--8.2\% for the multi-jet background.
 
\paragraph{Flavor tagging}
Uncertainties in the correction factors for the $b$-tagging
identification response are obtained by comparing the simulated event
samples with dedicated flavor-enriched samples in
data~\cite{ATL-PHYS-PUB-2016-012}. An additional term is included to
extrapolate the measured uncertainties to the high-\pt{} region of
interest. This term is calculated from simulated events by considering
variations of the quantities affecting the $b$-tagging performance
such as the impact parameter resolution, percentage of poorly measured
tracks, description of the detector material, and track multiplicity
per jet. The dominant effect on the uncertainty when extrapolating to
high \pt{} is related to the different tagging efficiency when
smearing the track impact parameters based on the resolution measured
in data and simulation.
 
Most of the \vrcjet{}-tagger flavor-tagging uncertainties can be
derived by propagating the \smallR{} jet uncertainties through the
DNN.  An additional uncertainty associated with $b$-tagging is
evaluated to take into account the use of $b$-tagging information in
the \vrcjet{}-tagger.  This is a \pt{}-dependent uncertainty in
the \vrcjet{}-tagging efficiency, considered separately for $V$-boson,
Higgs-boson, and top-quark tagging.  This uncertainty in the yields
ranges from 4.0 to 11.9\% for simulated samples and from 0.3 to 9.4\%
for the multi-jet background.
 
\subsection{Background modeling}
A theory cross-section uncertainty of 5.3\% is taken for the combined
small backgrounds, which are dominated by single-top-quark
processes~\cite{Kidonakis:2010ux}.
 
\paragraph{Multi-jet estimation}
The dominant multi-jet background is estimated using a data-driven
ABCD technique, as described in \Sect{\ref{sec:backgrounds}}.  To
quantify a closure uncertainty for this method, the difference between
the prediction and data in the one--$b$-tag validation regions is
propagated as an overall normalization uncertainty to the
corresponding two-- and three--$b$-tag signal regions. The impact of
including shape information in this uncertainty is negligible. To
allow potential differences in performance as a function of jet
multiplicity, the uncertainties are taken to be uncorrelated between
the regions with exactly two $b$-tags and at least three $b$-tags.
 
Another uncertainty is taken from the impact on the multi-jet
prediction of potential signal contamination in the validation
regions.  Detector-related uncertainties associated with all
backgrounds estimated using simulation, as well as modeling
uncertainties in \ttbar{} processes, are also propagated through the
multi-jet estimation via subtraction of non--multi-jet events in the
validation regions.  These uncertainties take into account differences
in both shape and normalization.
 
In addition to the systematic uncertainties, each bin of the multi-jet
prediction is assigned an uncertainty to account for statistical
uncertainties in the CRs propagated through the ABCD method.  Along
with the statistical uncertainty of the data in the SR, these tend to
have the largest impact on the sensitivity of the analysis.

\paragraph{\ttbar{} modeling}
For the \ttbar{} background, systematic uncertainties are considered
for variations in initial- and final-state radiation (ISR/FSR), choice
of parton shower, and choice of matrix-element generator. Each of
these sources of uncertainty are considered as separate nuisance
parameters in the likelihood fit.  These are evaluated using
alternative simulated \ttbar{} samples. The uncertainty in the
treatment of radiative effects is estimated by varying the NLO
radiation factor $h_\text{damp}$ and the factorization and
renormalization scales in a correlated way to produce more or less
radiation.  Alternative samples produced
with \AMCatNLOPYTHIAV{v2.3.3}{8.212} and \POWHERWIGV{v2}{7.0.1}~\cite{Herwig} are used to
evaluate generator and shower model uncertainties, respectively.  Due to the limited number of events in the
alternative \ttbar{} samples, the uncertainties are taken into account
after merging signal regions with two and at least three $b$-tagged
jets.  These uncertainties are in the range 1.4--33\% (13--51\%) for
the normalization of \ttlight{} (\tthf).  Because the predicted cross
sections of \ttlight\ and \tthf\ are not well known for the phase
space of the signal regions, separate normalization factors are
assigned to each of these two contributions and are allowed to float
freely in the profile likelihood fit.
 
 
\section{Statistical analysis}
\label{sec:StatisticalAnalysis}
The statistical analysis quantifies the probability of compatibility
between the measured data, expected SM background, and expected
signal.  The \LLH{} distributions for the twelve signal regions are
tested simultaneously for the presence of a VLQ signal.  Hypothesis
testing is performed using a modified frequentist method based on a
profile likelihood, taking into account systematic uncertainties as
nuisance parameters.  The statistical analysis is based on a binned
likelihood function $\mathcal{L}(\mu,\theta)$ constructed as the
product of Poisson probability terms over all bins.  In this function,
$\mu$ is a multiplicative factor applied to the predicted production
cross section times branching ratio of the signal, and $\theta$ is the
set of nuisance parameters, implemented in the likelihood function as
Gaussian or log-normal priors. In addition, there are two
unconstrained parameters in the fit, corresponding to the total
normalization of \ttlight{} and \tthf{}.
 
The test statistic $q_\mu$ is defined as the profile likelihood ratio:
$q_{\mu}=-2\mathrm{ln}(\mathcal{L}(\mu,\hat{\hat{\theta}}_\mu)/ \mathcal{L}(\hat{\mu},
\hat{\theta}))$,where $\hat{\mu}$ and $\hat{\theta}$ are the values of the
parameters that maximize the likelihood function (with the constraint
0$\leq\hat{\mu}\leq\mu$), and $\hat{\hat{\theta}}_\mu$ are the values
of the nuisance parameters that maximize the likelihood function for a
given value of $\mu$.  Upper limits on the signal production cross
section for each of the signal scenarios considered are derived by
using $q_\mu$ in the \CLs{} method~\cite{cls_2, cls}, where \CLs{} is
computed using the asymptotic approximation~\cite{cls_3}.  For a given
signal scenario, values of the production cross section that yield
$\CLs<0.05$ are excluded at $\geq$95\% confidence level (CL).

 
\section{Results}
 
Following the prescription described in~\Sect{\ref{sec:StatisticalAnalysis}},
the profile likelihood fit for the background-only hypothesis is
performed simultaneously in all signal regions.  The post-fit event
yields are given in \Tab{\ref{tab:results:FinalYields}}
and \Fig{\ref{fig:results:Summary}} shows a comparison between the
predicted and observed numbers of events in all signal regions both
before and after the fit.
 
The most notable shift in the post-fit yields is that of the \ttHF{}
normalization. The overall change in normalization is by a factor
slightly greater than two, which is achieved in the fit through a
shift of the \ttHF{} normalization factor, as well as through pulls of
systematic uncertainties, such as \ttbar{} modeling and jet energy
resolution uncertainties. The post-fit distributions of the \LLH{}
from each signal region are shown
in \Figrange{\ref{fig:results:VV_PostFitResult}}{\ref{fig:results:HH_PostFitResult}}.
 
\begin{figure}[!ht]
\centering
\subfloat[]{\includegraphics[width=0.75\textwidth]{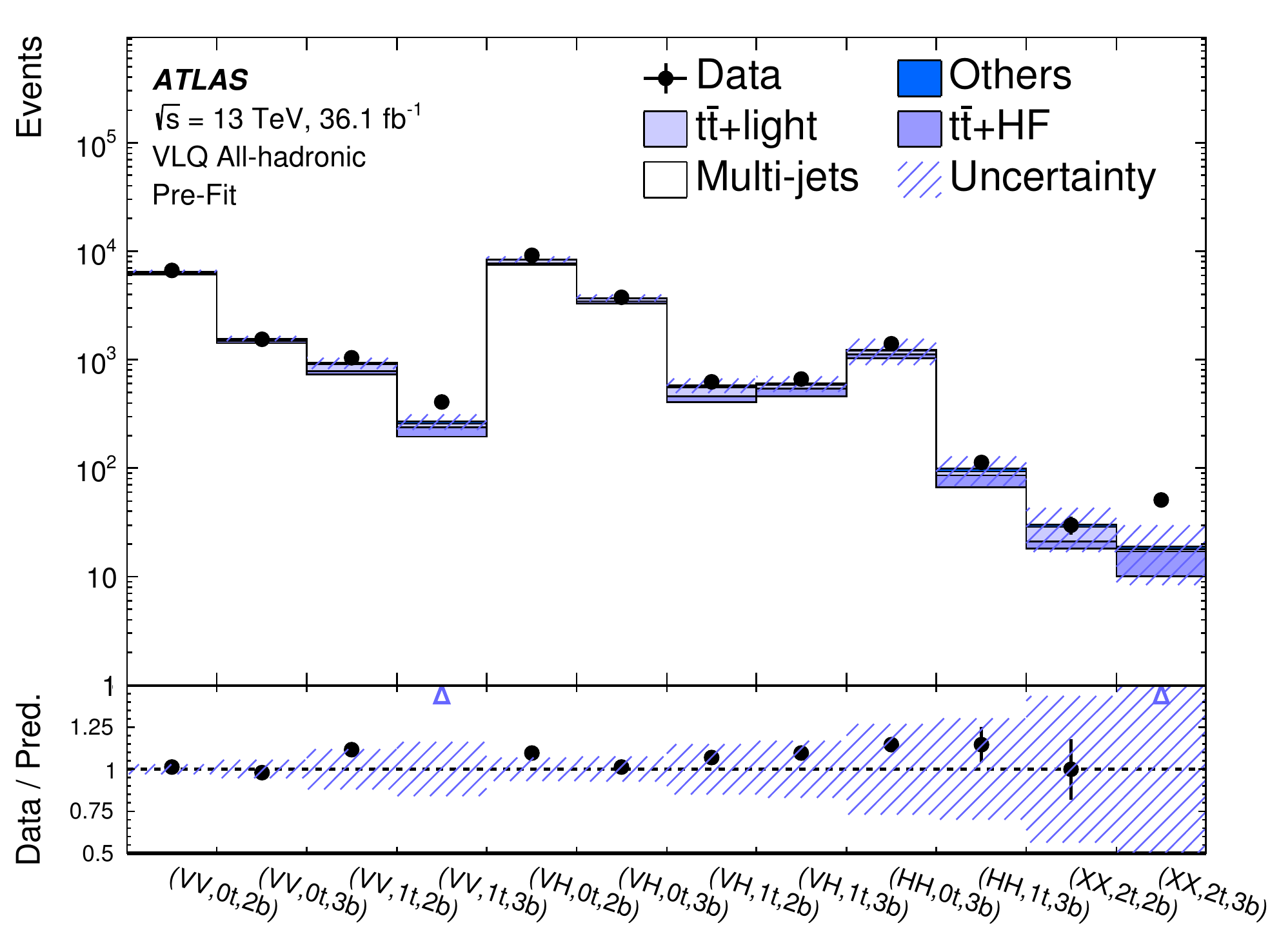}        } \\
\subfloat[]{\includegraphics[width=0.75\textwidth]{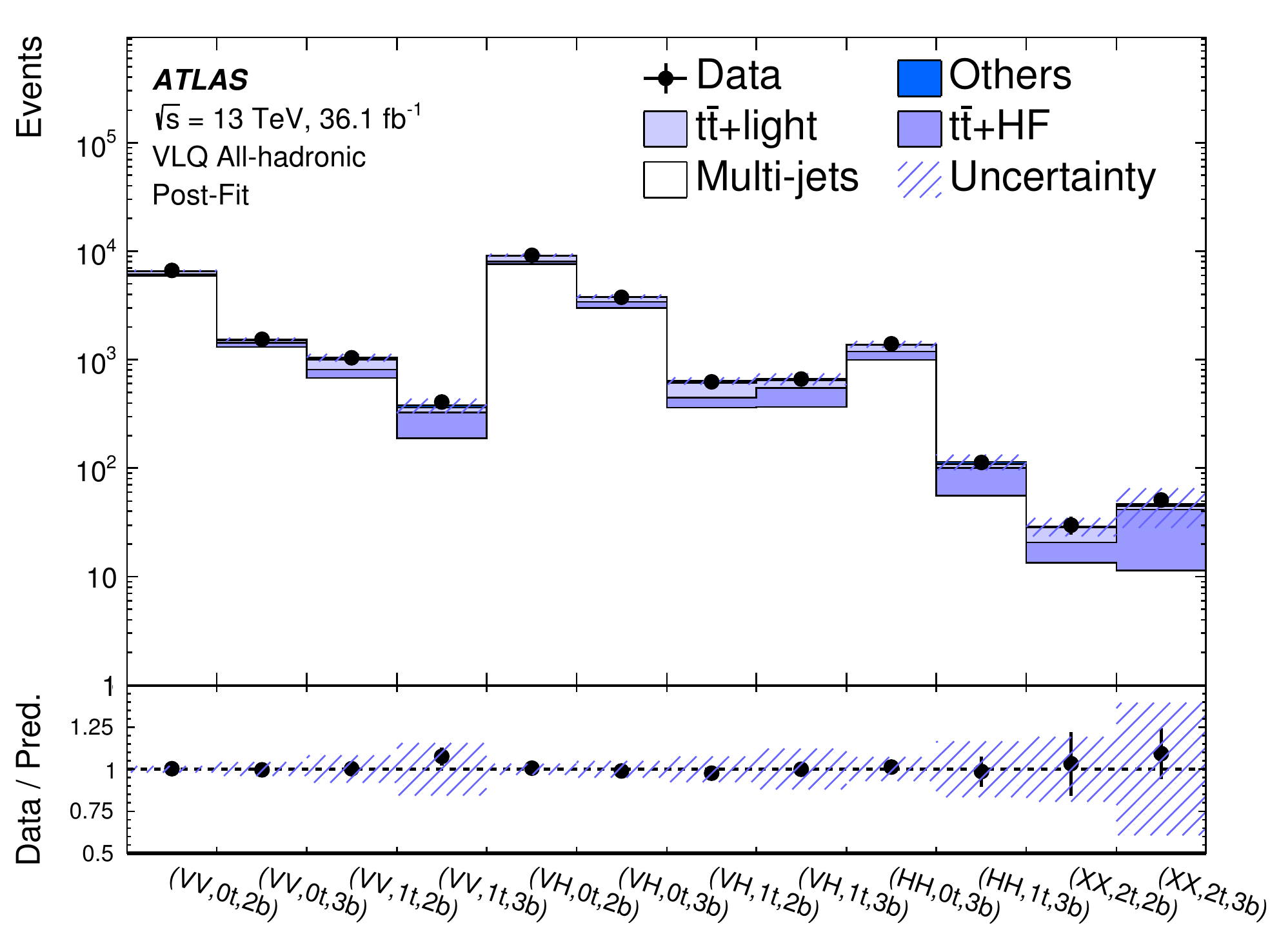}}
\caption{Comparison between data and prediction for the event yields (a) before and (b) after the fit to the
data under the background-only hypothesis. The figures show the
total normalization across all signal regions. The contribution
labeled `Others' is the combination of single-top-quark and \ttbarX{}
backgrounds.  The hatched area represents the total uncertainty of
the background.}
\label{fig:results:Summary}
\end{figure}
\begin{table}[htbp]
\small
\begin{center}
\caption{Event yields in all twelve signal regions after the fit to
data under the background-only hypothesis, as well as the predicted
signal event yields before the fit for a $B$ VLQ with a mass of
1~\TeV{}. The contribution labeled `Others' is the combination of
single-top-quark and \ttbarX{} backgrounds. The uncertainties include
statistical and systematic uncertainties. The uncertainties of the
individual background components can be larger than the uncertainty
on the sum of the backgrounds due to correlations.}
 
\label{tab:results:FinalYields}
\scriptsize
\begin{tabular}{ l
S[table-format=4(3), separate-uncertainty]
S[table-format=3(3), separate-uncertainty]
S[table-format=3(3), separate-uncertainty]
S[table-format=3.1(2), separate-uncertainty]
S[table-format=4(3), separate-uncertainty]
S[table-format=2.2(3), separate-uncertainty]
S[table-format=2.2(3), separate-uncertainty]
S[table-format=4, separate-uncertainty]}
\toprule
{Region}  &   {Multi-jet}   &  {\ttlight} & {\tthf} & {Others} & {Total background} & {$m_B$ = 1 TeV} & {$m_T$ = 1 TeV} & {Data}  \\
&    &   &  &  &  & {\BR($B \rightarrow Hb$) = 1} & {\BR($T \rightarrow Ht$) = 1} & \\
\midrule
\VVbb\     & 5890(190) & 380(170) & 230(90)  &  92(12) &  6590(110) & 8.0(10)     & 3.5(05)   & 6614 \\
\VVbbb\    & 1300(60)  &  80(40)  & 130(60)  &  31(8)  &  1540(40)  & 11.5(10)    & 3.8(06)   & 1534 \\
\VVtbb\    &  680(80)  & 190(90)  & 130(60)  &  41(11) &  1040(90)  & 2.2(04)     & 11.6(14)  & 1044 \\
\VVtbbb\   &  190(40)  &  40(26)  & 130(70)  &  16(5)  &   380(60)  & 3.1(04)     & 7.4(11)   & 409 \\
\VHbb\     & 7500(400) & 1000(500)& 500(210) & 129(15) &  9150(340) & 23.4(31)    & 1.33(033) & 9202 \\
\VHbbb\    & 3010(180) & 310(140) & 430(200) &  76(17) &  3820(170) & 70(6)       & 6.2(07)   & 3778 \\
\VHtbb\    &  360(60)  & 160(70)  &  80(40)  &  28(6)  &   640(50)  & 3.9(07)     & 6.1(08)   & 623 \\
\VHtbbb\   &  370(50)  & 100(60)  & 180(80)  &  19(5)  &   660(90)  & 18.2(22)    & 37.3(33)  & 662 \\
\HHbbb\    &  990(110) & 180(90)  & 200(100) &  19(5)  &  1390(110) & 77(6)       & 38(4)     & 1407 \\
\HHtbbb\   &   56(13)  &   8(5)   &  44(24)  & 6.4(16) &   115(16)  & 17.1(20)    & 39(4)     & 113 \\
\XXttbb\   &   13(4)   &   8(5)   &  7(5)    & 0.7(04) &    29(7)   & 0.17(010)   & 35(4)     & 30 \\
\XXttbbb\  &   11(7)   &   3(4)   &  30(19)  & 2.0(08) &    47(21)  & 2.4(05)     & 16.1(23)  & 51 \\
\bottomrule
\end{tabular}
\end{center}
\end{table}
 
\begin{figure}[!ht]
\centering
\subfloat[]{\includegraphics[width=0.4\textwidth]{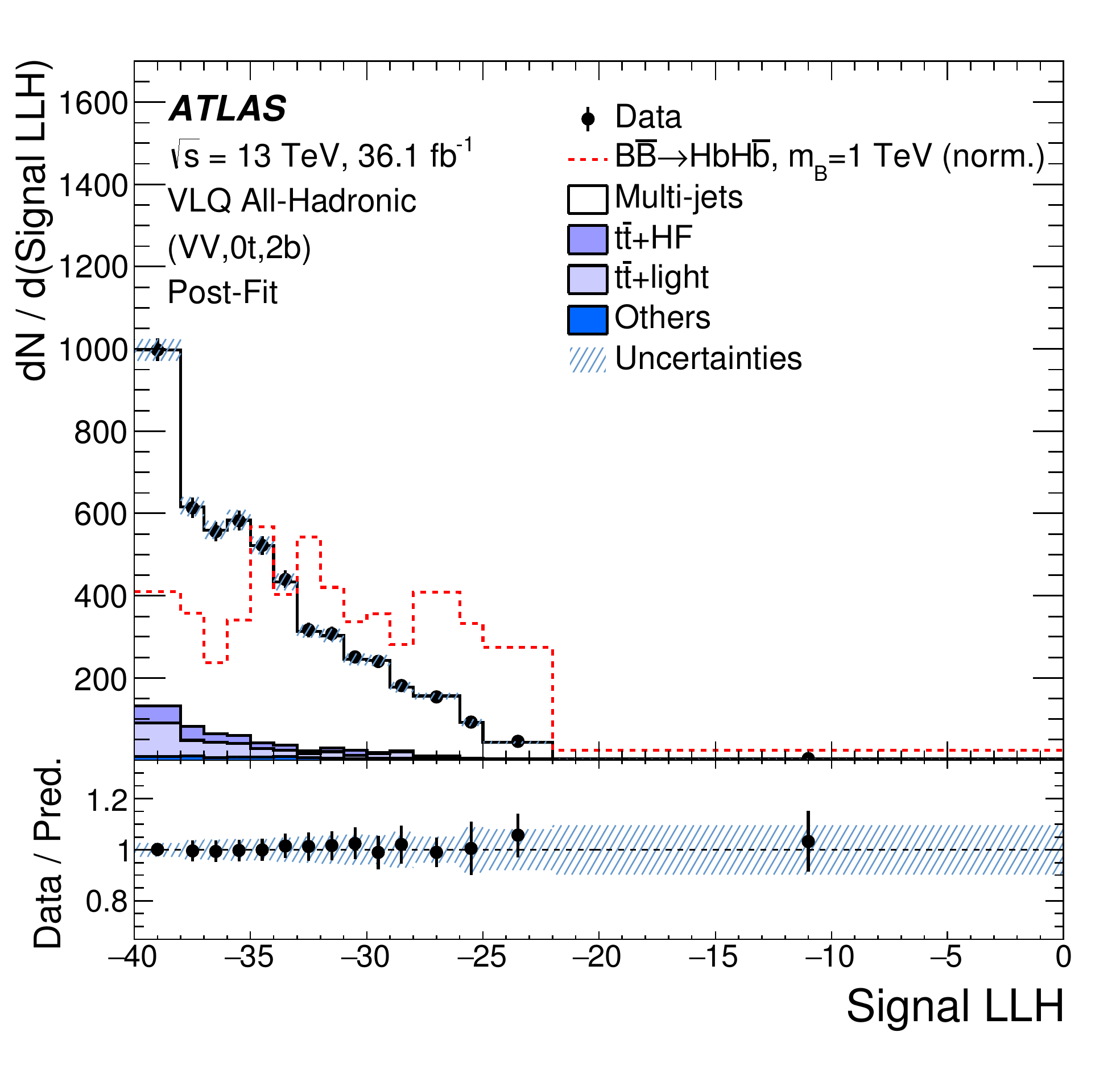}}
\subfloat[]{\includegraphics[width=0.4\textwidth]{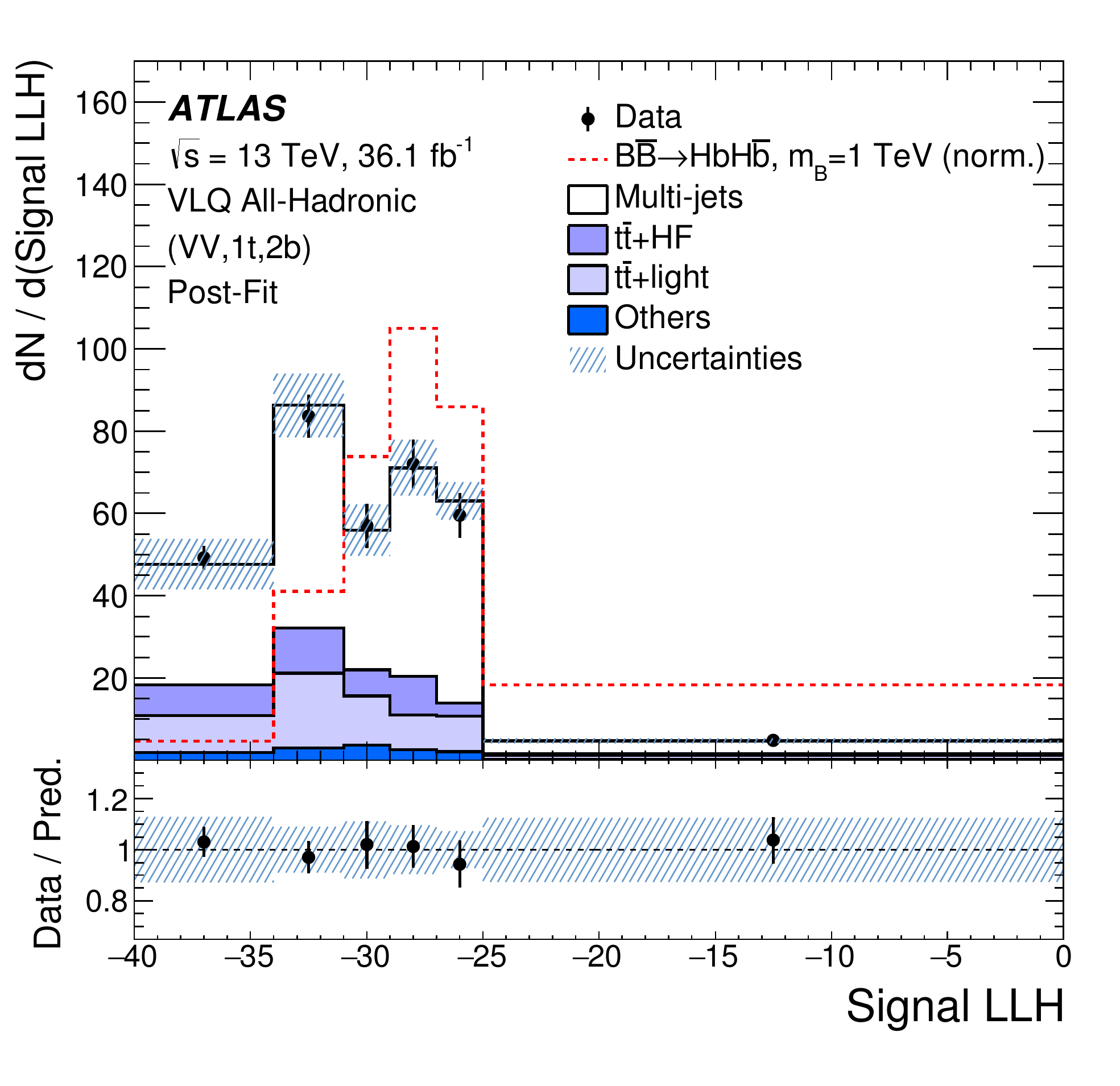}} \\
\subfloat[]{\includegraphics[width=0.4\textwidth]{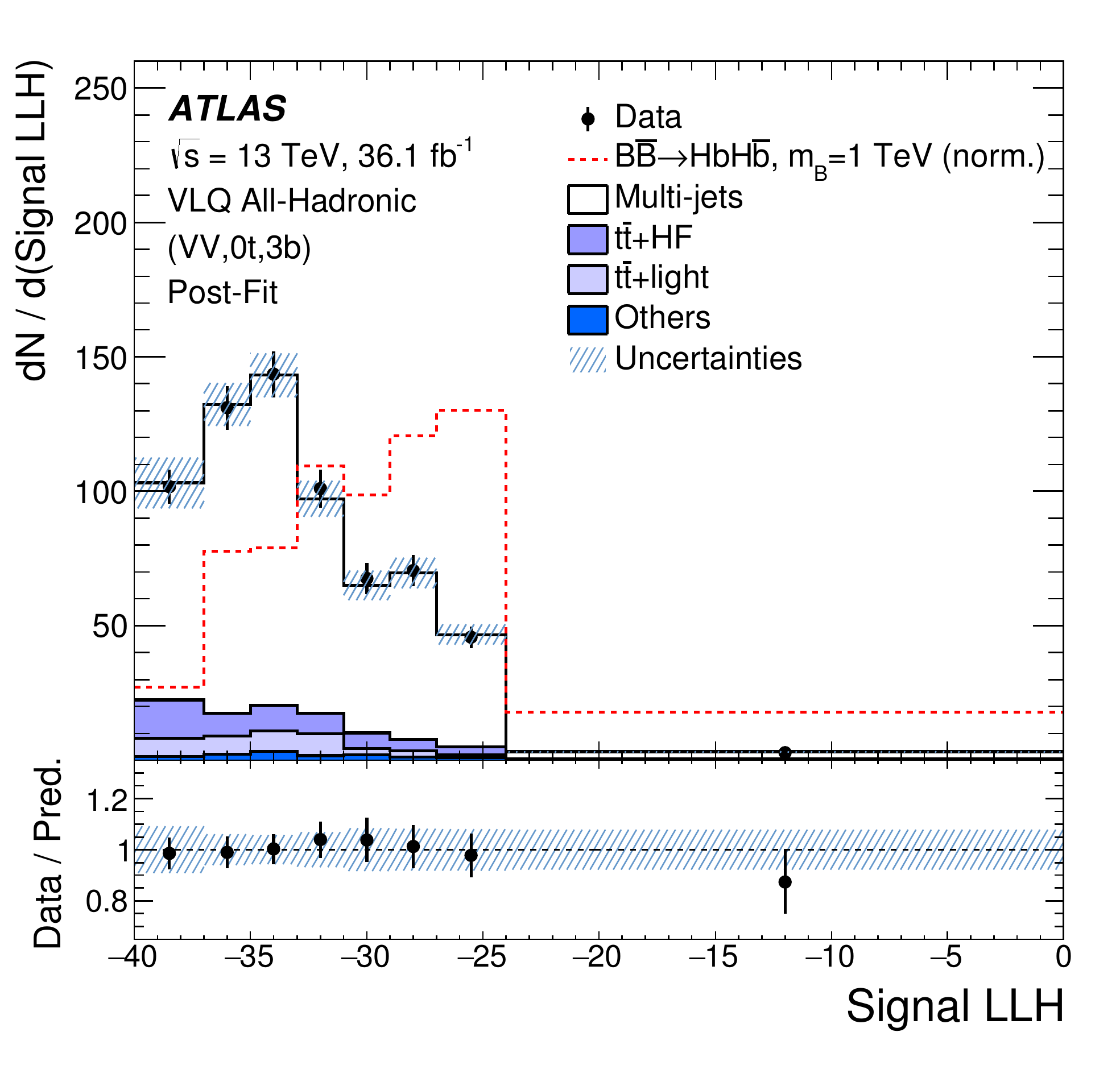}}
\subfloat[]{\includegraphics[width=0.4\textwidth]{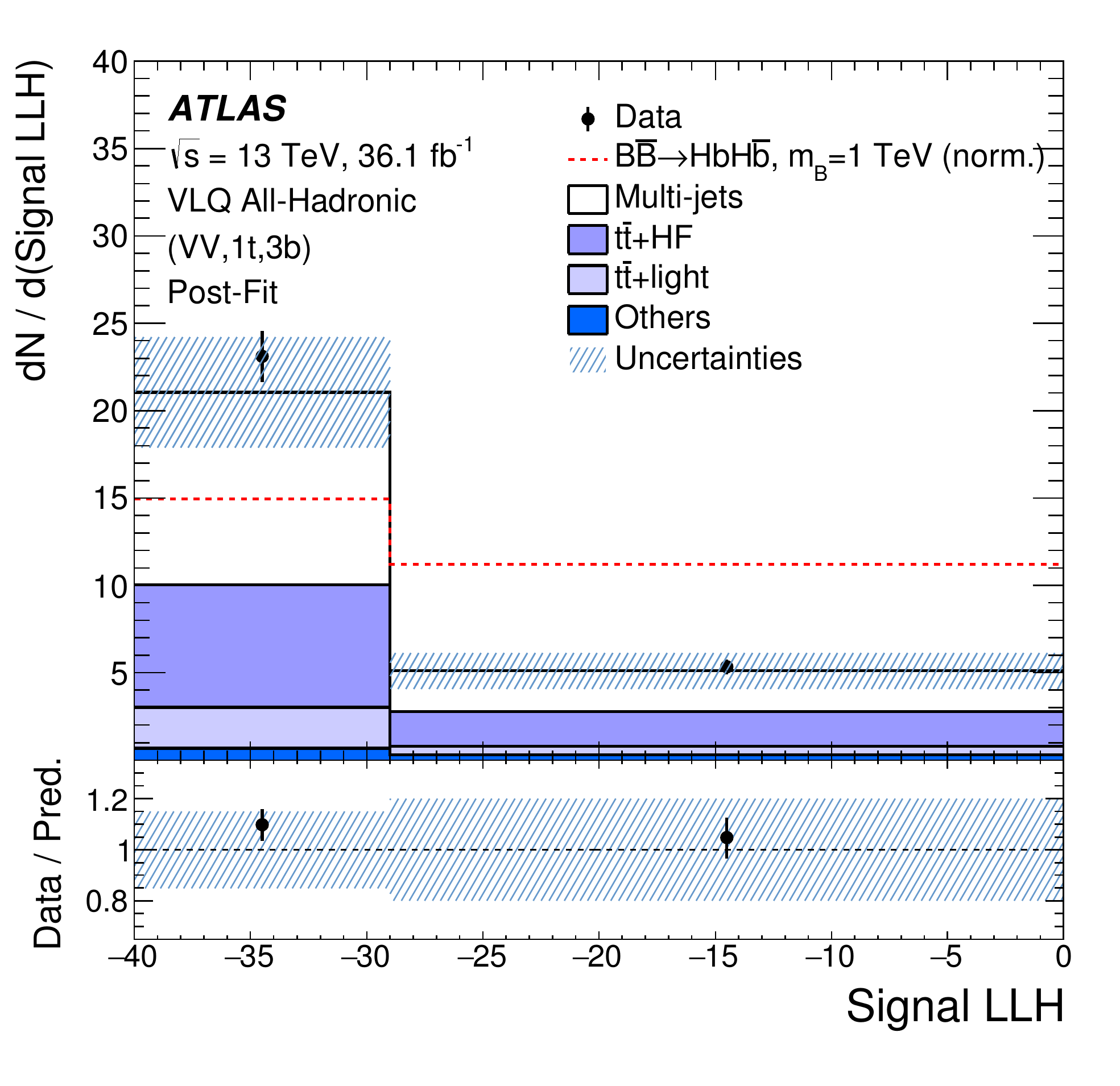}}
\caption{Comparison between data and prediction for the \LLH{} distribution after the fit to the data under the
background-only hypothesis for the (a) \VVbb, (b) \VVtbb,
(c) \VVbbb\ and (d) \VVtbbb\ signal regions. The contribution
labeled `Others' is the combination of single-top-quark and \ttbarX{}
backgrounds. The distributions show the number of events per width
of 1.0 in the $x$-axis. The hatched area represents the total
uncertainty of the background. The underflow and overflow are
included in the first and last bins, respectively. A hypothetical
signal for $\BR(B\rightarrow Hb)=$ 100\% and $m_B$ = 1 TeV is
shown overlaid, normalized to the integral of the total
background.}  \label{fig:results:VV_PostFitResult}
\end{figure}
 
\begin{figure}[!ht]
\centering
\subfloat[]{\includegraphics[width=0.4\textwidth]{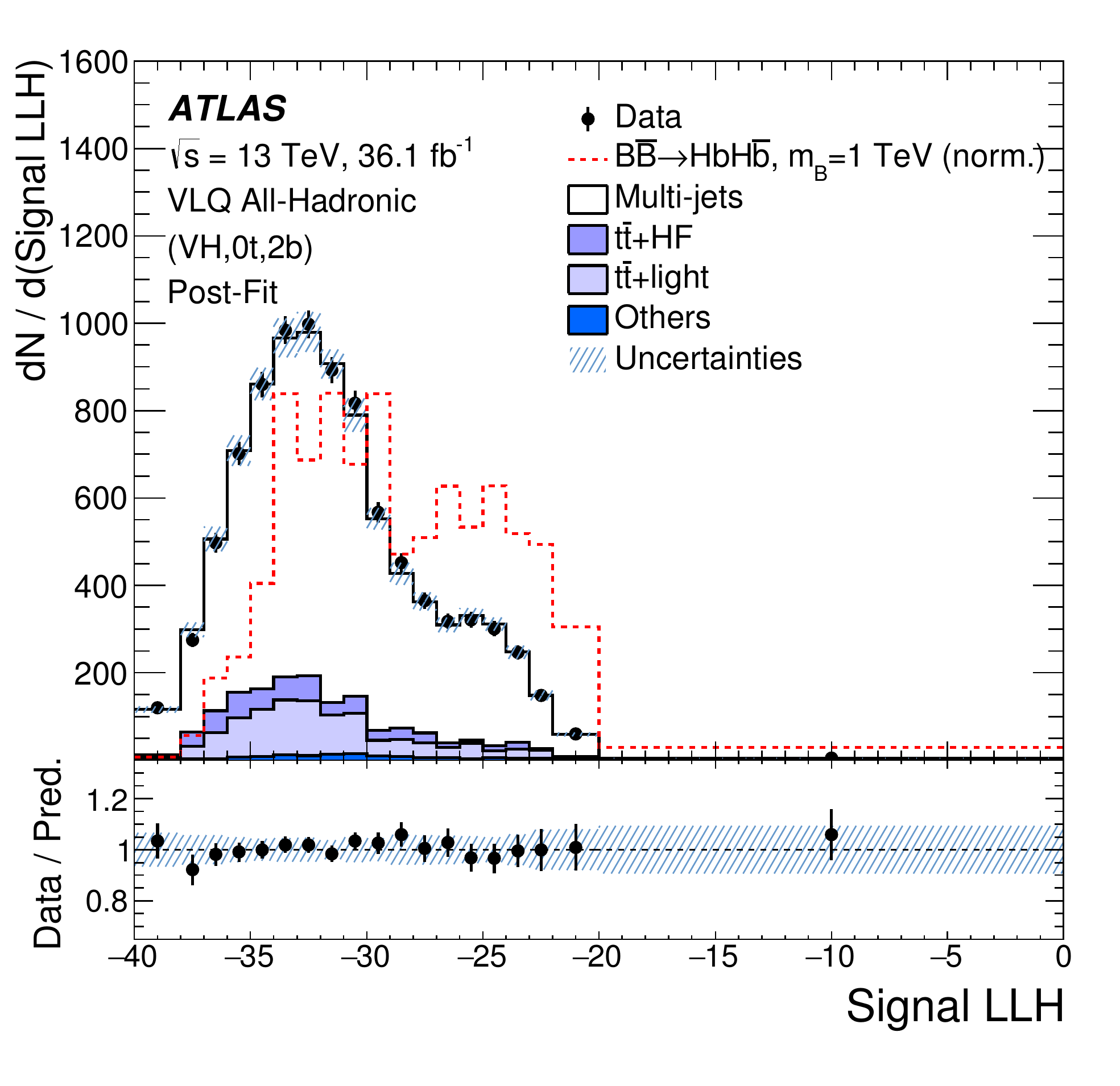}}
\subfloat[]{\includegraphics[width=0.4\textwidth]{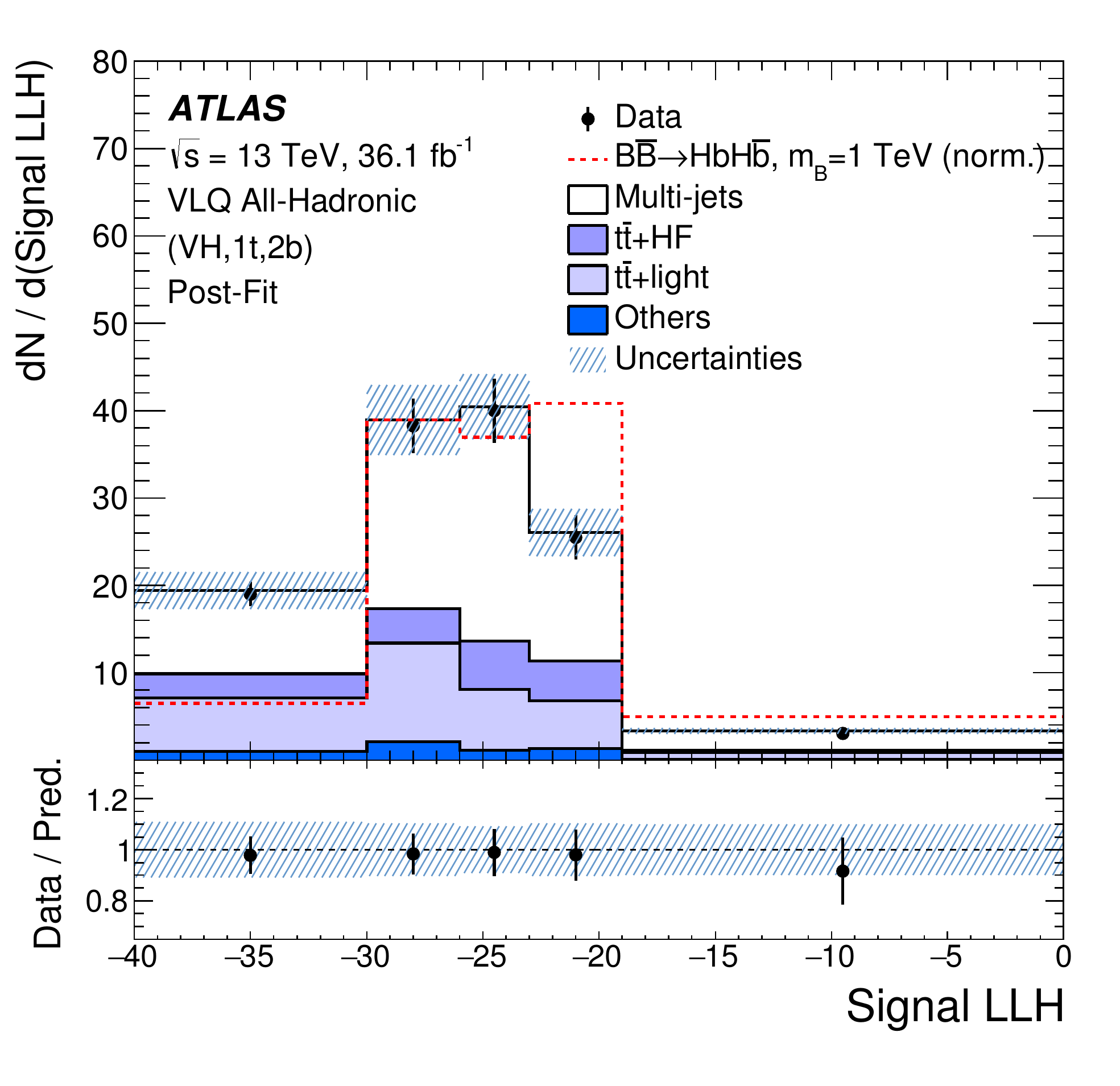}} \\
\subfloat[]{\includegraphics[width=0.4\textwidth]{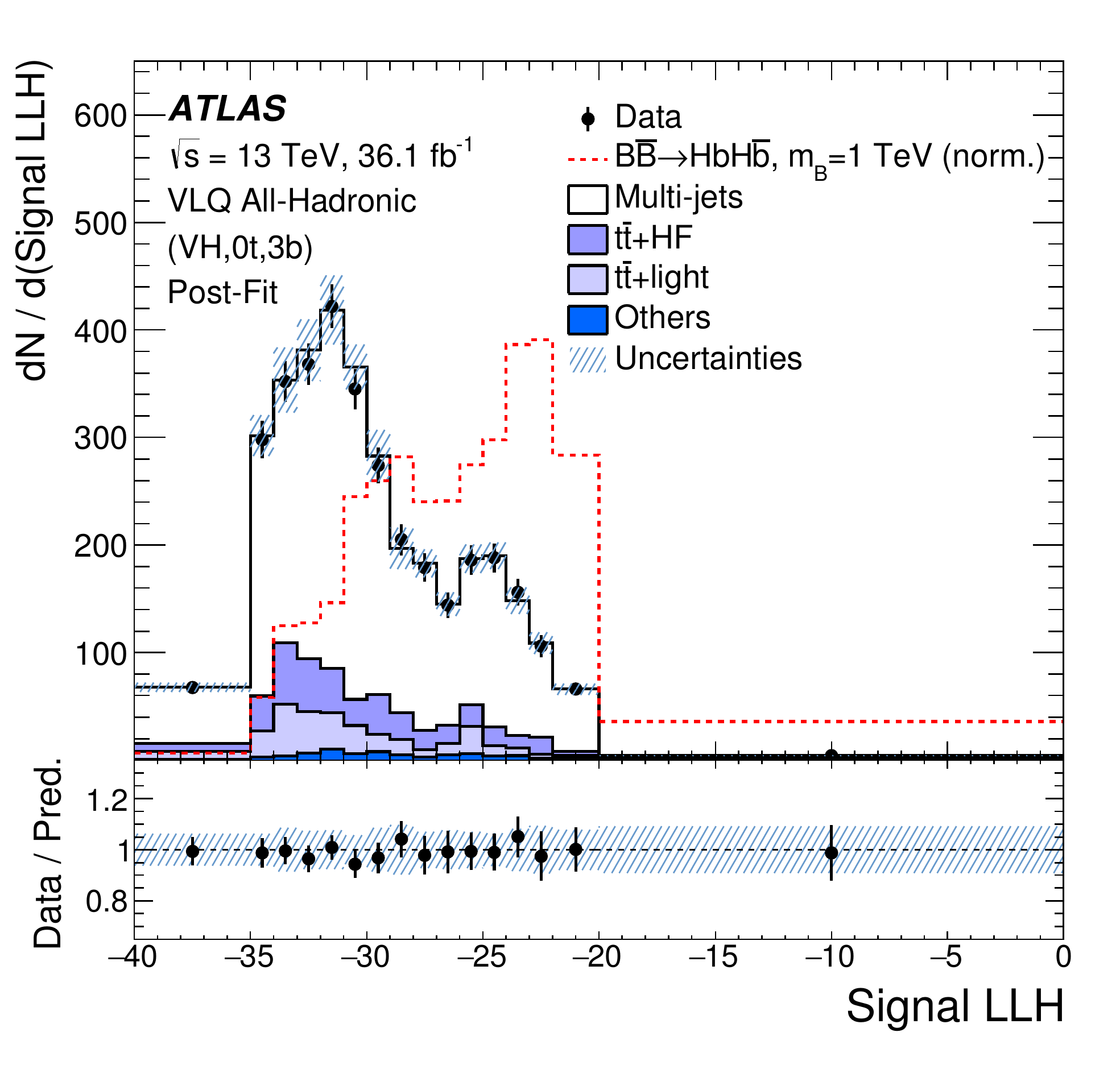}}
\subfloat[]{\includegraphics[width=0.4\textwidth]{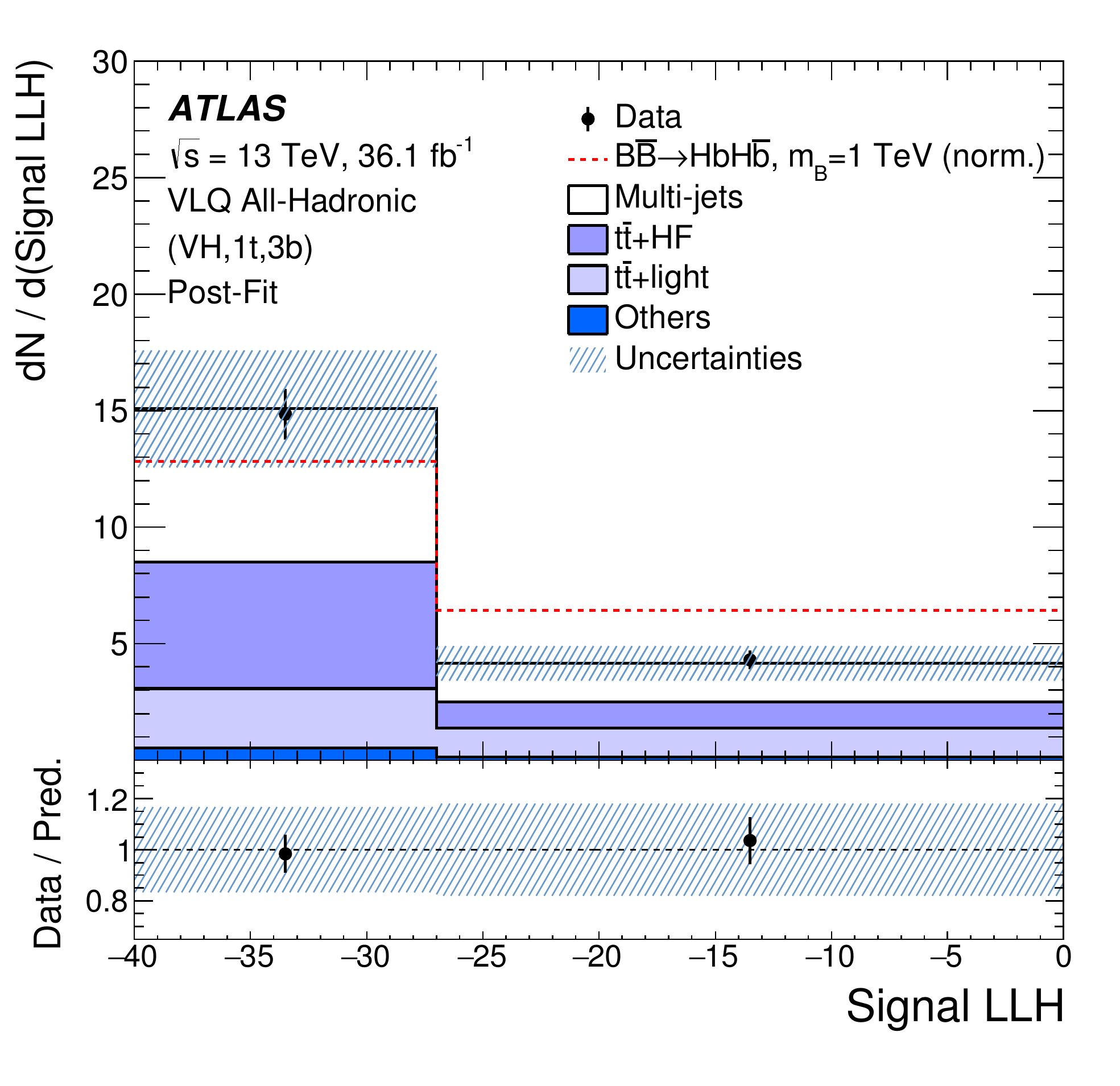}}
\caption{Comparison between data and prediction for the \LLH{} distribution after the fit to the data under the
background-only hypothesis for the (a) \VHbb, (b) \VHtbb,
(c) \VHbbb\ and (d) \VHtbbb\ signal regions. The contribution
labeled `Others' is the combination of single-top-quark and \ttbarX{}
backgrounds. The distributions show the number of events per width
of 1.0 in the $x$-axis. The hatched area represents the total
uncertainty of the background. The underflow and overflow are
included in the first and last bins, respectively. A hypothetical
signal for $\BR(B\rightarrow Hb)=$ 100\% and $m_B$ = 1 TeV is
shown overlaid, normalized to the integral of the total
background.}
\end{figure}
 
\begin{figure}[!ht]
\centering
\subfloat[]{\includegraphics[width=0.4\textwidth]{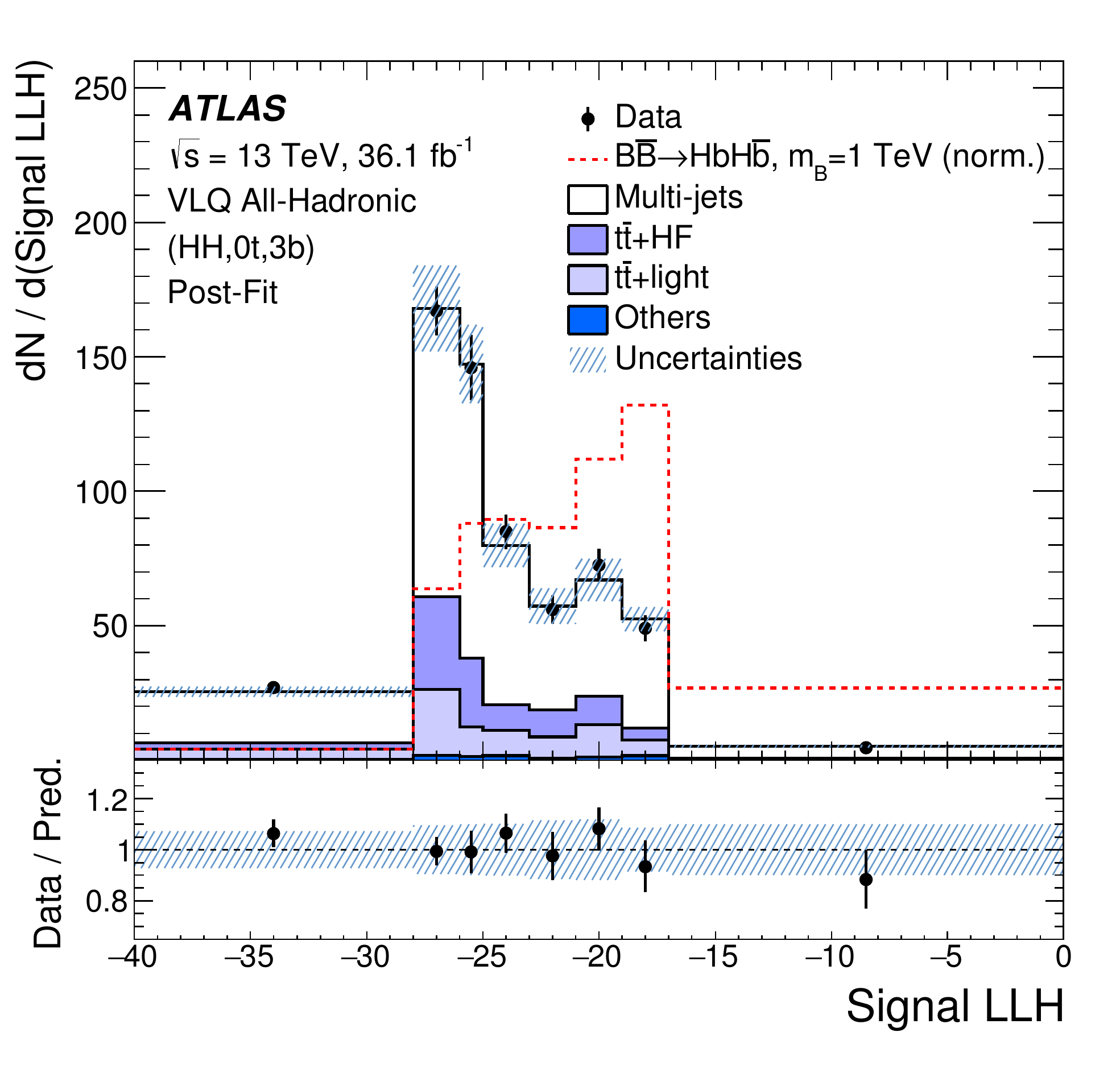}}
\subfloat[]{\includegraphics[width=0.4\textwidth]{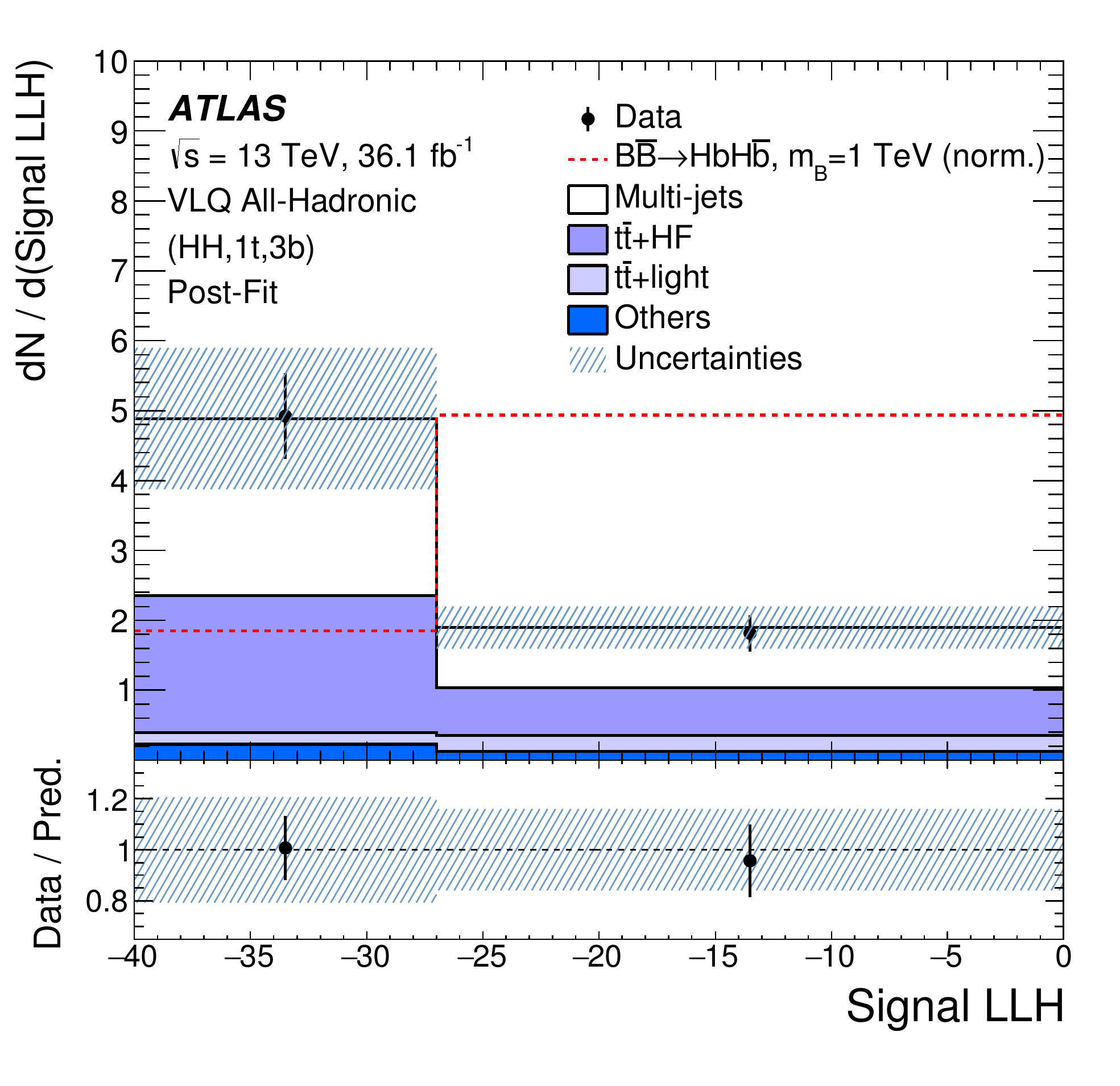}} \\
\subfloat[]{\includegraphics[width=0.4\textwidth]{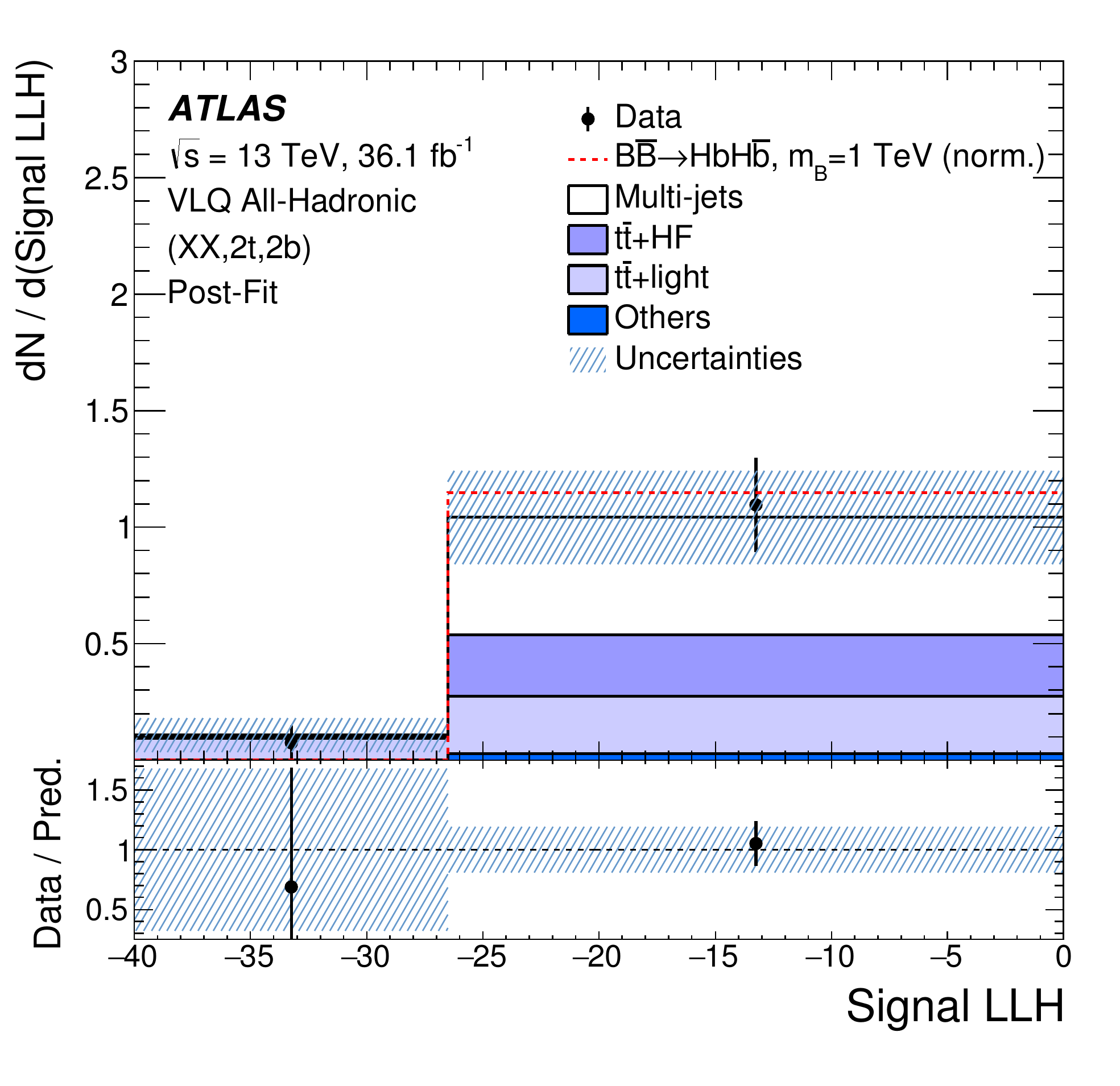}}
\subfloat[]{\includegraphics[width=0.4\textwidth]{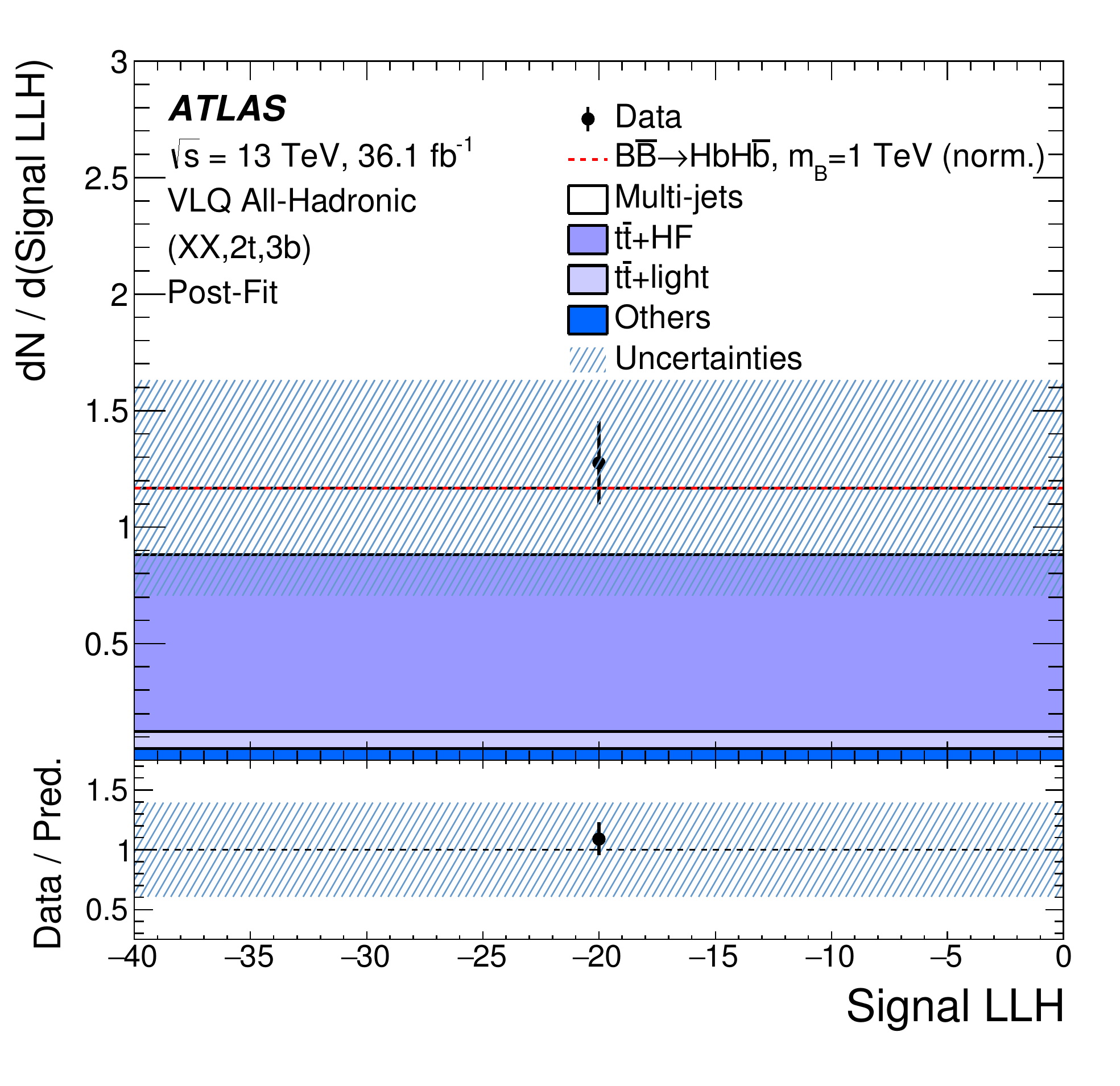}}
\caption{Comparison between data and prediction for the \LLH{} distribution after the fit to the data under the
background-only hypothesis for the (a) \HHbbb, (b) \HHtbbb,
(c) \XXttbb\ and (d) \XXttbbb\ signal regions. The contribution
labeled `Others' is the combination of single-top-quark and \ttbarX{}
backgrounds. The distributions show the number of events per width
of 1.0 in the $x$-axis. The hatched area represents the total
uncertainty of the background. The underflow and overflow are
included in the first and last bins, respectively. A hypothetical
signal for $\BR(B\rightarrow Hb)=$ 100\% and $m_B$ = 1 TeV is
shown overlaid, normalized to the integral of the total
background.}  \label{fig:results:HH_PostFitResult}
\end{figure}
 
No significant excess of signal-like events is observed, and the
analysis proceeds to set upper limits on the production cross section
of \TTbar{} and \BBbar{} events in various scenarios.  The sensitivity
is mainly limited by the statistical uncertainty in the signal regions
and in the control regions for the ABCD method.  For example, if only
statistical uncertainties and normalization factors are taken into
account, the expected (observed) cross-section limit for $\BBbar\to
HbH\bar b$ with $m_B=1$\,\TeV{} only changes by 5\% (11\%).
 
In a given scenario, a lower limit on the VLQ mass can be obtained by
comparing the cross-section limits with the predicted cross section as
a function of
mass~\cite{Czakon:2011xx}. \Fig{\ref{fig:results:LimitPure}} shows the
expected and observed upper limits on the \TTbar{} and \BBbar{} cross
section at 95\% CL as a function of the VLQ mass in the scenario where
the VLQ decays purely via the Higgs decay mode ($\TTbar\to HtH\bar t$
or $\BBbar \to HbH\bar b$), as well as in the benchmark scenario of
the ($B, Y$) doublet.  In this scenario, a $B$ VLQ will decay almost
equally into $Zb$ and $Hb$, although the exact branching ratios depend
on mass. For example, for $m_B = 1$\,\TeV{}, \BR$(B\to Zb) =0.51$
and \BR$(B\to Hb) = 0.49$~\cite{protos}.  Only contributions from the
$B$ VLQ are considered, so the limit is conservative.  In the case of
a ($B, Y$) doublet, $B$ masses below \ObsMassLimitBBS{} are excluded
at 95\% CL.
 
To evaluate the level of sensitivity of the results to the
weak-isospin of the VLQ, samples of VLQ events with masses of 700,
950, and 1200~\GeV{} were generated for an SU(2) doublet $T$ and $B$
quark and compared with the SU(2) singlet samples. Small differences
between the limits are observed in decay modes with bottom quarks in
the final state, where the SU(2) singlet produces a slightly weaker
limit due to the slightly lower average momenta of decay products in
the singlet final state. Thus, limits on the ($B, Y$) SU(2) doublet,
which are taken from scaling the SU(2) singlet samples to doublet
branching ratios, represent a slightly conservative limit. In final
states with top quarks, the SU(2) singlet produces a slightly stronger
limit, due to a slightly higher efficiency in VLQ DNN top-tagging.
Therefore, a limit on SU(2) doublets with $T$ VLQs are not included
here.
 
The largest difference between the observed and expected limits is
for \BR$(B\to Hb) = 1$ with a VLQ mass around 950~\GeV.  This results
from a deficit in data in the final two bins of the \HHbbb{} signal
region and the fact that the matrix element calculation for final
states with two bottom quarks has its maximum sensitivity for masses
near 900\,\GeV{}.

By reweighting the relative fractions of the three $T$ ($B$) decay
modes, it is possible to test all combinations of branching
ratios.  \Fig{\ref{fig:results:LimitGrid}} shows the lower limit on
the $T$ ($B$) mass as a function of \BR($T\to Ht$) versus
\BR($T\to Wb$) (\BR($B\to Hb$) versus \BR($B\to Wt$)). Each point on the figures
have a total branching ratio of 1, so \BR($T\to Zt$) (\BR($B\to Zb$))
make up the remaining branching ratio for $T$ ($B$). The expected and
observed limits on the VLQ mass for each corner of the branching-ratio
plane are listed in \Tab{\ref{tab:results:singleBRs}}.
 
\FloatBarrier
 
\begin{table}[htbp]
\small
\begin{center}
\caption{Expected and observed 95\% CL limits on the VLQ mass for \TTbar{} and \BBbar{} production.
Different branching ratios are presented for $T$ and $B$. }
\label{tab:results:singleBRs}
\begin{tabular}{ l r r}
\toprule
Branching Ratio           & Expected [GeV] & Observed [GeV] \\
\midrule
\BR($B\rightarrow Wt)=1$   &   730  &  710 \\
\BR($B\rightarrow Zb)=1$   &   910  &  710 \\
\BR($B\rightarrow Hb)=1$   &   970  & 1010 \\
SU(2) ($B, Y$) doublet    &   890  &  950 \\
\midrule
\BR($T\rightarrow Wb)=1$   &   790  &  650 \\
\BR($T\rightarrow Zt)=1$   &   780  &  650 \\
\BR($T\rightarrow Ht)=1$   &  1010  & 1010 \\
\bottomrule
\end{tabular}
\end{center}
\end{table}
 
\begin{figure}[!ht]
\centering
\subfloat[]{\includegraphics[width=0.4\textwidth]{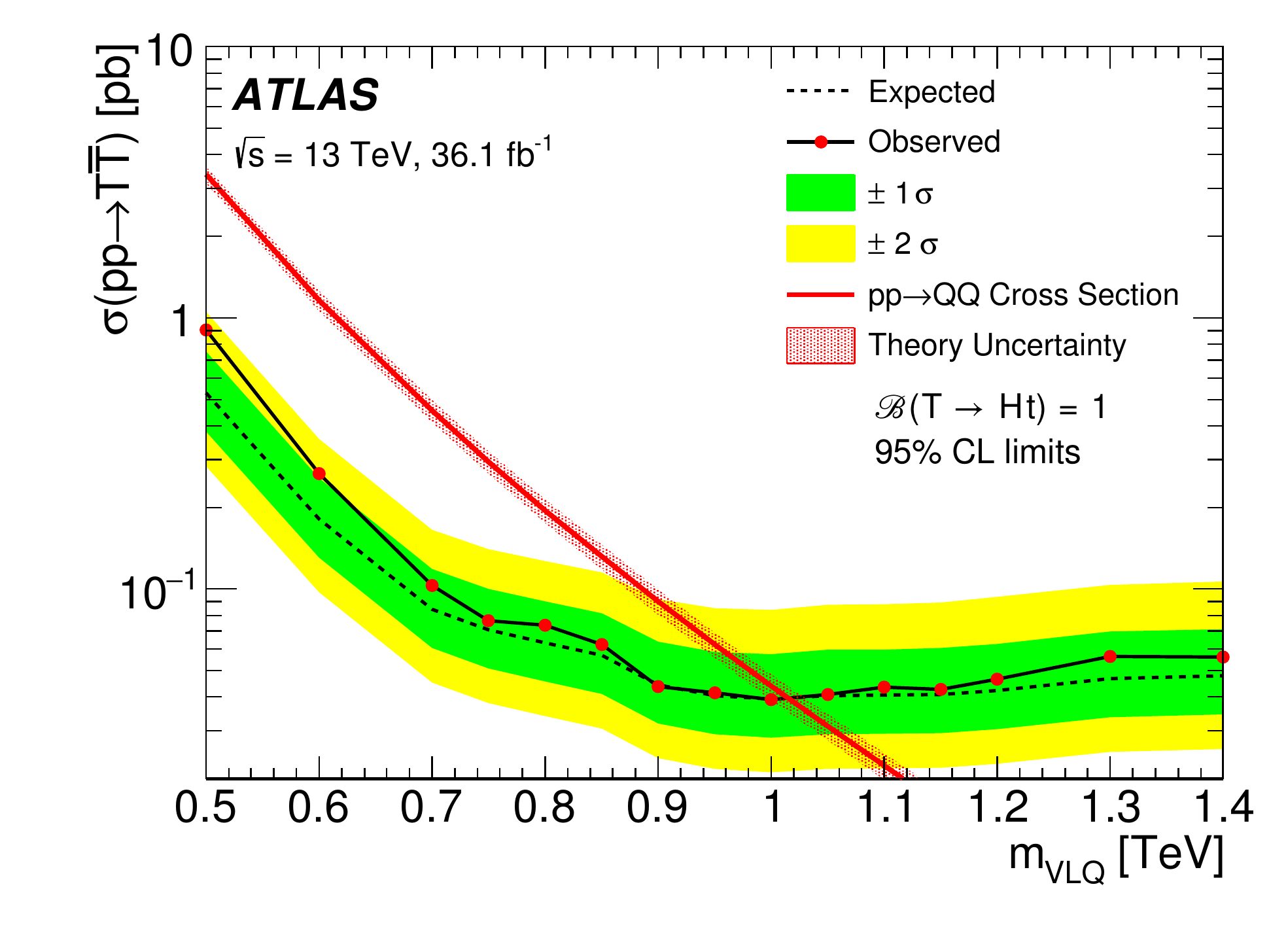}}
\subfloat[]{\includegraphics[width=0.4\textwidth]{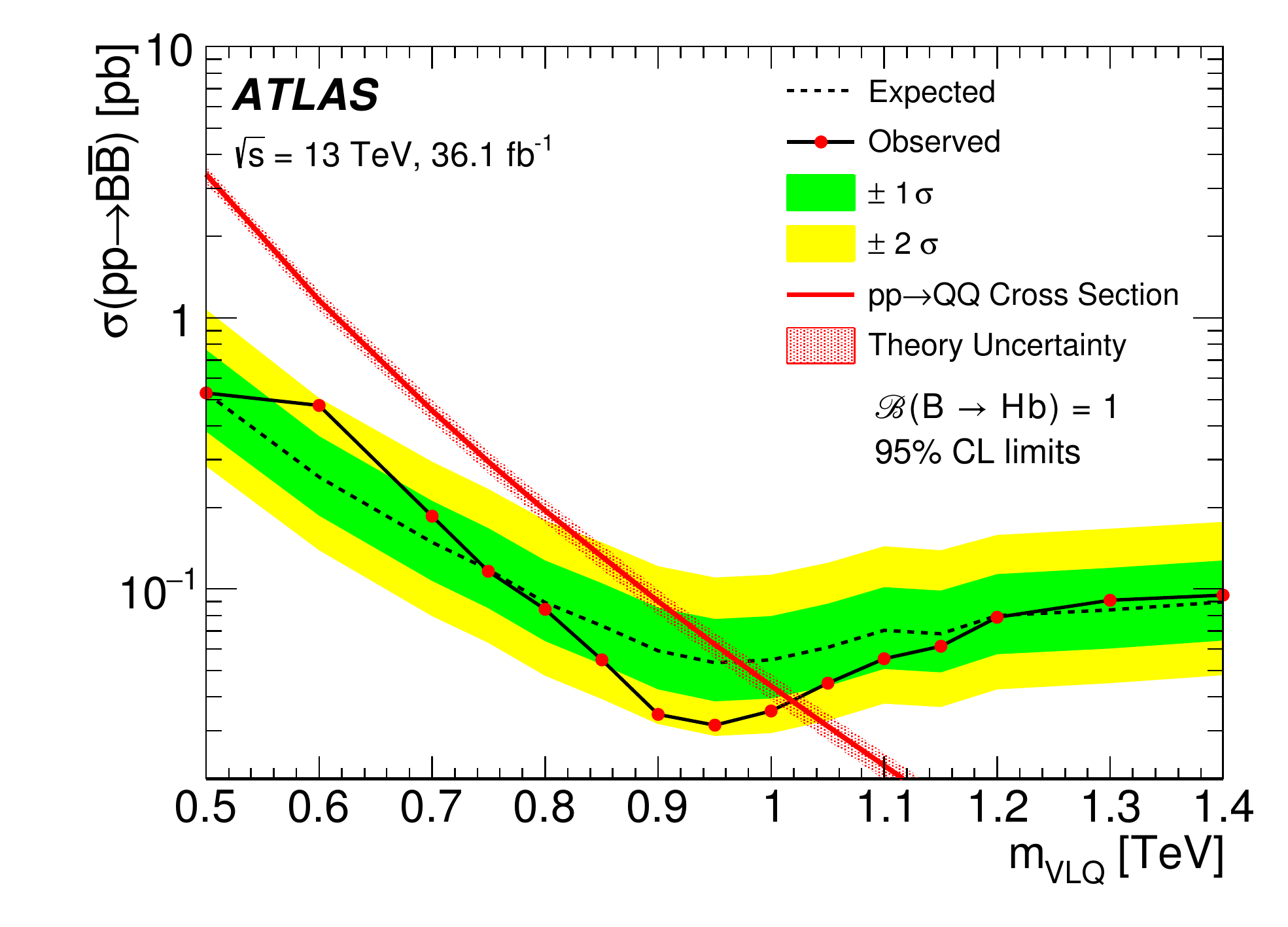}}\\
\subfloat[]{\includegraphics[width=0.4\textwidth]{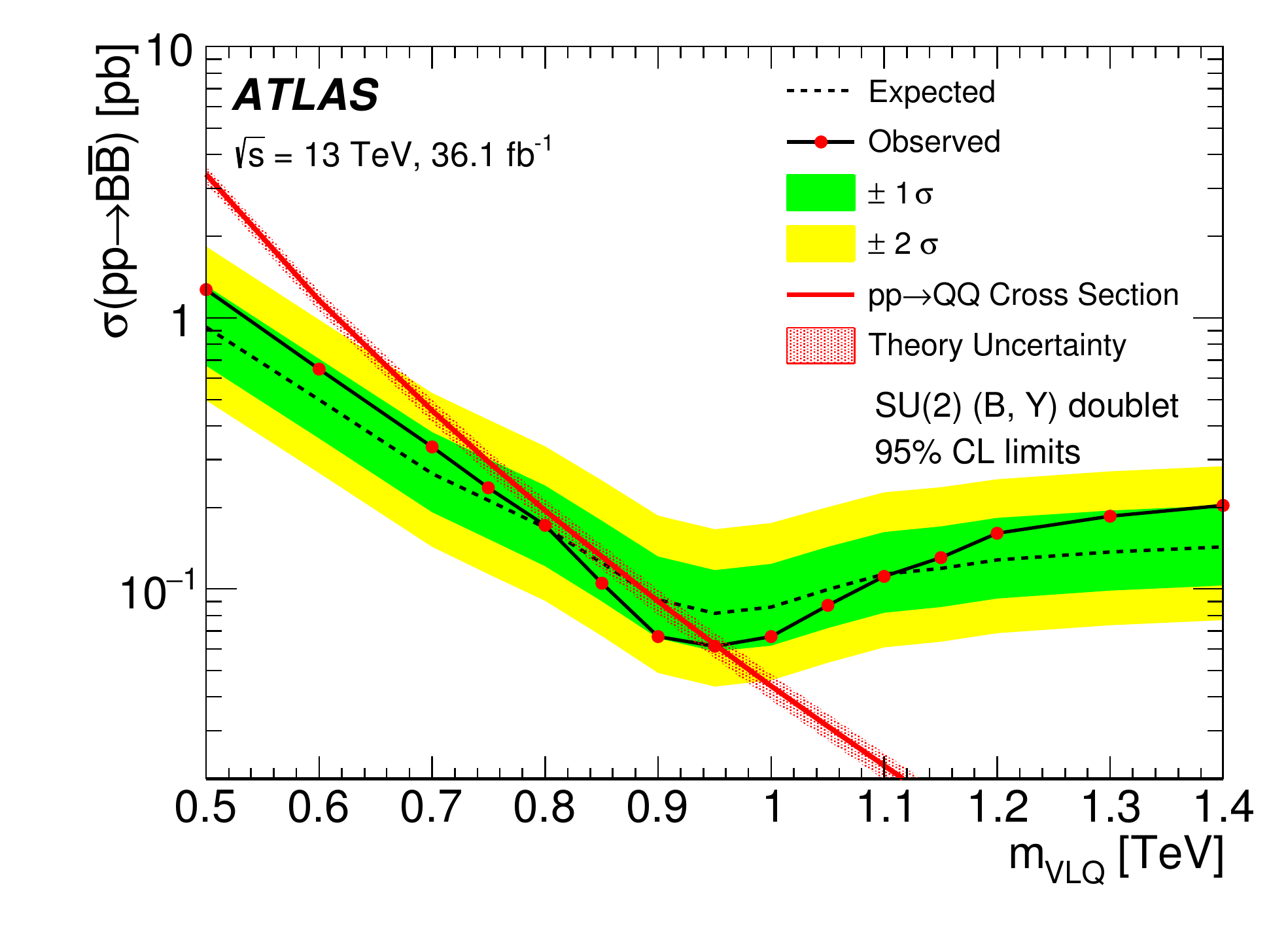}}\\
\caption{Expected upper limits at the 95\% CL on the (a) \TTbar{}
and (b) \BBbar{} cross section as a function of the VLQ mass
assuming \BR($T\rightarrow Ht)=1$ and \BR($B\rightarrow Hb)=1$,
respectively, as well as on (c) the \BBbar{} cross section with
the assumption of branching ratios consistent with a weak-isospin
doublet.  In the doublet case, only contributions from the $B$ VLQ
are considered, making the result conservative.  The green and
yellow bands correspond to $\pm$1 and $\pm$2 standard deviations
around the expected limit.  The thin red line and band show the
theoretical prediction and uncertainties, as described
in~\Sect{\ref{sec:Samples}}.}  \label{fig:results:LimitPure}
\end{figure}

\begin{figure}[!ht]
\centering
\subfloat[]{\includegraphics[width=0.4\textwidth]{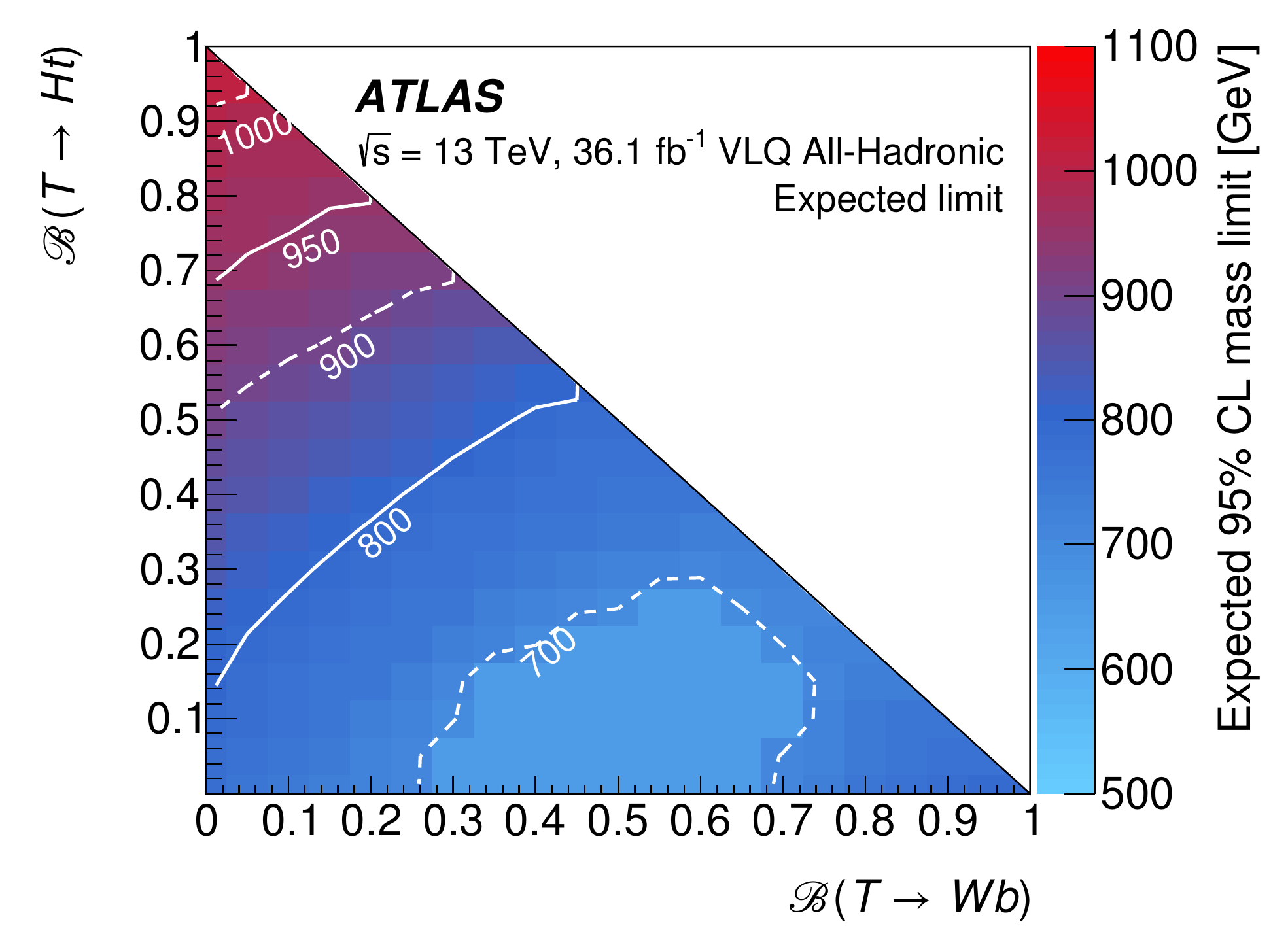}}
\subfloat[]{\includegraphics[width=0.4\textwidth]{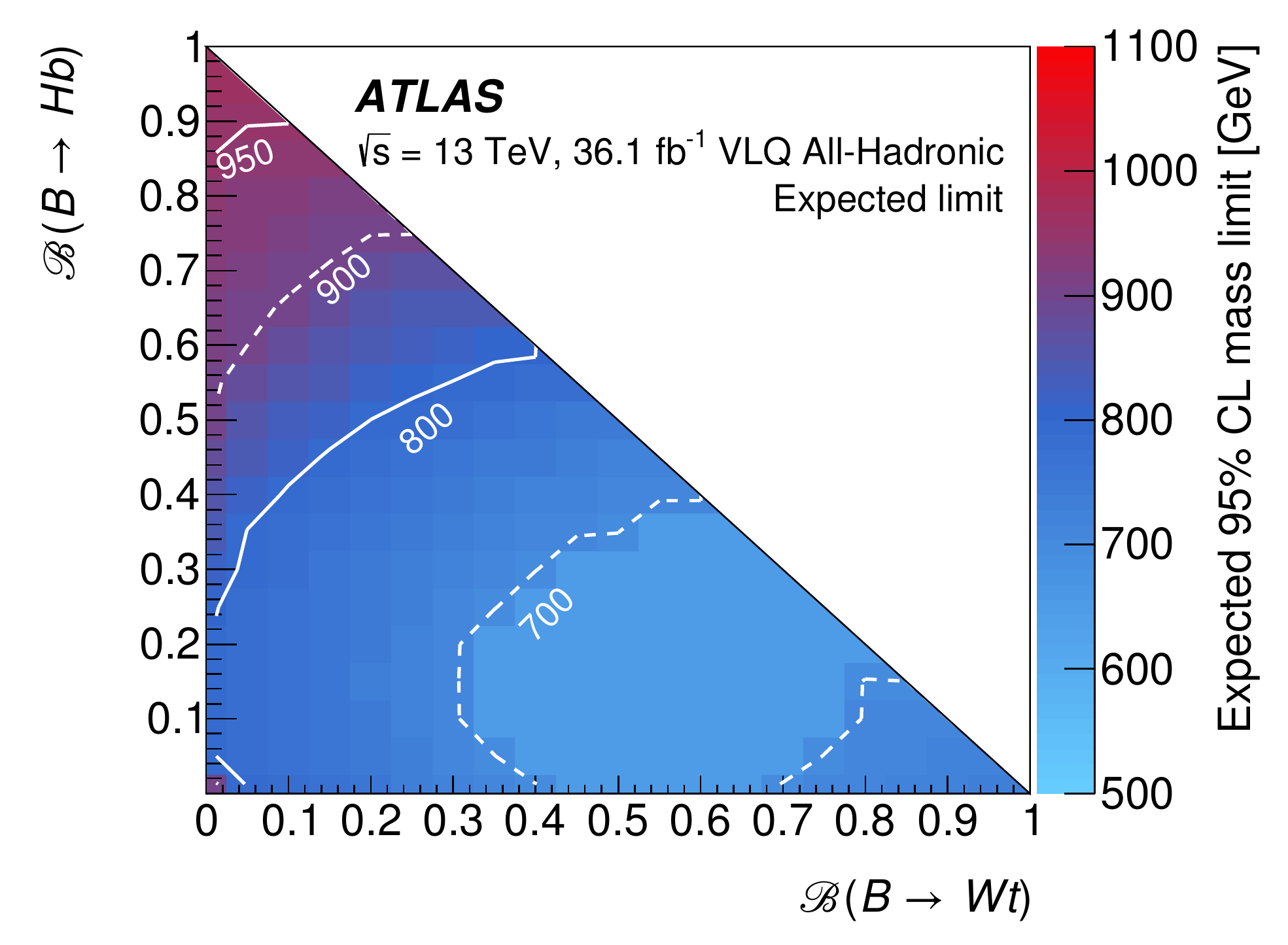}}\\
\subfloat[]{\includegraphics[width=0.4\textwidth]{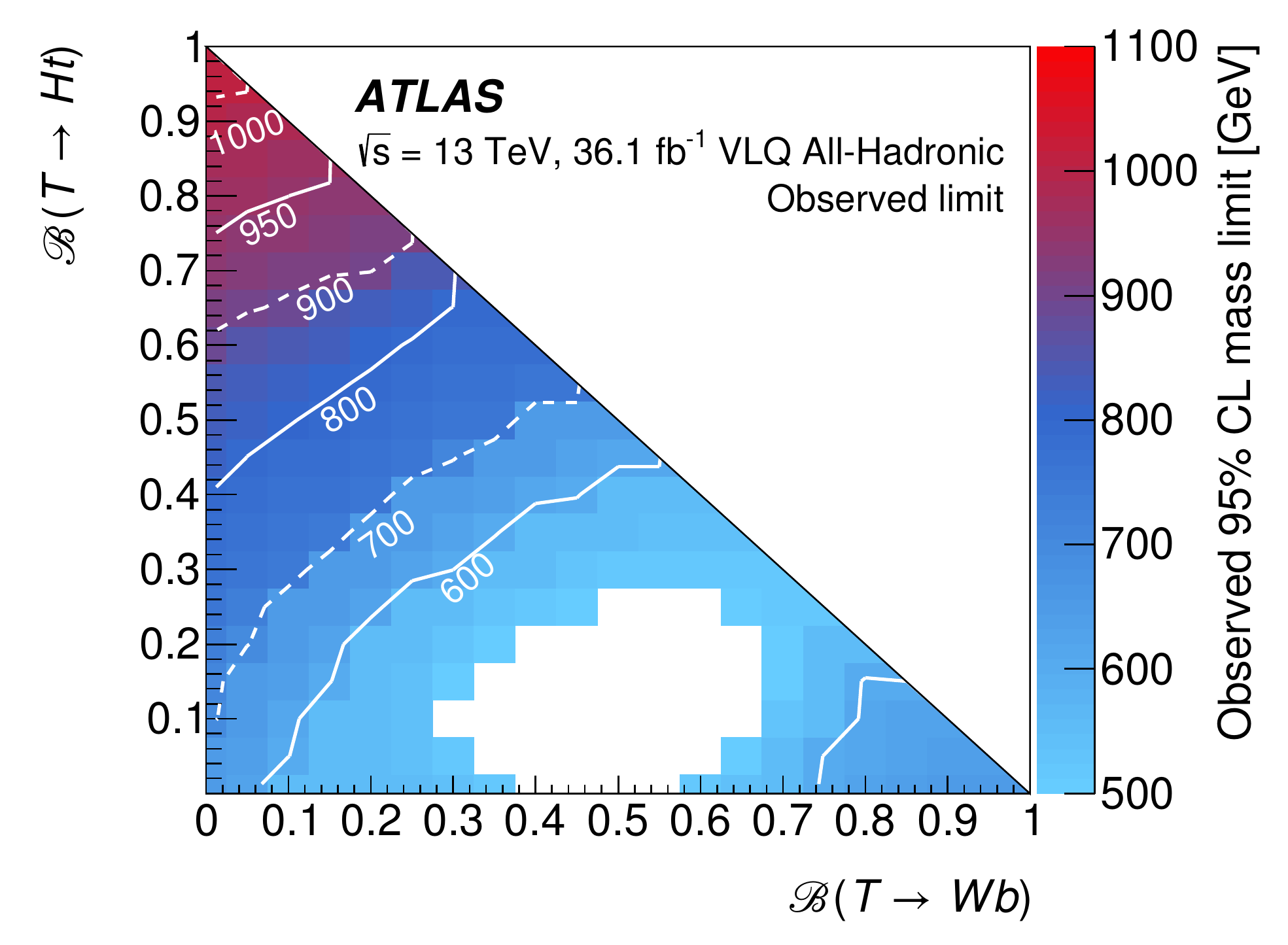}}
\subfloat[]{\includegraphics[width=0.4\textwidth]{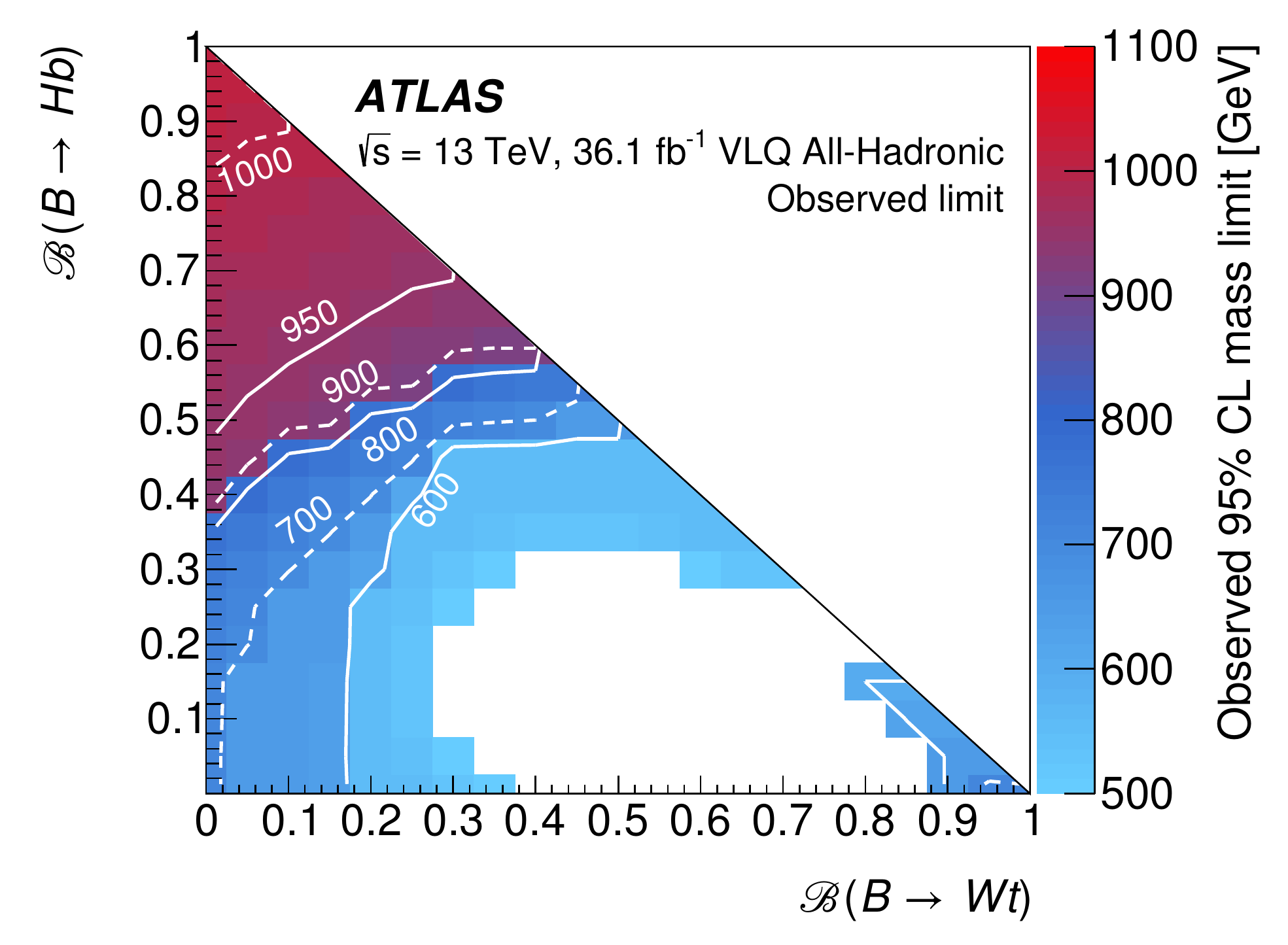}}\\
\caption{Expected and observed 95\% CL lower limits on (a,c) $m_T$
and (b,d) $m_B$ in the branching-ratio planes.  Contour lines,
alternating solid and dashed lines, are provided to show sensitivity
to different VLQ masses across the planes. Signal hypotheses are
considered in a mass range of 500--1400~\GeV, so the white space on
the observed limit figures corresponds to branching ratios where
there is no observed exclusion above a mass of 500~\GeV.}
\label{fig:results:LimitGrid}
\end{figure}
 
\afterpage{\clearpage}
 
\FloatBarrier
 
\section{Conclusion}
A search for pair production of vector-like quarks in the all-hadronic
final state is presented using \lumi{} of collision data
collected by the ATLAS detector at the LHC in 2015 and 2016. The analysis
selects events with high-\pt{} \smallR{} jets and multiple $b$-tags.
\SmallR{} jets are combined using a \vr{} clustering algorithm and then
classified with a neural network as a $V$-boson, Higgs-boson,
top-quark, or background jet.  A signal log-likelihood calculated via
the matrix element method is used as the final discriminant across
multiple categories based on the number of $V/H$-tags, top-tags, and
$b$-tags.  The analysis targets all third-generation decays of VLQs,
but it is particularly powerful for the $B\rightarrow Hb$ decay mode,
which is difficult to probe with leptonic final states. The observed
data are consistent with expected background and a 95\% CL limit is
placed on VLQ pair production as a function of the hypothetical VLQ
mass. The observed (expected) mass exclusion limit for a weak-isospin
($B, Y$) doublet $B$ is 950 (890)~\GeV, and the mass exclusion limits
for the pure decays $B\rightarrow Hb$ and $T\rightarrow Ht$, where
these results are strongest, are 1010 (970)~\GeV \;and 1010
(1010)~\GeV, respectively. Additionally, limits are placed across a
two-dimensional plane of branching ratio values of $Hb$ ($Ht$)
vs. $Wt$ ($Wb$) for $B$ ($T$) vector-like quarks.
 
\section*{Acknowledgments}
 
 
We thank CERN for the very successful operation of the LHC, as well as the
support staff from our institutions without whom ATLAS could not be
operated efficiently.
 
We acknowledge the support of ANPCyT, Argentina; YerPhI, Armenia; ARC, Australia; BMWFW and FWF, Austria; ANAS, Azerbaijan; SSTC, Belarus; CNPq and FAPESP, Brazil; NSERC, NRC and CFI, Canada; CERN; CONICYT, Chile; CAS, MOST and NSFC, China; COLCIENCIAS, Colombia; MSMT CR, MPO CR and VSC CR, Czech Republic; DNRF and DNSRC, Denmark; IN2P3-CNRS, CEA-DRF/IRFU, France; SRNSFG, Georgia; BMBF, HGF, and MPG, Germany; GSRT, Greece; RGC, Hong Kong SAR, China; ISF and Benoziyo Center, Israel; INFN, Italy; MEXT and JSPS, Japan; CNRST, Morocco; NWO, Netherlands; RCN, Norway; MNiSW and NCN, Poland; FCT, Portugal; MNE/IFA, Romania; MES of Russia and NRC KI, Russian Federation; JINR; MESTD, Serbia; MSSR, Slovakia; ARRS and MIZ\v{S}, Slovenia; DST/NRF, South Africa; MINECO, Spain; SRC and Wallenberg Foundation, Sweden; SERI, SNSF and Cantons of Bern and Geneva, Switzerland; MOST, Taiwan; TAEK, Turkey; STFC, United Kingdom; DOE and NSF, United States of America. In addition, individual groups and members have received support from BCKDF, the Canada Council, CANARIE, CRC, Compute Canada, FQRNT, and the Ontario Innovation Trust, Canada; EPLANET, ERC, ERDF, FP7, Horizon 2020 and Marie Sk{\l}odowska-Curie Actions, European Union; Investissements d'Avenir Labex and Idex, ANR, R{\'e}gion Auvergne and Fondation Partager le Savoir, France; DFG and AvH Foundation, Germany; Herakleitos, Thales and Aristeia programmes co-financed by EU-ESF and the Greek NSRF; BSF, GIF and Minerva, Israel; BRF, Norway; CERCA Programme Generalitat de Catalunya, Generalitat Valenciana, Spain; the Royal Society and Leverhulme Trust, United Kingdom.
 
The crucial computing support from all WLCG partners is acknowledged gratefully, in particular from CERN, the ATLAS Tier-1 facilities at TRIUMF (Canada), NDGF (Denmark, Norway, Sweden), CC-IN2P3 (France), KIT/GridKA (Germany), INFN-CNAF (Italy), NL-T1 (Netherlands), PIC (Spain), ASGC (Taiwan), RAL (UK) and BNL (USA), the Tier-2 facilities worldwide and large non-WLCG resource providers. Major contributors of computing resources are listed in Ref.~\cite{ATL-GEN-PUB-2016-002}.
 

\printbibliography

\clearpage
 
\clearpage
 
\begin{flushleft}
{\Large The ATLAS Collaboration}

\bigskip

M.~Aaboud$^\textrm{\scriptsize 34d}$,    
G.~Aad$^\textrm{\scriptsize 99}$,    
B.~Abbott$^\textrm{\scriptsize 124}$,    
O.~Abdinov$^\textrm{\scriptsize 13,*}$,    
B.~Abeloos$^\textrm{\scriptsize 128}$,    
D.K.~Abhayasinghe$^\textrm{\scriptsize 91}$,    
S.H.~Abidi$^\textrm{\scriptsize 164}$,    
O.S.~AbouZeid$^\textrm{\scriptsize 143}$,    
N.L.~Abraham$^\textrm{\scriptsize 153}$,    
H.~Abramowicz$^\textrm{\scriptsize 158}$
% ...
\bigskip
\\

% ...
$^{13}$Institute of Physics, Azerbaijan Academy of Sciences, Baku; Azerbaijan.\\
% ...
$^{34}$$^{(a)}$Facult\'e des Sciences Ain Chock, R\'eseau Universitaire de Physique des Hautes Energies - Universit\'e Hassan II, Casablanca;$^{(b)}$Centre National de l'Energie des Sciences Techniques Nucleaires (CNESTEN), Rabat;$^{(c)}$Facult\'e des Sciences Semlalia, Universit\'e Cadi Ayyad, LPHEA-Marrakech;$^{(d)}$Facult\'e des Sciences, Universit\'e Mohamed Premier and LPTPM, Oujda;$^{(e)}$Facult\'e des sciences, Universit\'e Mohammed V, Rabat; Morocco.\\
% ...
$^{91}$Department of Physics, Royal Holloway University of London, Egham; United Kingdom.\\
% ...
$^{99}$CPPM, Aix-Marseille Universit\'e, CNRS/IN2P3, Marseille; France.\\
% ...
$^{124}$Homer L. Dodge Department of Physics and Astronomy, University of Oklahoma, Norman OK; United States of America.\\
% ...
$^{128}$LAL, Universit\'e Paris-Sud, CNRS/IN2P3, Universit\'e Paris-Saclay, Orsay; France.\\
% ...
$^{143}$Santa Cruz Institute for Particle Physics, University of California Santa Cruz, Santa Cruz CA; United States of America.\\
% ...
$^{153}$Department of Physics and Astronomy, University of Sussex, Brighton; United Kingdom.\\
% ...
$^{158}$Raymond and Beverly Sackler School of Physics and Astronomy, Tel Aviv University, Tel Aviv; Israel.\\
% ...
$^{164}$Department of Physics, University of Toronto, Toronto ON; Canada.\\
% ...
$^{a}$ Also at  Department of Physics, University of Malaya, Kuala Lumpur; Malaysia.\\
$^{b}$ Also at Borough of Manhattan Community College, City University of New York, NY; United States of America.\\
$^{c}$ Also at Centre for High Performance Computing, CSIR Campus, Rosebank, Cape Town; South Africa.\\
$^{d}$ Also at CERN, Geneva; Switzerland.\\
$^{e}$ Also at CPPM, Aix-Marseille Universit\'e, CNRS/IN2P3, Marseille; France.\\
% ...
$^{*}$ Deceased

\end{flushleft}


\end{document}